\documentclass[11pt]{article}

\usepackage{a4wide}
\usepackage{multirow}
\usepackage{amssymb, bbm, bm}
\usepackage{amsthm}
\usepackage{amsmath}
\usepackage{epsfig}
\usepackage[round,authoryear]{natbib}
\usepackage[hyphens]{url}
\usepackage{hyperref}
\usepackage{esint}
\usepackage{subcaption}
\usepackage{arydshln}

\theoremstyle{definition}
\newtheorem{theorem}{Theorem}

\newtheorem{proposition}[theorem]{Proposition}
\newtheorem{remark}[theorem]{Remark}

\newtheorem{example}{Example}

\newcommand{\E}{\mathbb E}
\newcommand{\e}{\mathrm e}

\newcommand{\D}{\mathrm{d}}

\newcommand{\argmax}{\mathrm{arg max}}

\begin{document}
\title{Systemic Risk in Market Microstructure of Crude Oil and Gasoline Futures Prices: A Hawkes Flocking Model Approach}

\author{Hyun Jin Jang\thanks{School of Business Administration, Ulsan National Institute of Science and Technology (UNIST), Ulsan, Republic of Korea} \and
	Kiseop Lee\thanks{Department of Statistics, Purdue University, West Lafayette, Indiana, USA}
	\and
	Kyungsub Lee\thanks{Corresponding author. Department of Statistics, Yeungnam University, Gyeongsan, Republic of Korea (Email:ksublee@yu.ac.kr, Tel:+82-53-810-2324, Fax:+82-53-810-2036)}
}
\date{}

\maketitle

\begin{abstract}
	
	We propose the Hawkes flocking model that assesses systemic risk in high-frequency processes at the two perspectives -- endogeneity and interactivity.
	We examine the futures markets of WTI crude oil and gasoline for the past decade, and perform a comparative analysis with conditional value-at-risk as a benchmark measure.
	In terms of high-frequency structure, we derive the empirical findings.
	The endogenous systemic risk in WTI was significantly higher than that in gasoline, and the level at which gasoline affects WTI was constantly higher than in the opposite case. 
	Moreover, although the relative influence's degree was asymmetric, its difference has gradually reduced.
\end{abstract}

$\mathbf{keyword}$:
Systemic risk;
Hawkes process;
flocking;
WTI crude oil futures;
gasoline futures;
calibration;
branching ratio;
CoVaR

$\mathbf{JEL}$: G13, C13

\section{Introduction}
Over the past two decades, the systemic risk level has increased in financial markets due to the growth of securitization, hedge fund markets, and increase in intraday trading.
Recently, the emergence of innovative technologies has accelerated the paradigm shift of trading activities in financial markets. 
Traditional trading platforms such as phone conversations or clicks on a screen by humans has moved to automated trading by computers based on the ultra-low latency electronic system. 
The increased trading speed enables execution of orders within microseconds by the use of sophisticated algorithms; this is called high-frequency trading.
According to a report of the Commodity Futures Trading Commission (CFTC)\footnote{CFTC, ``Remarks of Chairman Timothy Massad before the Conference on the Evolving Structure of the US Treasury Market,'' October 21, 2015, at \url{http://www.cftc.gov/PressRoom/SpeechesTestimony/opamassad-30}.}, the volume of high-frequency trading in futures markets has grown remarkably over the past decade. 
It accounts for 80\% of foreign exchange futures, 67\% of interest rate futures, 62\% of equity futures, and 47\% of metals and energy futures trading volume.
In addition, flash crash events frequently occur in security markets which are attributed to high-frequency trading\footnote{For example, the DJIA index plunged roughly 1,100 points in the first five minutes of trading on August 24, 2015.}.
Such an environmental change in trading potentially allows large price movements within a short period of time as well as the rapid risk propagation to different assets, as mentioned in \cite{Miller&Shorter2016}.

In this context of high-frequency finance, we develop a novel Hawkes process-based model to examine the level of systemic risk that exists within and between price dynamics at the microscopic level.
The proposed model allows capturing contagious and clustered phenomena that can be investigated in the excessive volatile and correlated markets.
Studies related to systemic risk in high-frequency trading are discussed under various aspects.
\cite{Filimonov&Sornette2012} conduct event studies to investigate the changes in the systemic risk before and after the announcements of two extreme events: downgrading of Greece/Portugal and the flash crash event for the E-mini S\&P500 futures in 2010.
\cite{Hardimanetal2013} perform a similar analysis with \cite{Filimonov&Sornette2012} by taking power-law kernels. 
\cite{Chavez-Demoulin&McGill2012} compute intraday value-at-risk (VaR) for stocks in New York Stock Exchange (NYSE) using a peak-over-threshold model, and \cite{Jainetal2016} assess the extent to which a high-frequency system increases systemic risk in the Tokyo Stock Exchange.
\cite{Bormettietal2015} use a multivariate Hawkes process with a common factor that controls a large number of jumps in the transaction movement.
\cite{Calcagnileetal2018} compute the number of co-jumps occurring in Russell 3000 index stocks to measure the frequency of the collective instability at high-frequency.

On the other hand, there is little discussion on the increased systemic risk in energy markets associated with high-frequency trading. 
However, energy futures markets are no longer exceptional on this matter.
As noted in the beginning, almost the half of the trading volume in the energy markets is raised from high-frequency trading. Moreover, 
the CFTC examined how frequently flash events have occurred in the top-five most active futures contracts in 2015, that is, corn, gold, West Texas Intermediate (WTI) crude oil, E-mini S\&P 500 futures, and Euro FX\footnote{In this research, the flash crash is defined by the episodes in which a contract price moved at least 2\% within an hour, but returned to within 0.75\% of the original or starting price within that same hour.}. 
Among them, surprisingly, more than 35 similar intraday flash have occurred just for WTI crude oil futures\footnote{\url{https://www.cftc.gov/sites/default/files/idc/groups/public/@newsroom/documents/file/hourlyflashevents102115.pdf}}.
This result implies that WTI crude oil futures are utilized actively as instruments of algorithmic trading strategies.

In this study, we attempt to discover empirical evidences of the systemic risk level in the dynamics of the two futures prices of WTI crude oil and gasoline observed at the intraday transaction level over the past decade. 
The gasoline futures contracts are being traded most actively in the New York Mercantile Exchange (NYMEX) in the energy sector, following the WTI crude oil futures.
We consider two kinds of definitions for systemic risk in the market microstructure with the instability perspective.
The first view is the degree of instability that exists within a price process.
It is regarded as the term {\it endogeneity}, which is introduced in the earlier literature (e.g.,\citealp{Danielssonetal2012}; \citealp{Filimonov&Sornette2012}; \citealp{Hardimanetal2013}\footnote{In~\cite{Filimonov&Sornette2012} and \cite{Hardimanetal2013}, this is referred to ``reflexivity'' instead of endogeneity}). 
By estimating this level, we examine whether the trend of price decline leads to additional price decreases (or price rebounds). 
The second view is the degree of instability that exists between price processes caused by {\it interaction} between two different markets. 
In that point of view, we investigate how the change in one price affects to the change in the other price, and vice versa. In addition, we observe how micro-movements of prices in the two markets are likely to close to each other when the price difference widens or narrows.

Meanwhile, WTI crude oil and gasoline futures prices have maintained a strong dependence for a long time (\citealp{EIA2014}). 
From a macroeconomic perspective, the main causes of the price difference between crude oil and gasoline are refining costs and supply/demand balance of each product. 
Such comovement has been studied in terms of market cointegration in econometrics or flocking behavior.
When the two markets are co-integrated or have a flocking feature, the associated prices are closely correlated. Furthermore, one price could lead the other, while the reverse also occurs from time to time, or all prices in a system could follow the same behavior.

Cointegration refers to two or more non-stationary time series that are driven by one or more common non-stationary time series, proposed in the seminal works by \cite{Granger1981} and \cite{Engle&Granger1987}.
Many financial data series are known to exhibit the cointegration, 
for example, international stock markets (\citealp{Cerchi&Havenner1988}; \citealp{Taylor&Tonks1989}; \citealp{DuanPliska2004}),
foreign exchange rates (\citealp{Baillie&Bollerslev1989}; \citealp{Kellardetal2010}), 
futures and spot prices (\citealp{NgPirrong1994, NgPirrong1996}; \citealp{Maslyuka&Smyth2009}),
especially, crude oil, gasoline, and heating oil futures prices (\citealp{Serletis1992}; \citealp{Chiuetal2015}).
As a similiar manner, flocking is known to the collective motion of a large number of self-propelled entities. \cite{Reynolds1987} firstly proposed the breaking-through algorithm that makes it feasible to generate realistic computer simulation of flocking agents. 
The flocking behavior appears in many contexts of physics, biology, engineering, and human systems including financial markets
(\citealp{Rauchetal1995}; \citealp{Huepe&Aldana2008}; \citealp{HaKimLee2015}; \citealp{Fangetal2017}; among many others).
Even though comovement propensity in two or more dynamics has been discussed with different terms of cointegration or flocking in an amount of literature, at our knowledge there is no investigation of such tendency focusing on `microstructure' movements.

Based on the notions of systemic risk and comovement tendency in high-frequency markets, we propose a {\it Hawkes flocking model} that enables us to quantify systemic risk embedded in price structures at a microscopic level. 
This model addresses how to measure the extent of both endogeneity and interactivity.
By manipulating intensity processes that depend on the relative level of a couple of prices, the proposed model describes a feedback mechanism containing self/mutually-exciting features as well as the flocking behavior. 
Moreover, as a direct indicator of systemic risk, we formulate branching ratios, which are generally used in a Hawkes-based model to gauge how many additional jumps occur in the intensity process due to one exogenous event, and is employed for checking out the stability of the process. 
For the empirical analysis, we choose the nearest dated futures prices of WTI crude oil and gasoline that are collected at a transaction level in the time period of January 2007 to December 2016. 
During this period, prices plunged three times in both markets.

We also compute the systemic risk level by employing an existing methodology, that is, a conditional value-at-risk (CoVaR) approach, as a complementary measure of the proposed model. 
The concept of CoVaR is that the maximum loss can happen in an entity within a confidence level due to the effect of large loss from the other entity.
This was firstly proposed by \cite{Adrian&Brunnermeier2016} and generalized by \cite{Giulio&Ergun2013}. 
In this study, we adopt the CoVaR defined by \cite{Giulio&Ergun2013} and use 
a copula method to implement the CoVaR introduced by \cite{Reboredo&Ulgolini2015}. 
Then, we simulate the results from the CoVaR method and from the Hawkes flocking model and compare the evolution of systemic risk in the high-frequency markets of WTI crude oil and gasoline, which interplay actively. 

The main contribution of this study is twofold.
First, we develop a novel class of Hawkes-based model that assesses two types of systemic risk in high-frequency price processes: the endogenous systemic risk in a single process and interactive systemic risk in a couple of processes.
Second, we examine the existence of the systemic risk at a microscopic level via the futures markets of WTI crude oil and gasoline that are most liquid in the US energy sector.
Through the empirical test based on the proposed model, 
we obtain the following results. 
The overall systemic risk level that exists in the two futures markets was the highest just before the onset of the global credit crisis. 
For the past decade, the level of endogeneity in the WTI market was significantly higher than that in the gasoline market. 
In particular, the level at which gasoline price affects WTI price was steadily higher than in the opposite case. 
Although the two markets have been interactive, their relative influences, that is, from WTI to gasoline and vice versa, were very asymmetric, but the degree of the difference has been gradually reducing over the study period.

This paper is organized as follows.
In Section~\ref{Sec:Model}, we introduce the Hawkes flocking model and derive branching ratios to check the stability condition of the process.
Section~\ref{Sec:Test} presents the intraday transaction data for the two futures prices of WTI crude oil and gasoline from 2007 to 2016 along with estimate results under the proposed model using the maximum likelihood (ML) method.
Section~\ref{Sec:Comparison} presents a comparative analysis between the branching ratios of the proposed model and the CoVaR measure.
Section~\ref{Sec:Conclusion} concludes the paper, and
technical proofs and additional figures/tables are presented in the Appendix.

\section{The Hawkes Flocking Model}\label{Sec:Model}

In this section, we develop the Hawkes flocking model. The proposed model can be categorized to generalized Hawkes processes, which is introduced in the serial papers (\citealp{Hawkes1}; \citealp{Hawkes2}; \citealp{Hawkes&Oakes}).
As the Hawkes process is based on a class of multivariate counting processes, this model can account for the interaction of various types of Poisson-like events through its intensity process.

Because of their great flexibility and versatility, Hawkes-based models are very popular for modeling high-frequency finance. 
As a pioneering work, \cite{Bowsher2007} introduce a bivariate Hawkes processes to model the joint dynamics of trades and mid-price changes in NYSE stocks.
After that, many studies related to high-frequency finance have employed Hawkes processes (\citealp{Bacryetal2012}; \citealp{Bacryetal2013}; \citealp{DaFonseca&Zaatour2014}; \citealp{Bacry2015hawkes}; \citealp{Lee&Seo2017}).

\subsection{Model specification}
As noted in the introduction,
the model we propose captures the interaction between two highly correlated processes observed at the level of transaction data.
Consider the bivariate price processes that are defined by the differences between the two counting processes:
\begin{align*}
C_1 (t) &= \delta_1 (N_1^u (t) - N_1^d (t)) \\
C_2 (t) &= \delta_2 (N_2^u (t) - N_2^d (t))
\end{align*}
where processes $N_i^u(t)$ and $N_i^d(t)$ count the number of events for upward and downward movement in price $C_i(t)$ up to time $t$, respectively, and $\delta_i$ are tick sizes.

We present a system of the counting process $\bm{N}$ and its intensity process $\bm{\lambda}$ by employing the following matrix form 
$$ 
\bm{N}_t =
\begin{bmatrix} 
N_1^{u}(t) \\
N_1^{d}(t) \\
N_2^{u}(t) \\
N_2^{d}(t) 
\end{bmatrix}, \quad
\bm{\lambda}_t = 
\begin{bmatrix} 
\lambda_1^{u}(t) \\
\lambda_1^{d}(t) \\
\lambda_2^{u}(t) \\
\lambda_2^{d}(t) 
\end{bmatrix}
$$ 
where intensity process $\bm{\lambda}$ represents the expected number of arrivals of counting events over an infinitesimal time interval $\D t$ divided by $\D t$.
Let the intensity process be
\begin{equation}\label{Eq:excited}
\bm{\lambda}_t = \bm{\mu} + \int_{-\infty}^{t} \bm{h}(t-u) \D \bm{N}_u
\end{equation}
where $\bm{\mu} = [\mu_1, \mu_1, \mu_2, \mu_2]^{\top}$ is a constant vector, and $\bm{h}$ is a four-by-four matrix. 
Here, the vector $\bm{\mu}$ is called base intensity that accounts for the average frequency of exogenous events coming into this system, which is independent of the other asset's movement as well as its past movement. 
Parameters $\mu_1$ and $\mu_2$ are interpreted as the pressure of orders for buying or selling at price $C_1$ and $C_2$, respectively.
The matrix $\bm{h}$ is called a feedback kernel of the Hawkes process that decides the weight to be attributed to events $\D \bm{N}$ occurring at lag $u$ in the past.

We set the feedback kernel by
\begin{equation}\label{Eq:kernal}
\bm{h}(t-u) = \Phi(t-u) + {k}(t) \circ \Psi(t-u) 
\end{equation}
where $\Phi, {k}$, and $\Psi$ are four-by-four matrices, and ``$\circ$'' denotes the element-wise multiplication of matrices.
The kernel $\bm h$ consists of two components: 
\begin{enumerate}
\item[(i)] The matrix $\Phi$ controls the self/mutually-exciting patterns between the two prices, which is defined as 
\begin{equation}
\Phi(t) = \begin{bmatrix}
\alpha_{1s}\e^{-\beta_1 t} & \alpha_{1c}\e^{-\beta_1 t} & 0 & 0 \\
\alpha_{1c}\e^{-\beta_1 t} & \alpha_{1s}\e^{-\beta_1 t} & 0 & 0 \\
0 & 0 & \alpha_{2s}\e^{-\beta_2 t} & \alpha_{2c}\e^{-\beta_2 t} \\
0 & 0 & \alpha_{2c}\e^{-\beta_2 t} & \alpha_{2s}\e^{-\beta_2 t} 
\end{bmatrix}\label{Eq:phi}
\end{equation}
where non-negative constants $\alpha_{is}$ and $\alpha_{ic}$ denote the self/mutually-exciting terms, respectively,
and $\beta_i$ governs the speed of decay to the base intensity level of the $i$-th price process.

\item[(ii)] The matrix ${k} \circ \Psi$ controls the flocking phenomenon according to which two price movements interact, and where 
matrix ${k}$ is defined by an indicator function matrix such as
\begin{equation} 
{k}(t) = \begin{bmatrix}
\mathbbm{1}_{\{C_1(t) < C_2(t)\}} & \mathbbm{1}_{\{C_1(t) < C_2(t)\}} & \mathbbm{1}_{\{C_1(t) < C_2(t)\}} & \mathbbm{1}_{\{C_1(t) < C_2(t)\}} \\
\mathbbm{1}_{\{C_1(t) > C_2(t)\}} & \mathbbm{1}_{\{C_1(t) > C_2(t)\}} & \mathbbm{1}_{\{C_1(t) > C_2(t)\}} & \mathbbm{1}_{\{C_1(t) > C_2(t)\}} \\
\mathbbm{1}_{\{C_2(t) < C_1(t)\}} & \mathbbm{1}_{\{C_2(t) < C_1(t)\}} & \mathbbm{1}_{\{C_2(t) < C_1(t)\}} & \mathbbm{1}_{\{C_2(t) < C_1(t)\}} \\
\mathbbm{1}_{\{C_2(t) > C_1(t)\}} & \mathbbm{1}_{\{C_2(t) > C_1(t)\}} & \mathbbm{1}_{\{C_2(t) > C_1(t)\}} & \mathbbm{1}_{\{C_2(t) > C_1(t)\}} 
\end{bmatrix}, \label{eq:K}
\end{equation}
and the matrix $\Psi(t)$ is defined by
\begin{equation}
\Psi(t) = \begin{bmatrix}
0 & 0 & \alpha_{1w}\e^{-\beta_1 t} & \alpha_{1n}\e^{-\beta_1 t} \\
0 & 0 & \alpha_{1n}\e^{-\beta_1 t} & \alpha_{1w}\e^{-\beta_1 t} \\
\alpha_{2w}\e^{-\beta_2 t} & \alpha_{2n}\e^{-\beta_2 t} & 0 & 0 \\
\alpha_{2n}\e^{-\beta_2 t} & \alpha_{2w}\e^{-\beta_2 t} & 0 & 0 
\end{bmatrix}\label{eq:phi_nw}
\end{equation}
where non-negative constants $\alpha_{iw}$ and $\alpha_{in}$ denote the flocking exciting terms. 

\end{enumerate}

In part (i), the exponential decaying setup in the bivariate Hawkes process with symmetric $\alpha$'s shows a prototypical model for high-frequency finance. 
In part (ii), $k$ deals with the flocking phenomenon while considering an additional term under which intensity $\lambda_1^u$ increases only when $C_1$ is less than $C_2$
and $\lambda_1^d$ increases only when $C_1$ is larger than $C_2$.
$\Psi$ captures the degree of a flocking phenomenon.
The parameter $\alpha_{iw}$ or $\alpha_{in}$ is triggered depending on whether the difference between the two prices is narrowed ($n$) or widened ($w$).

\begin{figure}
\begin{center}
\includegraphics[width=1\textwidth]{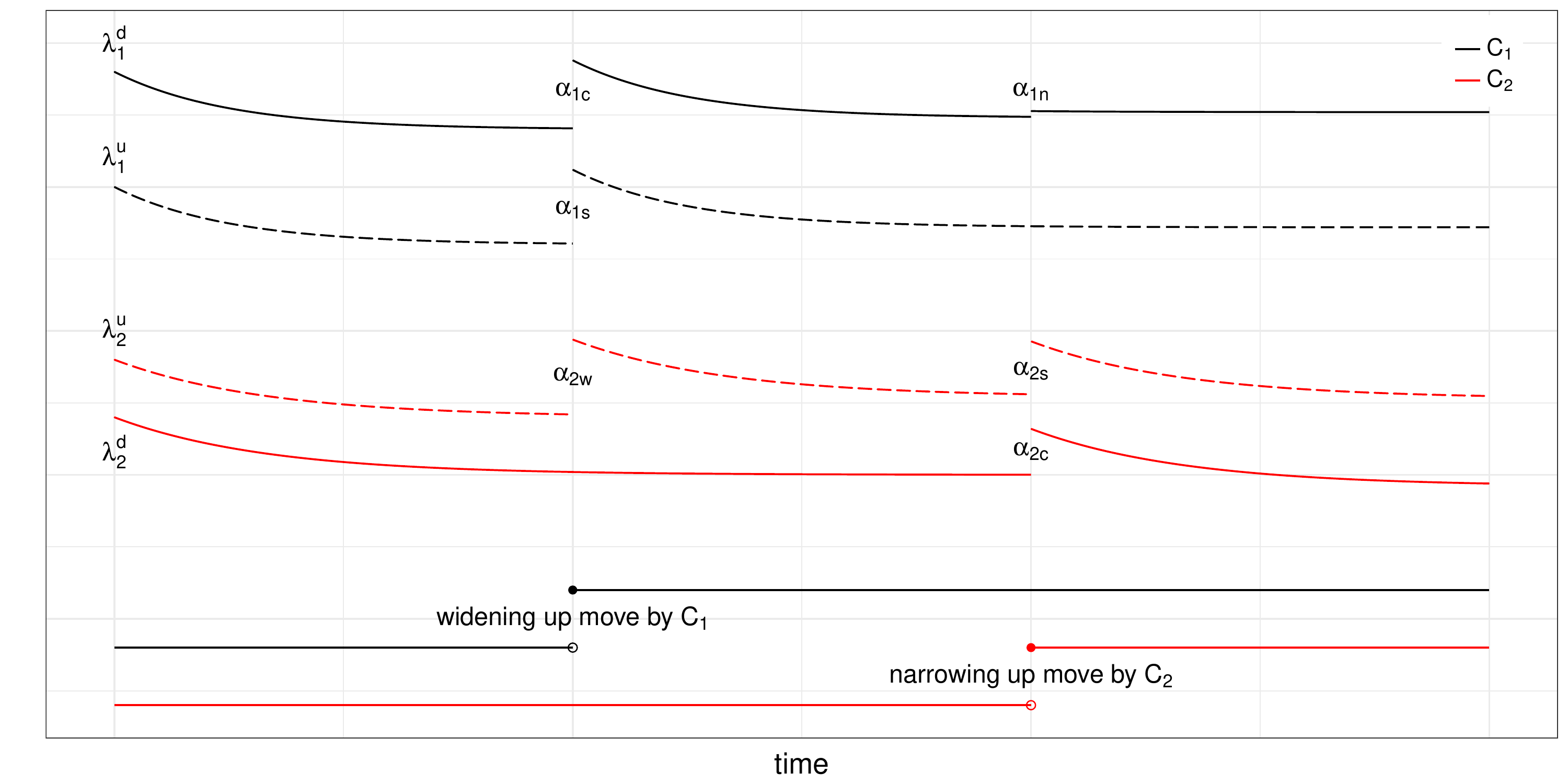}
\end{center}
\caption{Illustration of the idea on the Hawkes flocking model}
\label{Fig:explanation}
\end{figure}

Figure~\ref{Fig:explanation} shows a descriptive idea on the intensity movements in a Hawkes flocking model.
In this figure, the paths represent the dynamics of two prices and the associated four intensities\footnote{This is for an illustrative purpose and the actual values of prices and intensities may be different from the figure.}.
The black straight line is for price movement $C_1$. The black curved line and curved dashed line represent the intensities for upward and downward movement, respectively. 
Accordingly, the red lines are for price $C_2$ and its intensities. 

Assume $C_1 > C_2$.
Suppose that a widening upward jump of $C_1$ occurs. Then, three simultaneous jumps in intensities are instantly activated, $\lambda_1^u$, $\lambda_1^d$ and $\lambda_2^u$.
The jumps in $\lambda_1^u$ and $\lambda_1^d$ are due to self/mutually-exciting tendency in the individual Hawkes model and jump sizes are given by $\alpha_{1s}$ and $\alpha_{1c}$, respectively.
The jump in $\lambda_2^u$ is due to the flocking feature to accelerate a upward movement of $C_2$ resulting from the jump in $C_1$
and a jump size is given as $\alpha_{2w}$.

Later on, a narrowing upward movement of $C_2$ occurs. In a similar way, simultaneous jumps arrive in intensities, $\lambda_2^u$, $\lambda_2^u$ and $\lambda_1^d$.
The jumps in $\lambda_2^u$ and $\lambda_2^d$ are due to self/mutually-exciting propensity,
and the jump in $\lambda_1^d$ is due to the flocking mechanism attributed to a narrowing event.
Note that jump sizes of $\lambda_1^d$ and $\alpha_{1n}$, are intentionally expressed as quite small in the figure,
since the jumps in intensities due to the narrowing event are close to zero in the empirical analysis. 

\begin{remark}[The role of $k$]
We simply put $\Phi_{ij} = 0$ and $\Psi_{ij} = 1$ for all $i, j$.
Then
$$ \lambda_{1}^{u}(t) = \mu_1 + \int_{-\infty}^{t} \mathbbm{1}_{ \{C_1(u) < C_2(u)\} } \D \left( N_1^{u} + N_1^{d} + N_2^{u} + N_2^{d} \right)(u)$$
and this implies that the intensity of the up movement of $C_1$ changes (typically increases) when $C_1(t)$ is less than $C_2(t)$.

When $C_1(t) < C_2(t)$, the intensity of the up movement of $C_1$ increases, and $C_1$ and $C_2$ tend to be close to each other, indicating a flocking phenomenon.
With similar arguments, when $C_1(t) < C_2(t)$, 
the up movement of $C_1$ and down movement of $C_2$ tend to increase and
when $C_1(t) > C_2(t)$, 
the down movement of $C_1$ and up movement of $C_2$ tend to increase.
\end{remark}

\begin{remark}[Comparison between $\Phi$ and $\Psi$]
The matrices $\Phi$ and $\Psi$ have zero components in positions that do not overlap each other.
This ensures each role of the matrix is separated: 
$\Phi$ only affects the movements in the individual price process,
and $\Psi$ only controls the effect of the new information from another prices process.
Therefore, the test of the significance of $\Psi$ verifies the existence of interactions between the two prices, especially through the sign of $C_1 - C_2$.

Through combination with ${k}$, $\alpha_{in}$, and $\alpha_{iw}$ respectively represent the effects of the narrowing and widening events of price difference on intensities.
To see this, we simply put $\Phi_{ij} = 0$, and then
$$ \lambda_{1}^{u}(t) = \mu_1 + \int_{-\infty}^{t} \mathbbm{1}_{ \{C_1(u) < C_2(u)\} } \e^{-\beta_1 (t-u)} \left(\alpha_{1w} \D N_2^{u}(u) + \alpha_{1n} \D N_2^{d}(u)\right). $$
When $C_1(u) < C_2(u)$, the increase of $N_2^{u}(u)$ is the price difference widening event, and
the occurrence of $N_2^{d}$ is the price difference narrowing event.
\end{remark}

From the setup for the components in the kernel matrix $\bm{h}$, 
the proposed model can be expressed with a differential form using a Markov property.
Let the decay parameter for $C_1$ and $C_2$ be fixed as $\beta_1$ and $\beta_2$, respectively. 
Then, the intensity process satisfies the system of stochastic differential equations such as
$$ \D \bm \lambda_t = \bm \beta \circ (\bm \mu - \bm \lambda_t) \D t 
+ \bm \alpha \D \bm N_t $$
where $\bm \alpha$ and $\bm \beta$ are given as
\begin{align*}
\bm \alpha = \begin{bmatrix}
\alpha_{1s} & \alpha_{1c} & \mathbbm{1}_{\{C_1(t) < C_2(t)\}} \alpha_{1w} & \mathbbm{1}_{\{C_1(t) < C_2(t)\}} \alpha_{1n} \\
\alpha_{1c} & \alpha_{1s} & \mathbbm{1}_{\{C_1(t) > C_2(t)\}} \alpha_{1n} & \mathbbm{1}_{\{C_1(t) > C_2(t)\}} \alpha_{1w} \\
\mathbbm{1}_{\{C_2(t) < C_1(t)\}} \alpha_{2w} & \mathbbm{1}_{\{C_2(t) < C_1(t)\}} \alpha_{2n} & \alpha_{2s} & \alpha_{2c} \\
\mathbbm{1}_{\{C_2(t) > C_1(t)\}} \alpha_{2n} & \mathbbm{1}_{\{C_2(t) > C_1(t)\}} \alpha_{2w} & \alpha_{2c} & \alpha_{2s} 
\end{bmatrix}, \quad
\bm \beta = \begin{bmatrix} \beta_1 \\ \beta_1 \\ \beta_2 \\ \beta_2 \end{bmatrix}.
\end{align*}
This model can be understood as follows. Market orders (buy or sell) arrive with intensity $\bm \mu$, and the arrival intensity jumps by the amount of $\bm \alpha$ instantly when an arrival event occurs, and then it decays to the base intensity level $\bm \mu$ with the speed of $\bm \beta$. 

This model can be extended to the multi-dimensional case as the Remark~\ref{Rem:Ext} shows.
\begin{remark}\label{Rem:Ext}(The multi-dimensional Hawkes flocking model)
Consider $m$-dimensional price processes such that
\begin{equation}
C_1(t) = N_1^u(t) - N_1^d(t), \,\,\,\cdots,\,\,\, C_m(t) = N_m^u(t) - N_m^d(t),
\end{equation}
where processes $N_i^u(t)$ and $N_i^d(t)$ have intensity processes $\lambda_i^u(t)$ and $\lambda_i^d(t)$, respectively, for each $i=1,\cdots, m$.
For a system of $2m$-dimensional counting process $\bm{N}$ and its intensity process $\bm{\lambda}$ 
$$ 
\bm{N}_t =
\begin{bmatrix} 
N_1^{u}(t) \\
N_1^{d}(t) \\
\vdots \\
N_m^{u}(t) \\
N_m^{d}(t) 
\end{bmatrix}, \quad
\bm{\lambda}_t = 
\begin{bmatrix} 
\lambda_1^{u}(t) \\
\lambda_1^{d}(t) \\
\vdots \\
\lambda_m^{u}(t) \\
\lambda_m^{d}(t) 
\end{bmatrix}
$$ 
we can extend the intensity process in \eqref{Eq:excited} with the feedback kernel in \eqref{Eq:kernal} to the multi-dimensional version as follows.
\begin{enumerate}
\item[(i)] The matrix $\Phi$ in \eqref{Eq:phi} is given as the $2m$-by-$2m$ sized form, 
\begin{equation*}
\Phi(t) = \begin{bmatrix}
\alpha_{1s}\e^{-\beta_1 t} & \alpha_{1c}\e^{-\beta_1 t} & \cdots &0 & 0 \\
\alpha_{1c}\e^{-\beta_1 t} & \alpha_{1s}\e^{-\beta_1 t} & \cdots &0 & 0 \\
\cdots & \cdots & \cdots & \cdots & \cdots \\
0 & 0 & \cdots &\alpha_{ms}\e^{-\beta_m t} & \alpha_{mc}\e^{-\beta_m t} \\
0 & 0 & \cdots &\alpha_{mc}\e^{-\beta_m t} & \alpha_{ms}\e^{-\beta_m t} 
\end{bmatrix}
\end{equation*}

\item[(ii)] The matrix ${k}\circ\Psi$ in \eqref{eq:K} and \eqref{eq:phi_nw} is also given as the $2m$-by-$2m$ sized form such that
\begin{equation*} 
{k}(t) = \begin{bmatrix}
\mathbbm{1}_{\{C_1(t) < \bar{C}(t)\}} & \mathbbm{1}_{\{C_1(t) < \bar{C}(t)\}} & \cdots &\mathbbm{1}_{\{C_1(t) < \bar{C}(t)\}} & \mathbbm{1}_{\{C_1(t) < \bar{C}(t)\}} \\
\mathbbm{1}_{\{C_1(t) > \bar{C}(t)\}} & \mathbbm{1}_{\{C_1(t) > \bar{C}(t)\}} & \cdots &\mathbbm{1}_{\{C_1(t) > \bar{C}(t)\}} & \mathbbm{1}_{\{C_1(t) > \bar{C}(t)\}} \\
\cdots & \cdots & \cdots & \cdots & \cdots \\
\mathbbm{1}_{\{C_m(t) < \bar{C}(t)\}} & \mathbbm{1}_{\{C_m(t) < \bar{C}(t)\}} & \cdots &\mathbbm{1}_{\{C_m(t) < \bar{C}(t)\}} & \mathbbm{1}_{\{C_m(t) < \bar{C}(t)\}} \\
\mathbbm{1}_{\{C_m(t) > \bar{C}(t)\}} & \mathbbm{1}_{\{C_m(t) > \bar{C}(t)\}} & \cdots &\mathbbm{1}_{\{C_m(t) > \bar{C}(t)\}} & \mathbbm{1}_{\{C_m(t) > \bar{C}(t)\}} 
\end{bmatrix}, 
\end{equation*}
where $\bar{C}(t)$ is given as the average of $C_i(t)$'s such as $$\bar{C}(t) = \frac{C_1(t)+ \cdots + C_m(t)}{m},$$
and 
\begin{equation*}
\Psi(t) = \begin{bmatrix}
0 & 0 & \cdots & \alpha_{1w}\e^{-\beta_1 t} & \alpha_{1n}\e^{-\beta_1 t} \\
0 & 0 & \cdots & \alpha_{1n}\e^{-\beta_1 t} & \alpha_{1w}\e^{-\beta_1 t} \\
\cdots & \cdots & \cdots & \cdots & \cdots \\
\alpha_{mw}\e^{-\beta_m t} & \alpha_{mn}\e^{-\beta_m t} & \cdots & 0 & 0 \\
\alpha_{mn}\e^{-\beta_m t} & \alpha_{mw}\e^{-\beta_m t} & \cdots & 0 & 0 
\end{bmatrix} 
\end{equation*}

\end{enumerate} 

\end{remark}

\subsection{Stability condition}
The proposed process can be shown to be well defined and to admit a version with stationary increments under the condition holds of which all eigenvalues of the matrix $$\int_0^{\infty} \bm h(\tau) d\tau$$ are less than one 
(\citealp{Hawkes2}; \citealp{Bacryetal2013}). The matrix is called a branching matrix, which is used by, e.g. \cite{Filimonov&Sornette2012}, \cite{Hardimanetal2013}.
Such terminology is related to the ancestor-offspring argument. 
The immigrant ancestor arrives at the system exogenously at a basic intensity rate $\bm{\mu}$,
and this ancestor event produces internally offspring arrivals in the system with the intensity $h$ that relies on the ancestor's arrivals.
Both ancestor and offspring arrivals can increase likelihood of occurrence for additional events in the system.
Satisfying the stability condition indicates that each ancestor arrival generates ``less than one offspring event'' on average, and hence the process can be stable. Otherwise, the process can diverge to an infinite value within a finite time.

From this argument, we investigate stability of the Hawkes flocking process by computing spectral radius, which is defined by the largest absolute value among the eigenvalues of the branching matrix, and checking out that the spectral radius is less than one.
Since the branching matrix still depends on the original process $\bm N_t$, unlike the pure Hawkes model\footnote{In the Hawkes process, the feedback kernel is usually given as a deterministic function.}, the stability condition gets to have a stochastic form. 
To make the stability condition be feasible to implement in computation, we take account of approximation of the branching matrix by taking unconditional expectation on it. 
For the proposed model, we set a branching matrix $M$ with a component
\begin{equation*}
M_{i, j} = \int_0^\infty | \Phi_{ij}(t) + \E[k_{ij}(t)] \Psi_{ij}(t) | \D t,\,\,\, \text{for}\,\,\, 1\leq i, j \leq 4. 
\end{equation*}
It implies the expected level of the number of offspring events caused by one ancestor event arriving at the rate $\bm \mu$.

For the expectation $\E[k_{ij}(t)]$, we set the probabilities of $\{C_1(t) < C_2(t)\}$ and $\{C_1(t) > C_2(t)\}$ by $p\leq 0.5$ and $1-p$, respectively\footnote{This model does not consider occurrence for the case $\{C_1(t) = C_2(t)\}$.}. Thus, we have 
\begin{equation}\label{Eq:branchmatrix}
M= \begin{bmatrix}
\frac{\alpha_{1s}}{\beta_1} & \frac{\alpha_{1c}}{\beta_1} & p\frac{\alpha_{1w}}{\beta_1} & p\frac{\alpha_{1n}}{\beta_1} \\[6pt]
\frac{\alpha_{1c}}{\beta_1} & \frac{\alpha_{1s}}{\beta_1} & (1-p)\frac{\alpha_{1n}}{\beta_1} & (1-p)\frac{\alpha_{1w}}{\beta_1} \\[6pt]
(1-p)\frac{\alpha_{2w}}{\beta_2} & (1-p)\frac{\alpha_{2n}}{\beta_2} & \frac{\alpha_{2s}}{\beta_2} & \frac{\alpha_{2c}}{\beta_2} \\[6pt]
p\frac{\alpha_{2n}}{\beta_2} & p\frac{\alpha_{2w}}{\beta_2} & \frac{\alpha_{2c}}{\beta_2} & \frac{\alpha_{2s}}{\beta_2}
\end{bmatrix}.
\end{equation}
Since it is empirically shown that $p$ is quite close to 0.5, we may assume that both probabilities
are identical.
Under the setup, we obtain the spectral radius of the branching matrix $M$ such that
\begin{equation} \label{Eq:spectral}
\rho_{M} = \frac{1}{2}\left(a+ \sqrt{a^2 + 4(b-c)}\right),
\end{equation}
where $a, b,$ and $c$ are given by 
\begin{align*}
a &= \frac{\alpha_{1s} + \alpha_{1c}}{\beta_1} + \frac{\alpha_{2s} + \alpha_{2c}}{\beta_2}, \\
b &= \frac{\alpha_{1n}\alpha_{2n} + \alpha_{1n}\alpha_{2w} + \alpha_{1w} \alpha_{2n} + \alpha_{1w}\alpha_{2w}}{4\beta_1\beta_2}, \\
c &= \frac{\alpha_{1s}\alpha_{2s} + \alpha_{1s}\alpha_{2c} + \alpha_{1c}\alpha_{2s} + \alpha_{1c}\alpha_{2c}}{\beta_1\beta_2},
\end{align*}
and hereafter we call $\rho_M$ by a branching ratio.

The branching ratio $\rho_M$ measures how fast the increased feedback kernel due to arrivals generated from internal and external sources is shrinking as time goes. 
If it shrinks quickly, it infers that instability in microscopic price dynamics stays low. 
Oppositely, if it does shrink reluctantly, it may infer that instability stays high. 
In this context, we may find a relevance of the branching ratio with a systemic risk level in price dynamics.
More discussion on the branching ratio with a systemic risk indicator will be in Section~\ref{Sec:Relevance}.

To summarize, the Hawkes flocking model has the following characteristics.
Each price process has a self/mutually-exciting terms that are affected by its original changes in tick price dynamics.
In addition, both prices incorporate a flocking feature describing as propensity of two dynamics moving together.
Such phenomenon emerges when narrowing and widening events between two prices occur.
We obtain a direct mapping between an original process $\bm N_t$ and a feedback kernel with a branching matrix $M$, in which a main event occurs exogenously with basic intensity $\bm \mu$ and may give rise to additional following events endogenously with intensity $M$ on average. 

\section{Application to Empirical Data}\label{Sec:Test}

\begin{figure}[h]
\begin{center}
\includegraphics[width=\textwidth]{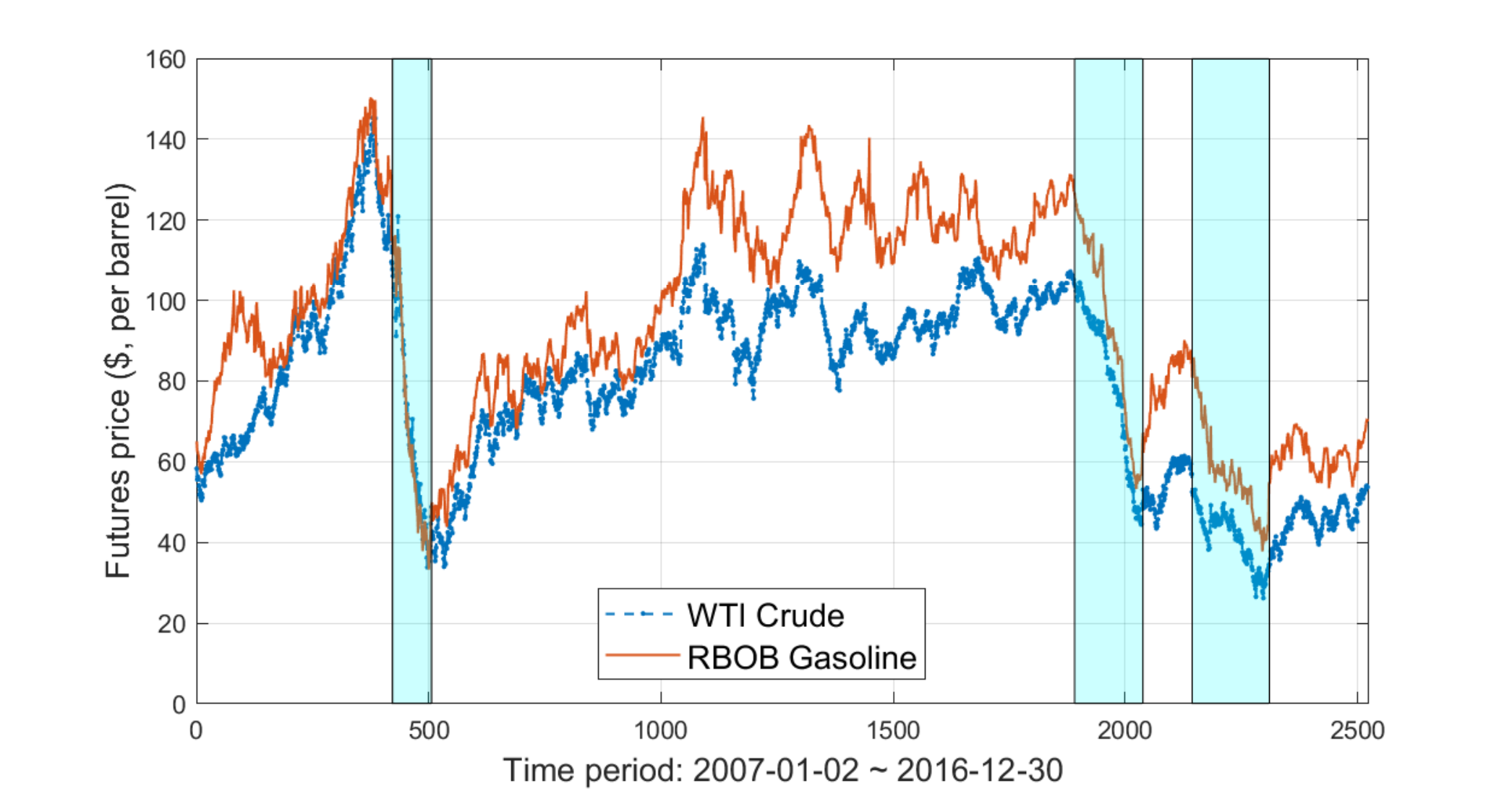}
\end{center}
\caption{The history of daily closing prices for WTI crude oil futures (black) and RBOB gasoline futures (orange) from January 1, 2007 to December 30, 2016. The cyan-shaded areas represent (i) September 2008 -- December 2008 (ii) July 2014 -- January 2015 (iii) July 2015 -- February 2016, in order. }
\label{Fig:Price_history}
\end{figure}

In this section we examine the stylized facts of two major oil-related energy prices in the US.
Figure~\ref{Fig:Price_history} indicates the time series of daily closing prices for WTI crude oil futures and RBOB\footnote{RBOB stands for reformulated gasoline blendstock for oxygen blending.} gasoline futures from 2007 to 2016. 
In this section, we use the gasoline price multiplied by a factor 42 because crude oil price is quoted per barrel, whereas gasoline price is quoted per gallon\footnote{1 barrel = 42 gallons}. 
Each series is created by connecting the prices of the nearest maturity contracts. 

As shown in Figure~\ref{Fig:Price_history}, there were a three times of significant drops in both series -- from September 2008 to December 2008; from July 2014 to January 2015; from July 2015 to February 2016 -- representing a cyan-shaded area in order. 
We investigate the change in parameters of the Hawkes flocking model, especially focusing on the three specified periods when the two prices plunged, and we compare them with the CoVaR values in the following section. 
The data considered in this test are tick size change data in milliseconds for the WTI crude oil and gasoline futures listed in NYMEX from 2007 to 2016.

\subsection{Estimation method}\label{Subsect:simul}

This section explains the ML estimation method used for the empirical study and demonstrates simulation to show the ML estimator's goodness-of-fit to the Hawkes flocking model.
By following the idea in the algorithm by \cite{Ogata1978, Ogata1981}, we perform the ML method using the log-likelihood function represented by conditional intensities as follows.
\begin{align}
L(\theta) =& \sum_{j=1}^{N_1^u(T)} \log\lambda_1^u (t_{1,j}^u) + \sum_{j=1}^{N_1^d(T)} \log\lambda_1^d (t_{1,j}^d) + \sum_{j=1}^{N_2^u(T)} \log\lambda_2^u (t_{2,j}^u) + \sum_{j=1}^{N_2^d(T)} \log\lambda_2^d (t_{2,j}^d) \nonumber \\
&- \int_0^T \left(\lambda_1^u(u) + \lambda_1^d(u) + \lambda_2^u(u) + \lambda_2^d(u) \right) \D u \label{Eq:likelihood}
\end{align}
where $\lambda_{i}^{\cdot}$ indicates the left-continuous versions of the conditional intensity processes, and
$t_{i,j}^{\cdot}$ indicates the associated event times.
The parameter set $\theta = (\mu_i, \beta_i, \alpha_{is}, \alpha_{ic}, \alpha_{in}, \alpha_{iw})$ for all $i=1, 2$ is estimated by maximizing the log-likelihood function numerically.

Since the estimation proceeds numerically with twelve parameters without the assurance of the convexity of the log-likelihood function,
it is not mathematically guaranteed to find the global maximum.
It is therefore worthwhile to check under various situations whether the numerical optimizer can find the correct estimates.
Using the present parameters, 500 sample paths are generated for two price processes under the Hawkes flocking model.
The presumed values are presented in the column titled ``True'' in Table~\ref{Table:simul}.
The column ``Mean'' presents the means of the estimates of 500 estimation procedures, which are quite close to the true values.
Since the above results show that the numerical optimizer that \cite{Henningsen} used works well, we apply this procedure in empirical studies.

\begin{table}[h]
\caption{Simulation using ML estimation for a Hawkes flocking model with 500 sample paths}\label{Table:simul}
\centering
\begin{tabular}{cccc|ccc|ccc}
\hline
& True & Mean & Std. & True & Mean & Std. & True & Mean & Std.\\
\hline
$\mu_1$ & 0.0800 & 0.0803 & 0.0039 & 0.0500 & 0.0503 & 0.0025 & 0.1000 & 0.1000 & 0.0074\\
$\alpha_{1n}$ & 0.0000 & -0.0009 & 0.0124 & 0.2000 & 0.2001 & 0.0206 & 0.3000 & 0.3015 & 0.0324 \\
$\alpha_{1w}$ & 0.2000 & 0.2007 & 0.0142 & 0.3500 & 0.3509 & 0.0200 & 0.3500 & 0.3501 & 0.0247 \\
$\alpha_{1s}$ & 0.4000 & 0.4013 & 0.0083 & 0.1500 & 0.1505 & 0.0140 & 0.2000 & 0.1988 & 0.0176 \\
$\alpha_{1c}$ & 0.0000 & -0.0004 & 0.0141 & 0.4000 & 0.4005 & 0.0168 & 0.2000 & 0.2005 & 0.0157 \\
$\beta_1$ & 0.6000 & 0.6008 & 0.0212 & 1.0500 & 1.0513 & 0.0368 & 0.9000 & 0.9011 & 0.0402 \\
$\mu_2$ & 0.0500 & 0.0500 & 0.0023 & 0.0700 & 0.0702 & 0.0027 & 0.1200 & 0.1207 & 0.0072 \\
$\alpha_{2n}$ & 0.0000 & -0.0001 & 0.0105 & 0.3500 & 0.3495 & 0.0269 & 0.0000 & 0.0010 & 0.0215 \\
$\alpha_{2w}$ & 0.1000 & 0.1009 & 0.0115 & 0.1000 & 0.1001 & 0.0162 & 0.1000 & 0.1003 & 0.0216 \\
$\alpha_{2s}$ & 0.5000 & 0.5014 & 0.0242 & 0.4500 & 0.4505 & 0.0213 & 0.3000 & 0.2995 & 0.0224 \\
$\alpha_{2c}$ & 0.3000 & 0.3005 & 0.0178 & 0.2500 & 0.2497 & 0.0142 & 0.6000 & 0.6006 & 0.0282 \\
$\beta_2$ & 1.2000 & 1.2028 & 0.0446 & 1.3000 & 1.2999 & 0.0465 & 1.1500 & 1.1519 & 0.0467 \\
\hline
\end{tabular}
\end{table}

\subsection{Data}
Data on WTI crude oil and gasoline futures' trade prices are obtained from the database of Tickdatamarket (\url{www.tickdatamarket.com}).
We choose the futures data on ``Light Sweet Crude Oil'' for WTI crude oil (referred to as ``CL'' henceforth) and ``RBOB Gasoline'' for gasoline (referred to as ``RB'' henceforth) in the energy sector of North America.
We consider the data for 10 years from January 1, 2007 to December 30, 2016, and each year has twelve delivery months.
To construct a single time series over that period for each futures contract, we select data with the nearest delivery month from the observation date, which is usually the most liquid among contracts with maturity longer than one month.

To ensure that raw data can be more feasibly applied to the proposed model, a data wrangling procedure is needed.
Without loss of generality, we transform the raw data into a more appropriate format in terms of the following aspects.

\smallskip\noindent
{\it (i) Different price level}: The two futures prices of CL and RB have different price levels.
For instance, the CL price at January 13, 2016 with maturity in January 2016 is around \$31, but the RB price is around \$1.07 per gallon, which is \$44.94 per barrel. 
Minimum tick sizes are also different. 
The tick size is 0.01 for CL and 0.0001 for RB.
Therefore, we need to adjust prices to similar levels. 
For adjustment, we consider the sample mean ratio, $\bar X/\bar Y$
where $\bar X$ implies the sample mean of prices, for example, of the day.

\begin{figure}
\begin{center}
\includegraphics[width=7cm]{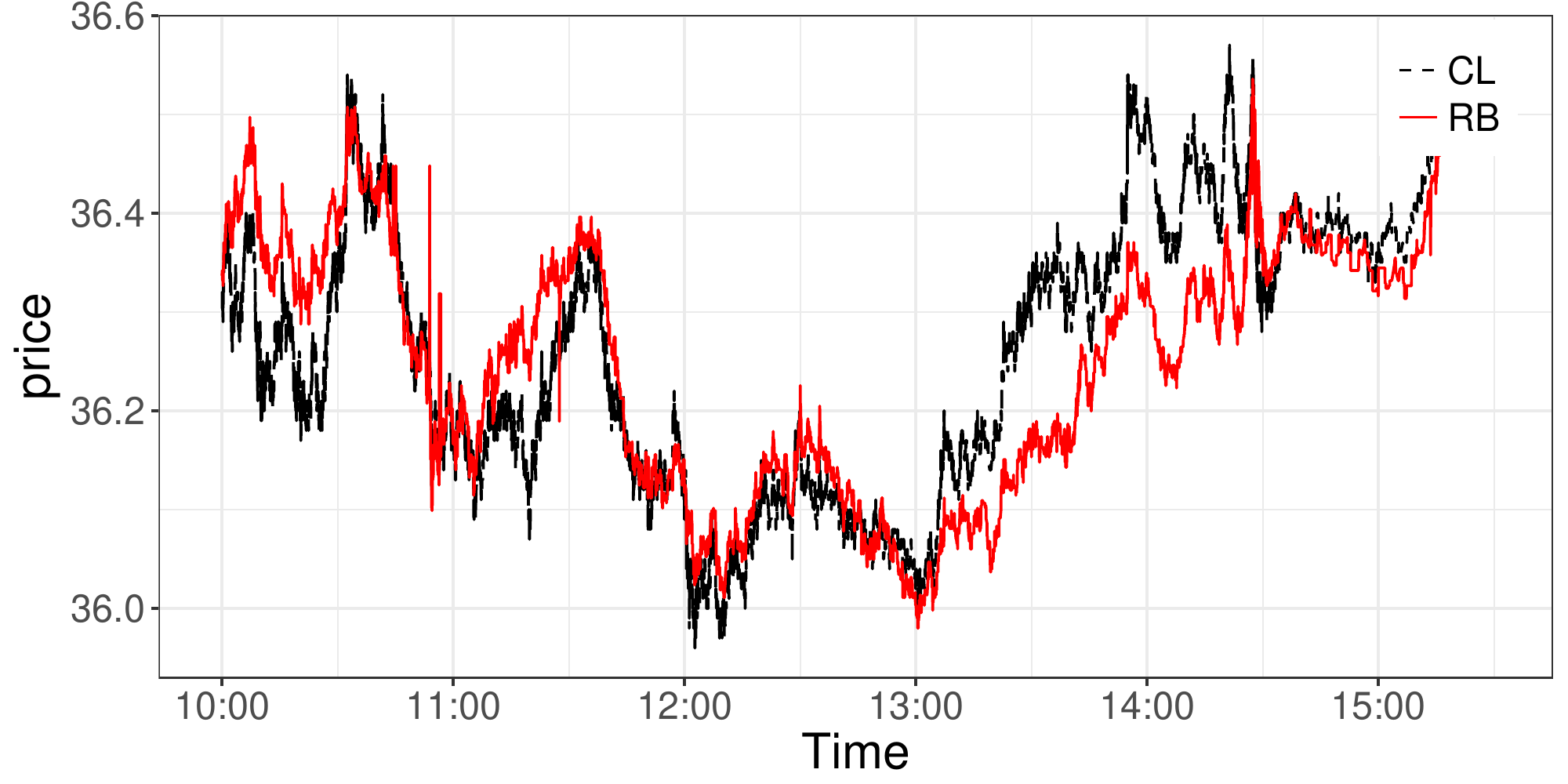}
\quad
\includegraphics[width=7cm]{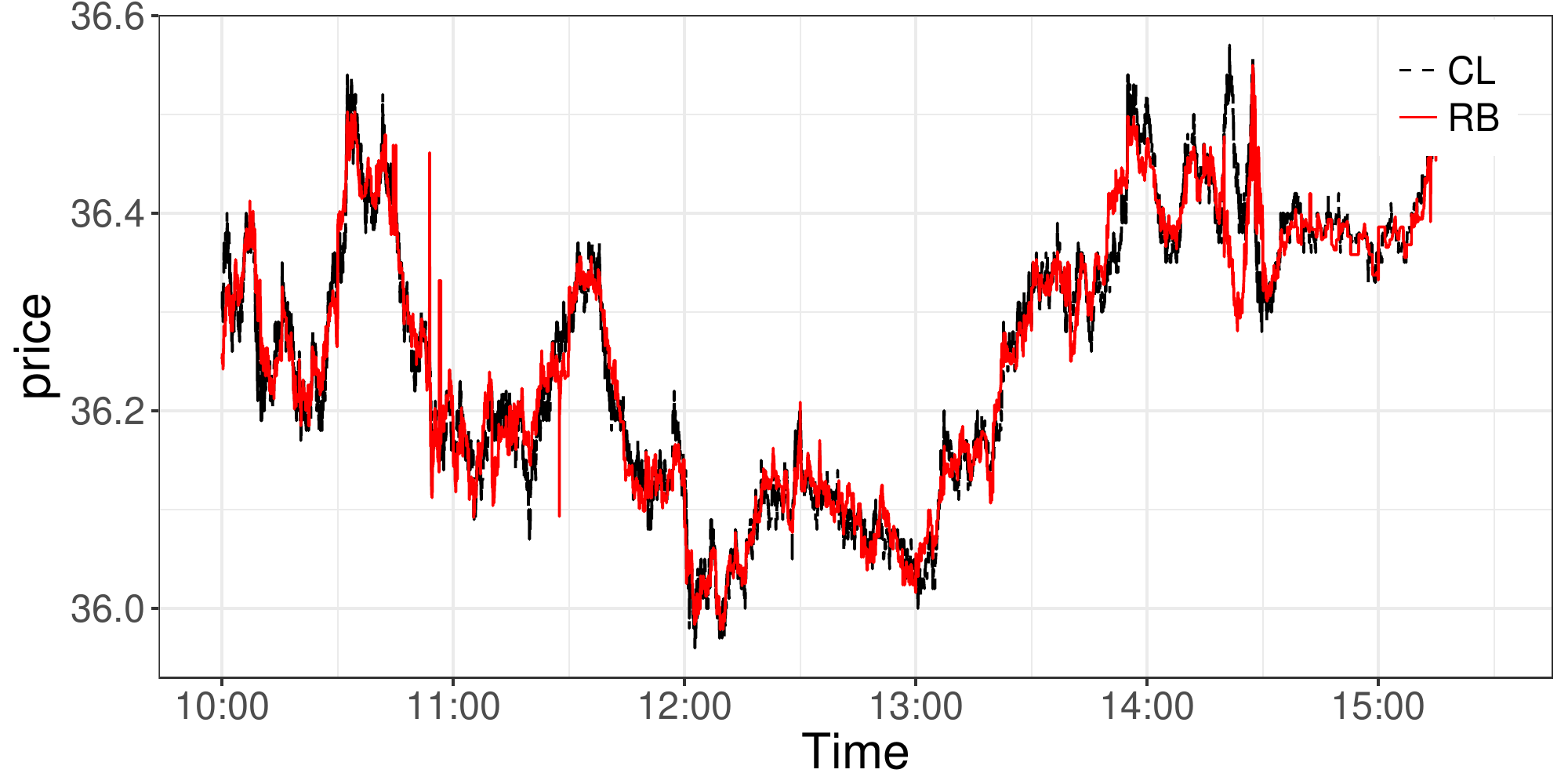}
\end{center}
\caption{Before (left) and after adjustment (right) for CL and RB prices traded at March 15, 2016}\label{Fig:adjust}
\end{figure}

Let $X$ and $Y$ be the original price processes.
Define $C_1 = X$, $C_2 = (\bar X/\bar Y) Y$.
However, the adjusted prices on a daily basis are not applicable for the Hawkes flocking model as presented in the left of Figure~\ref{Fig:adjust}, where $C_1$ denotes CL and $C_2$ denotes RB.
Data for March 15, 2016 with maturity March 2016 were used.
In the early part of the day, it is almost always $C_1 < C_2$, but in the later part, it is almost always $C_2 > C_1$.
Thus, the sample mean should be computed under a shorter time interval.
For example, we calculate the sample mean every 10 minutes, and the prices are adjusted during each 10-minute interval.
Then the price processes are more applicable, as in the right of Figure~\ref{Fig:adjust}. 

\smallskip\noindent
{\it (ii) Multiple price changes in unit time}:
The minimum resolution time of the data is one second, and multiple price changes can be observed in one second.
In this case, we assume that each change occurs at the equi-distant time interval that divides one second with the same number of observations.

\smallskip\noindent
{\it (iii) Simultaneous changes in the two prices}:
The probability that the two prices change at the same time is almost zero in our model, but simultaneous changes in both prices can be recorded in practice.
In this case, one price change was assumed to have occurred slightly earlier than the other change.
Since this simultaneous jump can be observed several times a day,
we consider that one jump to have occurred before the other.

\subsection{Robustness test for calibration}\label{SubSec:Calibration}
From the time series of the CL and RB futures prices after data pre-processing,
we estimate the model parameters using the ML estimator presented in Section~\ref{Subsect:simul}.
The estimation is performed for up to 10 years.
We investigate the dynamics of all parameters for the proposed model over time, 
especially focusing on the periods of plunges in the CL and RB prices.

Before implementing model calibration, we conduct a test for the relation between the two parts in the intensity kernel $\bm h$: a self/mutually-exciting factor and a flocking factor.
We consider two models with different kernels: (i) a symmetric Hawkes model without a flocking term versus (ii) a Hawkes flocking model. Then, we compare the parameter estimation results derived from each model.

First, we assume that CL and RB futures prices follow a symmetric Hawkes process that has an exponential decaying kernel without a flocking term, that is
\begin{equation}
\bm{\lambda}_t = \bm{\mu} + \int_{-\infty}^{t} \Phi(t-u) \D \bm{N}_u
\end{equation}
where the matrix $\Phi$ is symmetric with parameters $\alpha_{is}, \alpha_{ic}$, and $\beta_i$ for $i=1,2$, as defined in \eqref{Eq:phi}.
In this setup, since the flocking phenomenon between the CL and RB futures prices is not implicated, the two processes are independent.

Second, we assume that CL and RB futures prices follow the Hawkes flocking process, that is,
\begin{equation}
\bm{\lambda}_t = \bm{\mu} + \int_{-\infty}^{t} \Phi(t-u) + {k}(u) \circ \Psi(t-u) \D \bm{N}_u
\end{equation}
where the matrix ${k}$ and $\Psi$ are defined in \eqref{eq:K} and \eqref{eq:phi_nw}, respectively,
with additional parameters $\alpha_{in}, \alpha_{iw}$ for $i=1,2$.

\begin{example}\label{Ex:sc}
[{\it Multicollinearity for $\alpha_s$ and $\alpha_c$}].
We estimate the parameters $\alpha_s$ and $\alpha_c$ from CL and RB futures prices under each model assumption. 
The test is performed on a daily basis on the futures prices with maturity in February 2016, and the observation period is from January 4 to February 22, 2016. 
Figure~\ref{Fig:alpha} illustrates the results of $\alpha_s$ and $\alpha_c$ for CL and RB under the symmetric Hawkes model (red solid line) and the Hawkes flocking model (black dashed line).

We find that the estimates under different model assumptions are almost identical over the observed period. 
This implies that ${k}(u) \circ \Psi(t-u)$ in the Hawkes flocking model does not affect the existing self/mutually-exciting parts $\Phi(t-u)$.
This concept can be considered as being similar to the non-existence of multicollinearity in linear regression. 

\begin{figure}[h]
\centering
\begin{subfigure}[b]{0.48\textwidth}
\includegraphics[width=\textwidth]{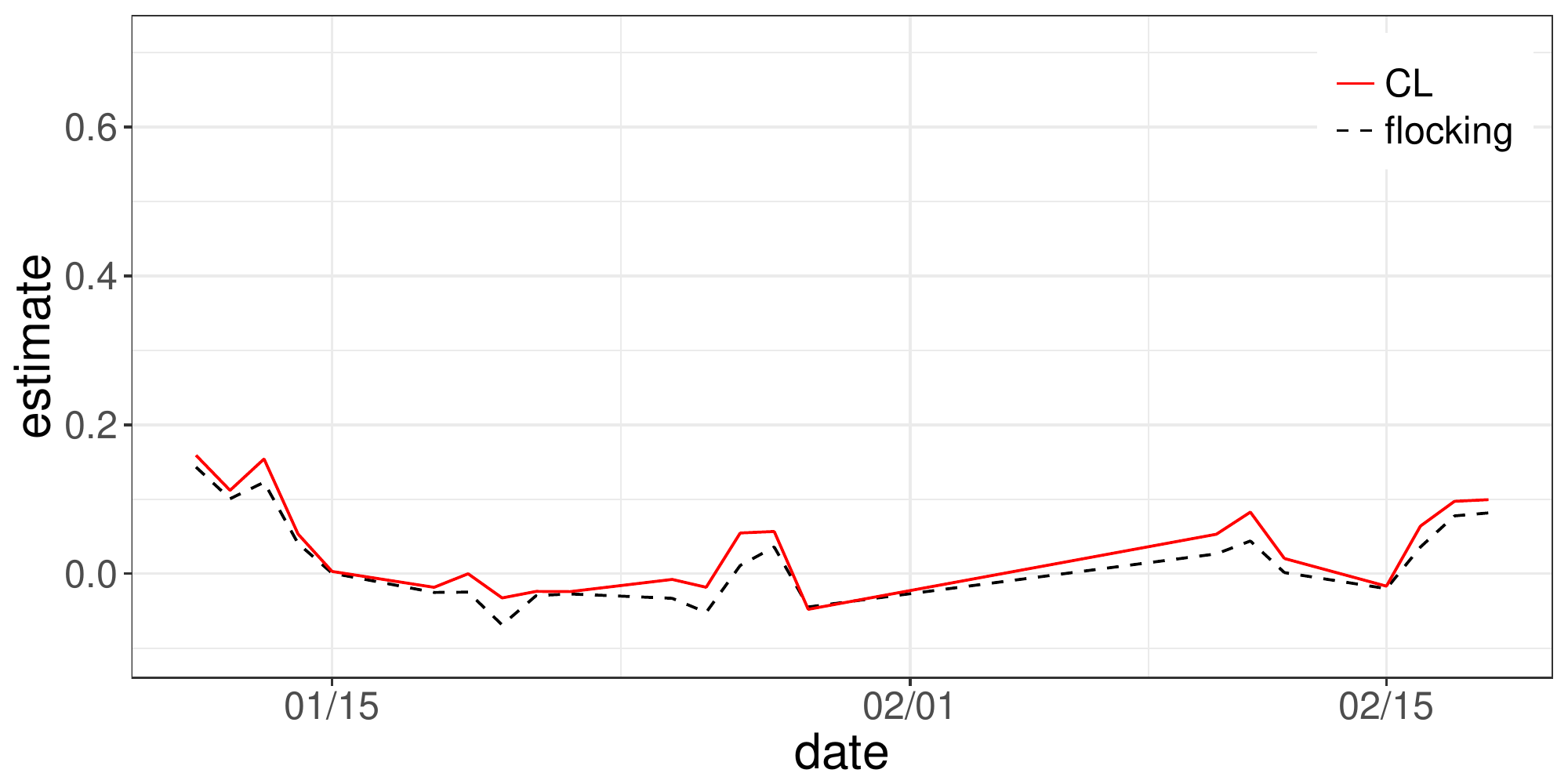}
\caption{$\alpha_s$ in CL}
\end{subfigure}
\begin{subfigure}[b]{0.48\textwidth}
\includegraphics[width=\textwidth]{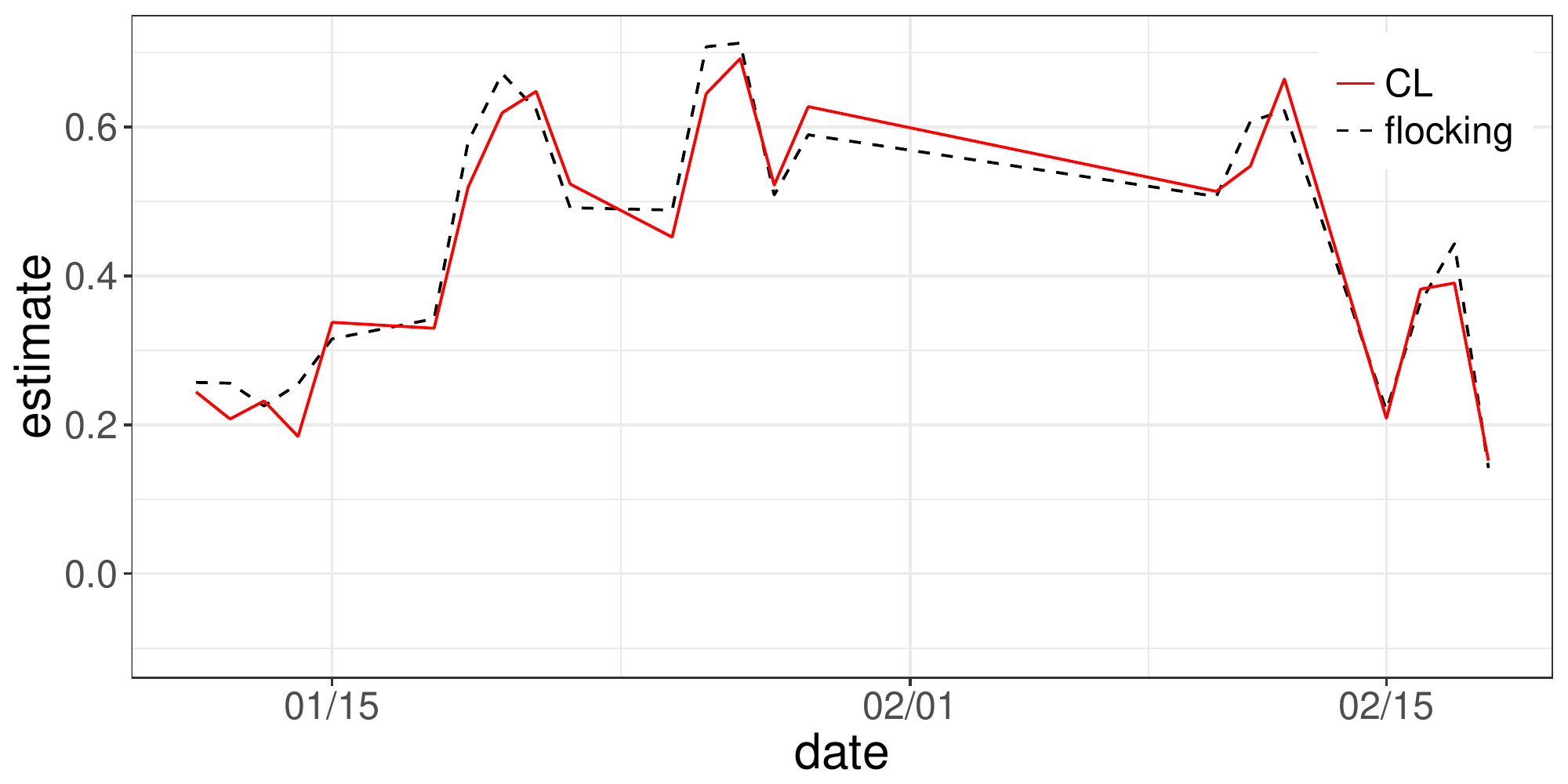}
\caption{$\alpha_c$ in CL}
\end{subfigure}
\begin{subfigure}[b]{0.48\textwidth} 
\includegraphics[width=\textwidth]{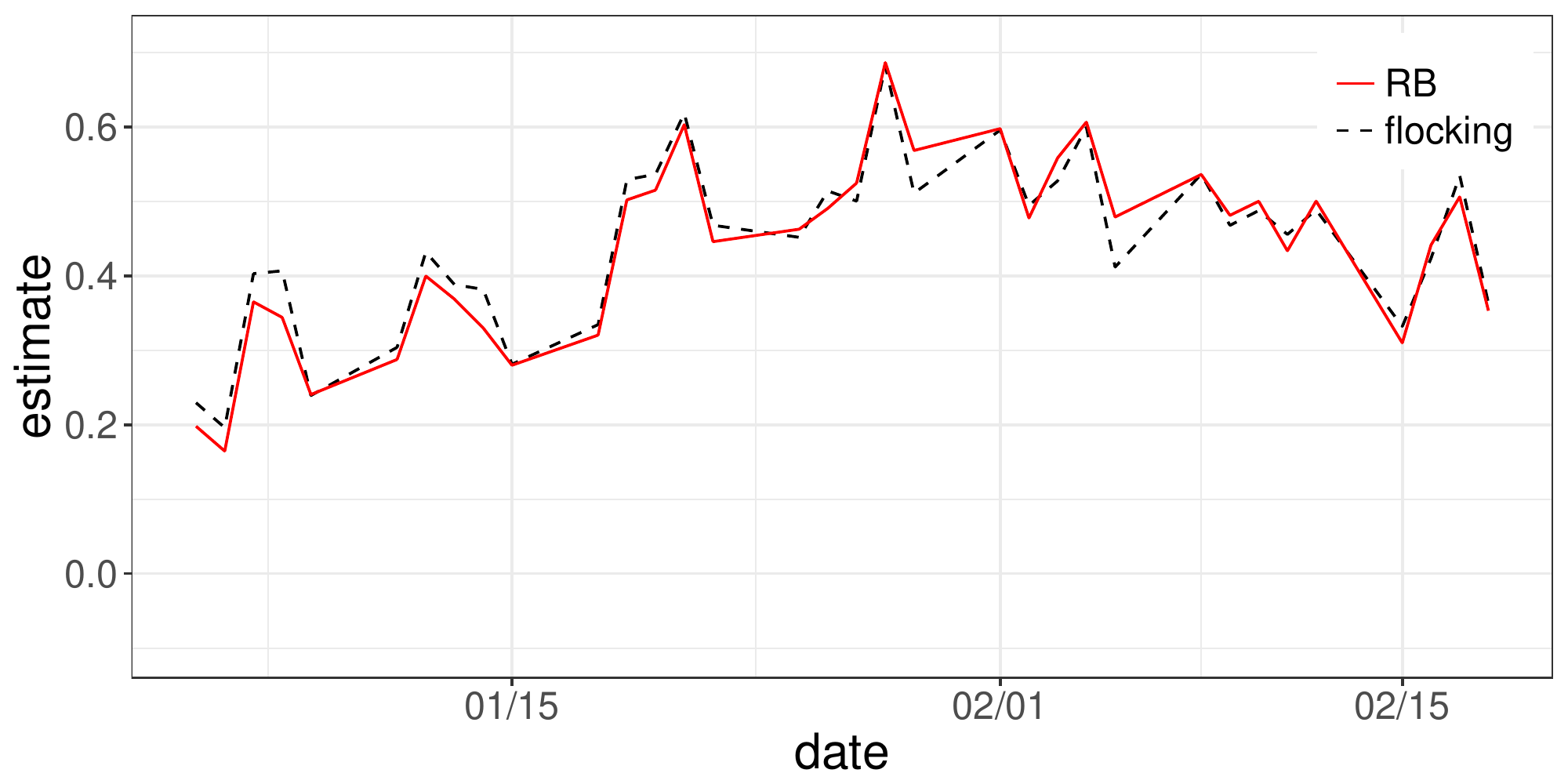}
\caption{$\alpha_s$ in RB}
\end{subfigure}
\begin{subfigure}[b]{0.48\textwidth}
\includegraphics[width=\textwidth]{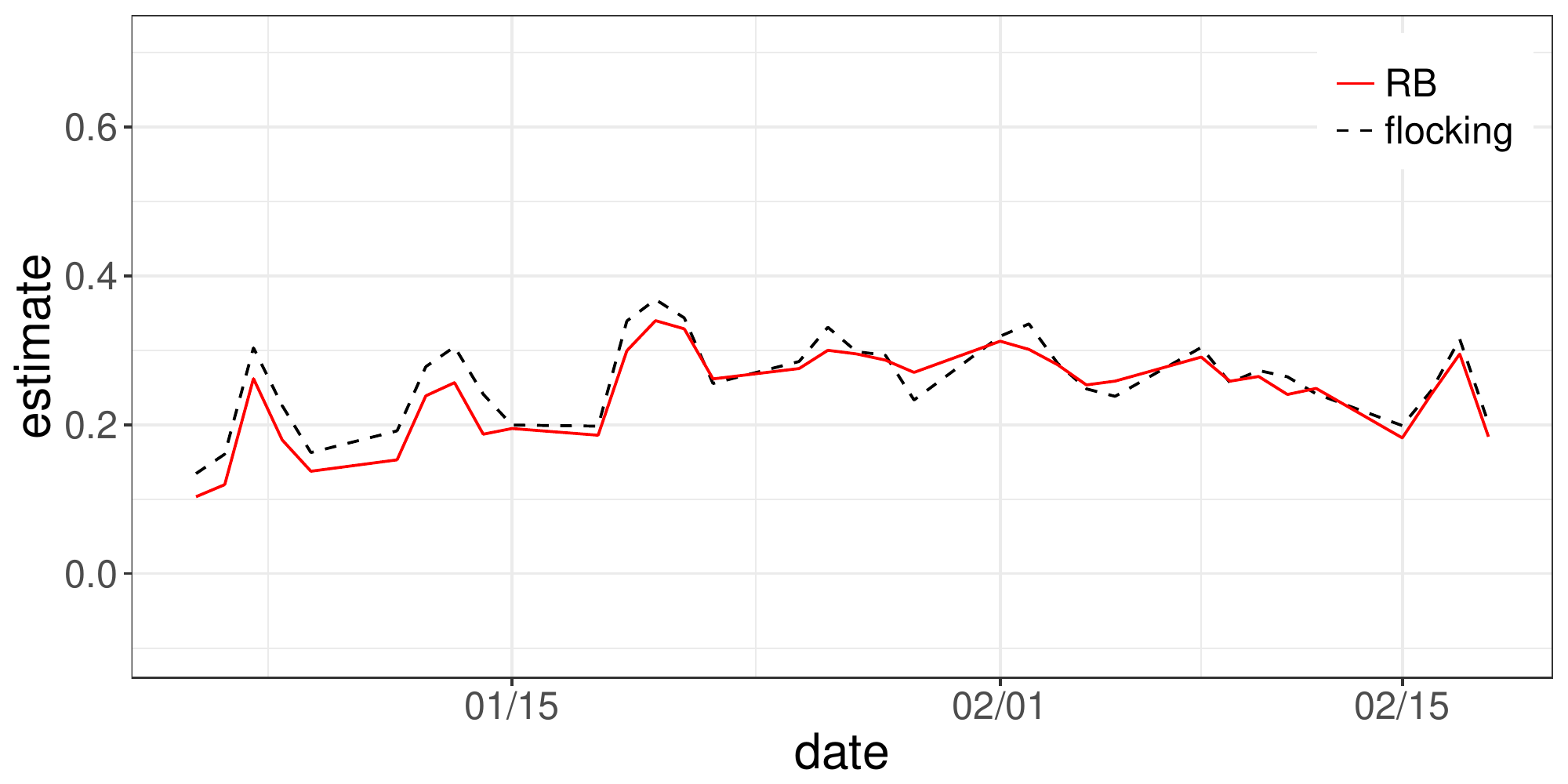}
\caption{$\alpha_c$ in RB}
\end{subfigure}

\caption{Comparison of estimates of $\alpha_{s}$ and $\alpha_c$ under a symmetric Hawkes model (red solid line) and the Hawkes flocking model (black dotted line) for CL and RB futures prices (with maturity in February 2016) from January 4 to February 22, 2016}
\label{Fig:alpha}
\end{figure}
\end{example}

\begin{example}\label{Ex:mubeta}
[{\it Multicollinearity for $\mu$ and $\beta$}].
We conduct the same test as conducted in Example~\ref{Ex:sc} on the base intensity parameter $\mu_i$ and the resilience speed to the base intensity $\beta_i$ over the same sample period. 
Figures~\ref{Fig:mu} and \ref{Fig:beta} show the results of parameters $\mu_i$ and $\beta_i$ based on the two models, respectively.

For $\mu$, we find that both CL and RB futures prices of the symmetric Hawkes model have larger $\mu$ than those of the Hawkes flocking model.
The reason is that additional fluctuations in price processes due to the flocking phenomenon are considered as a part of exogenous dynamics, and hence they are inherent to $\mu$
under the symmetric Hawkes models where flocking terms, $\alpha_{n}$ and $\alpha_{w}$, do not exist.

For $\beta$,
we see that results from the symmetric Hawkes and Hawkes flocking models are very similar for the CL futures price.
Meanwhile, the $\beta$ in the Hawkes flocking model is slightly larger than $\beta$ in the symmetric Hawkes model for the RB futures price.
This implies that ``$\beta$ due to $\alpha_{n}$ and $\alpha_{w}$'' is larger than ``$\beta$ due to $\alpha_s$ and $\alpha_c$'' in RB.
Note that a large $\beta$ implies weak persistence.
We do not rule the possibility out that $\beta$ depends on $\alpha$'s but merely consider unified $\beta$ for model parsimony.
\begin{figure}[h]
\begin{center}
\includegraphics[width=0.48\textwidth]{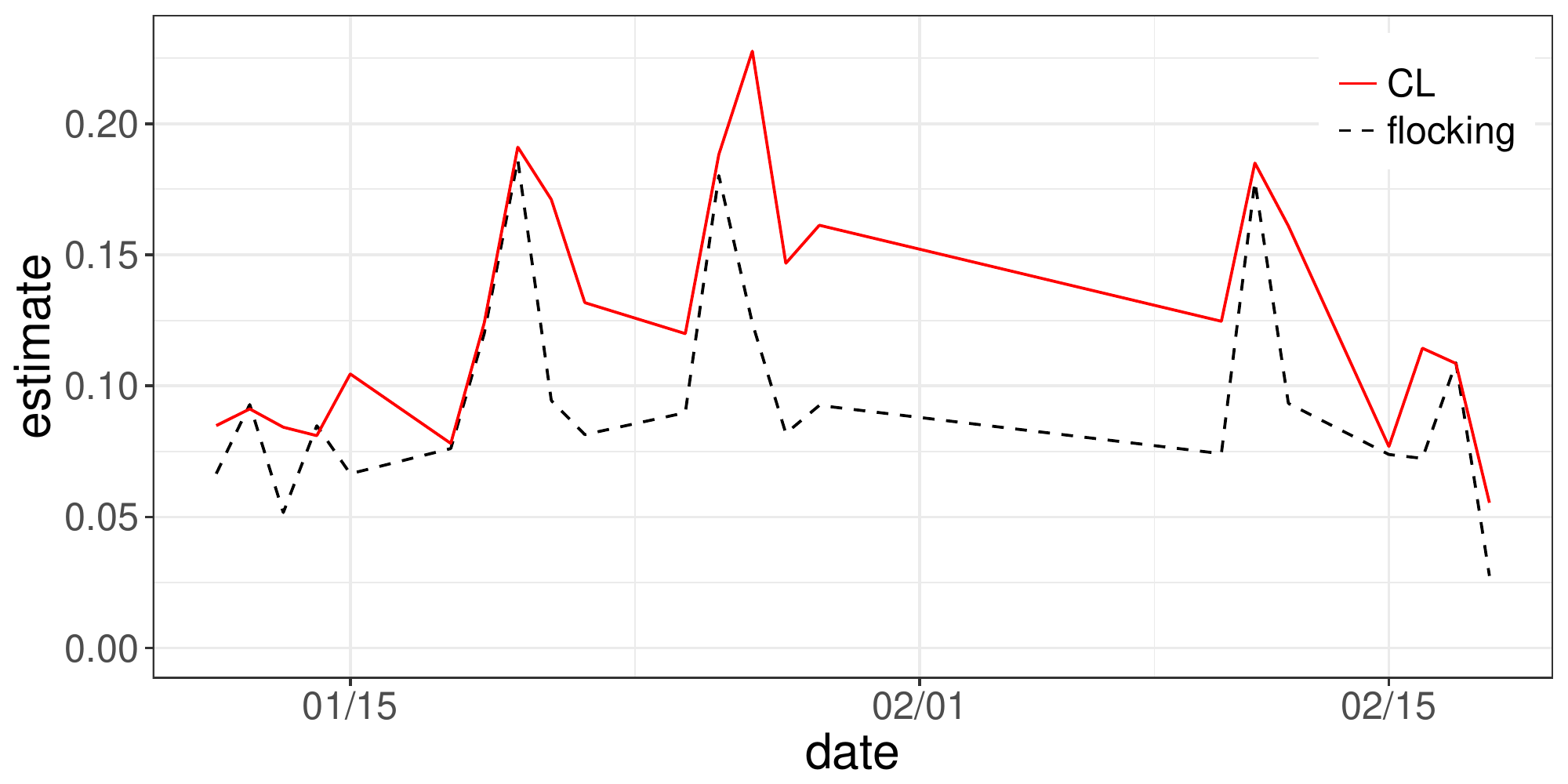}\quad
\includegraphics[width=0.48\textwidth]{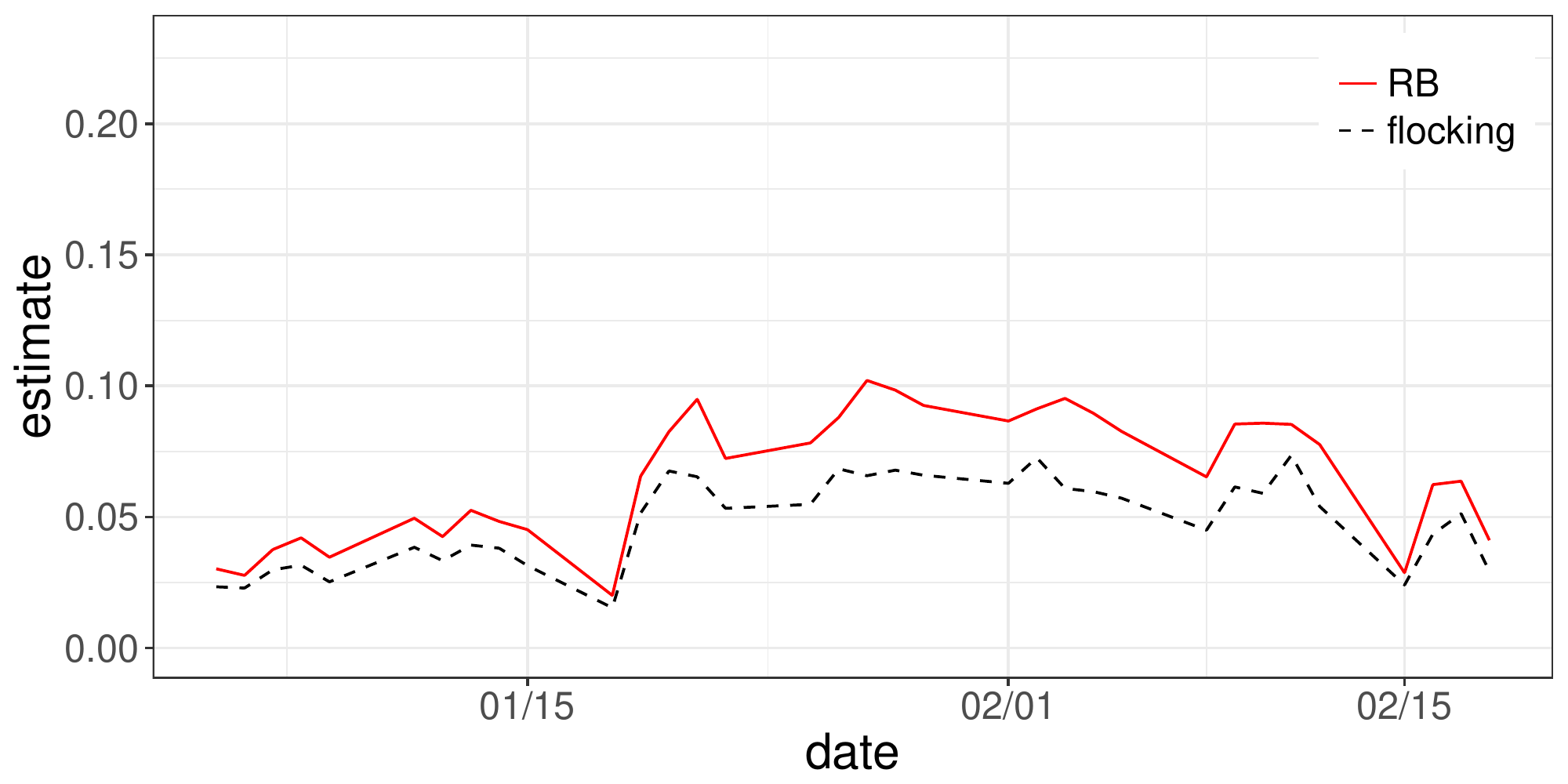}
\end{center}
\caption{Comparison of estimates of $\mu_1$ and $\mu_2$ under the symmetric Hawkes model (red line) and the Hawkes flocking model (black dotted line) for CL and RB futures prices (with maturity in February 2016) from January 4 to February 22, 2016}
\label{Fig:mu}
\begin{center}
\includegraphics[width=0.48\textwidth]{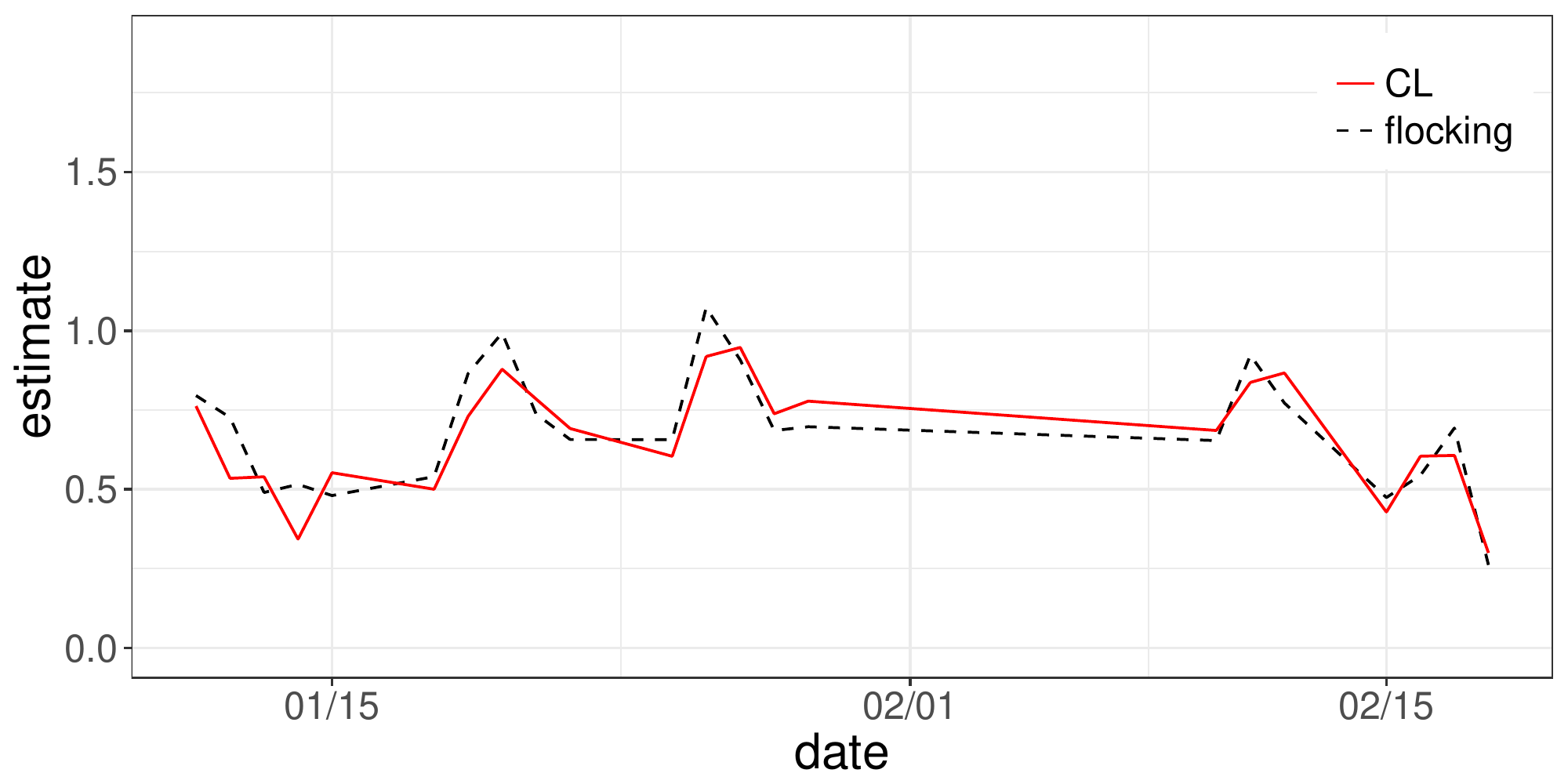}\quad
\includegraphics[width=0.48\textwidth]{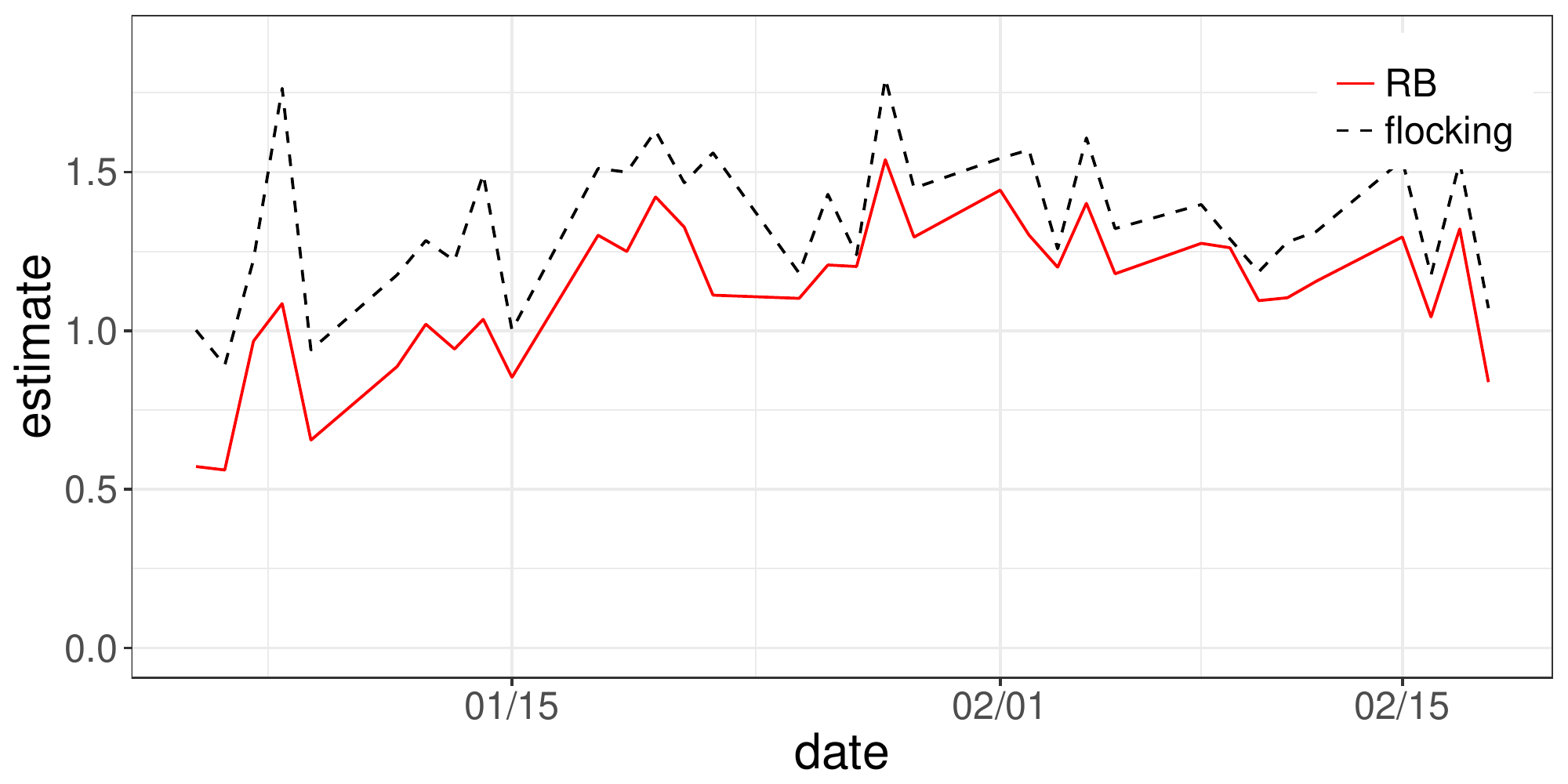}
\end{center}
\caption{Comparison of estimates of $\beta_1$ and $\beta_2$ under the symmetric Hawkes model (red line) and the Hawkes flocking model (black dotted line) for CL and RB futures prices (with maturity in February 2016) from January 4 to February 22, 2016}
\label{Fig:beta}
\end{figure}
\end{example}
More relevant examples for other sample periods appear in Appendix~\ref{App:B}.

\subsection{Calibration}\label{Sec:Calibration}
\begin{figure}[h]
\begin{center}
\includegraphics[width=0.48\textwidth]{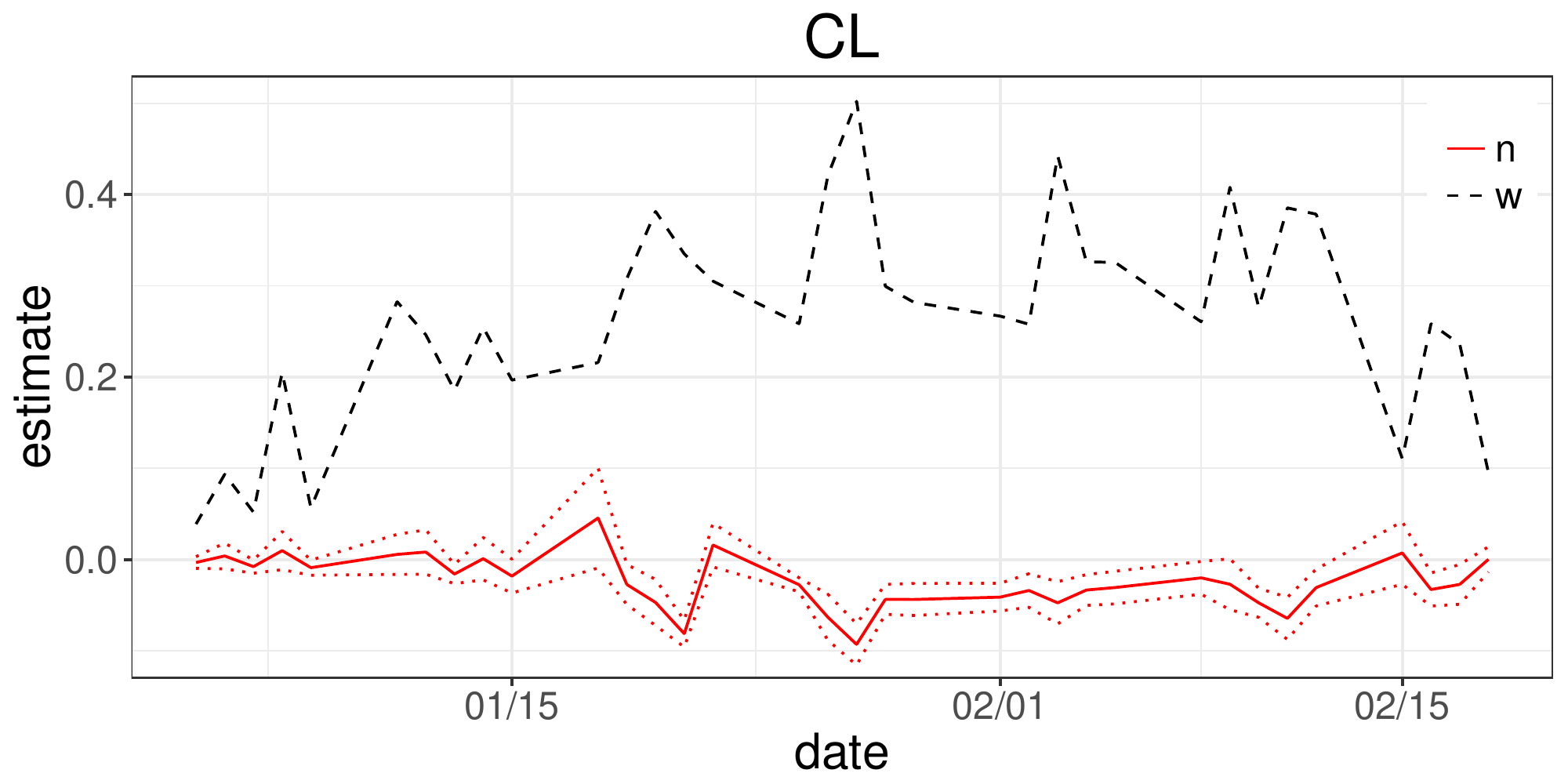}\quad
\includegraphics[width=0.48\textwidth]{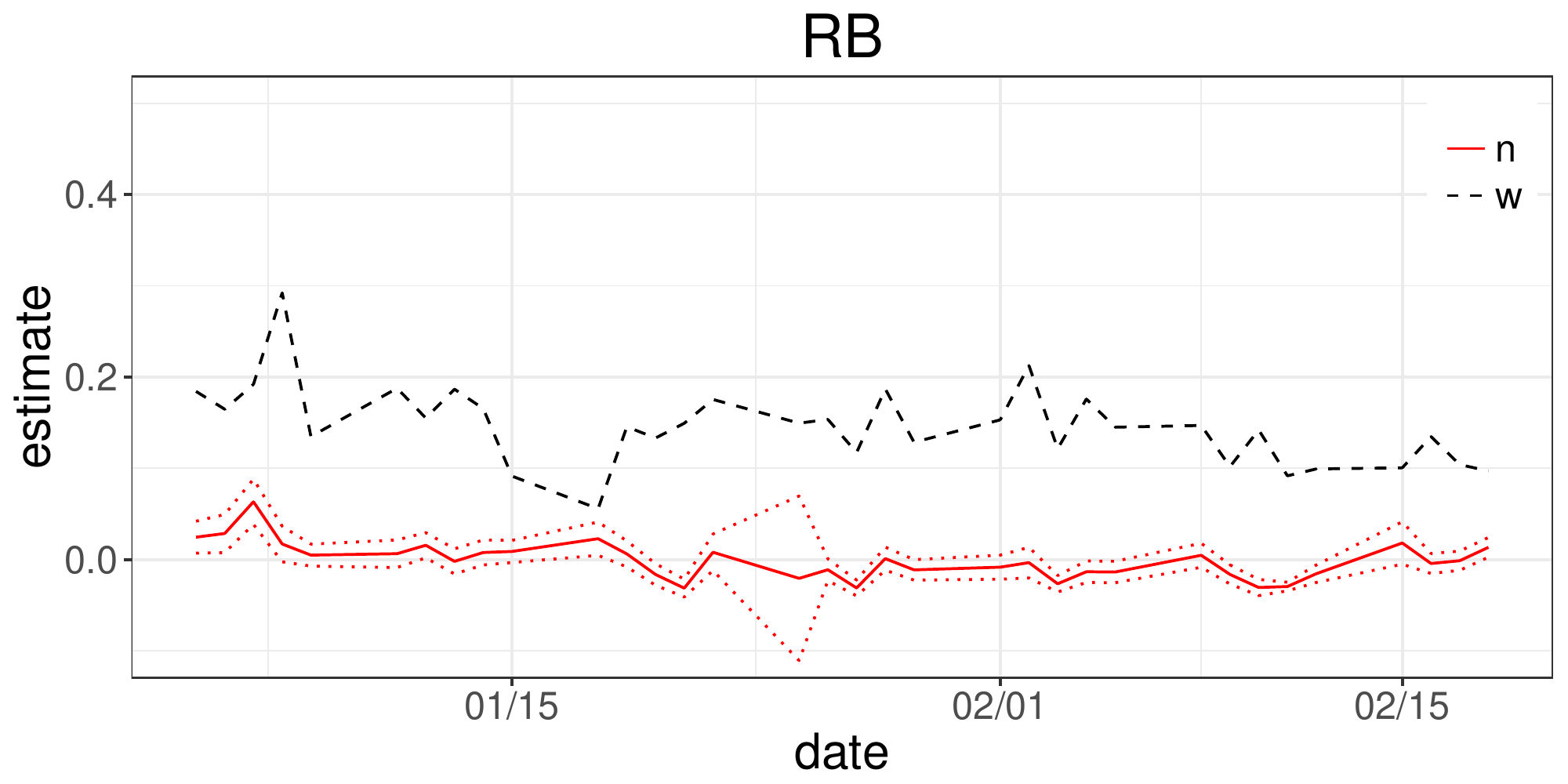}
\end{center}
\caption{Comparison of $\alpha_n$ and $\alpha_w$ for CL (left) and RB (right) futures prices (with maturity in February 2016) from January 4 to February 22, 2016}
\label{Fig:alpha_nw}
\end{figure}

Based on the argument of the model's robustness check, we conduct calibration for all 12 parameters in the Hawkes flocking model on the partial period in 2016, where the results are presented in Tables~\ref{Table:flock_RBCL} and \ref{Table:mean_estimates}.
In particular, we investigate significance of flocking parameters $\alpha_n$ and $\alpha_w$ with more attention.
Figure~\ref{Fig:alpha_nw} compares $\alpha_n$ and $\alpha_w$ under the Hawkes flocking model over the same sample period.
The narrowing events' parameters $\alpha_{in}$ are depicted with plus minus two times of standard errors (dotted lines) to check the significance of the estimates.
The result shows that $\alpha_{in}$ are close to zero in the selected time period 
and this means that the price difference of narrowing events does not substantially affect the intensities.

On the other hand, $\alpha_{iw}$ are significant. This means that the widening event significantly increases intensities associated with the flocking
so that the two price processes tend to converge toward each other after widening events.
The standard errors of $\alpha_{iw}$ are omitted for clarity of the graph
but the estimates of $\alpha_{iw}$ are statistically significant for all time period.
For the graph, selected maturity for the futures is in February 2016 and estimates are computed on a daily basis.
In addition, $\alpha_w$ in CL is larger than that in RB.

\begin{figure}[h]
\begin{center}
\includegraphics[width=0.48\textwidth]{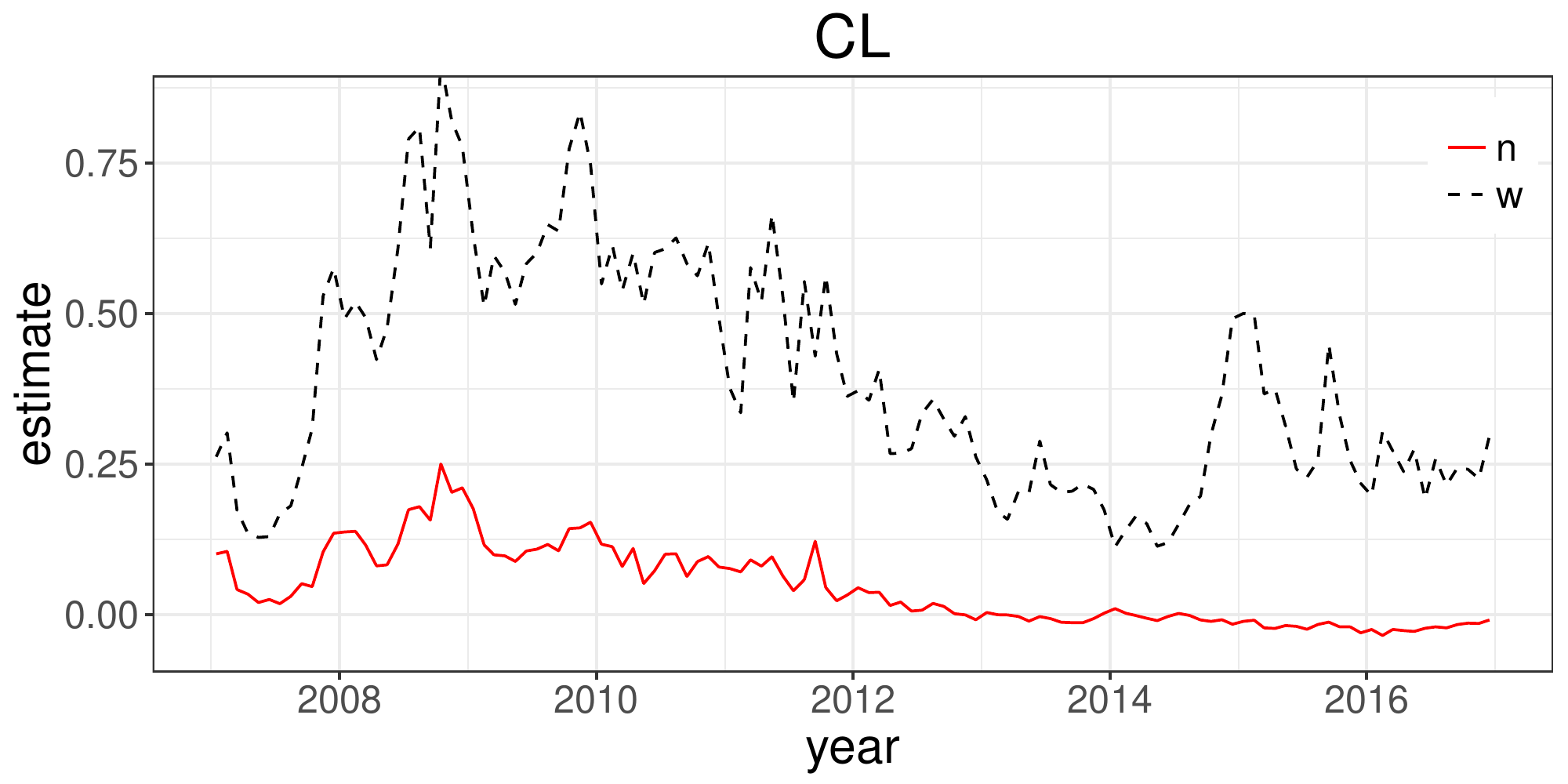}\quad
\includegraphics[width=0.48\textwidth]{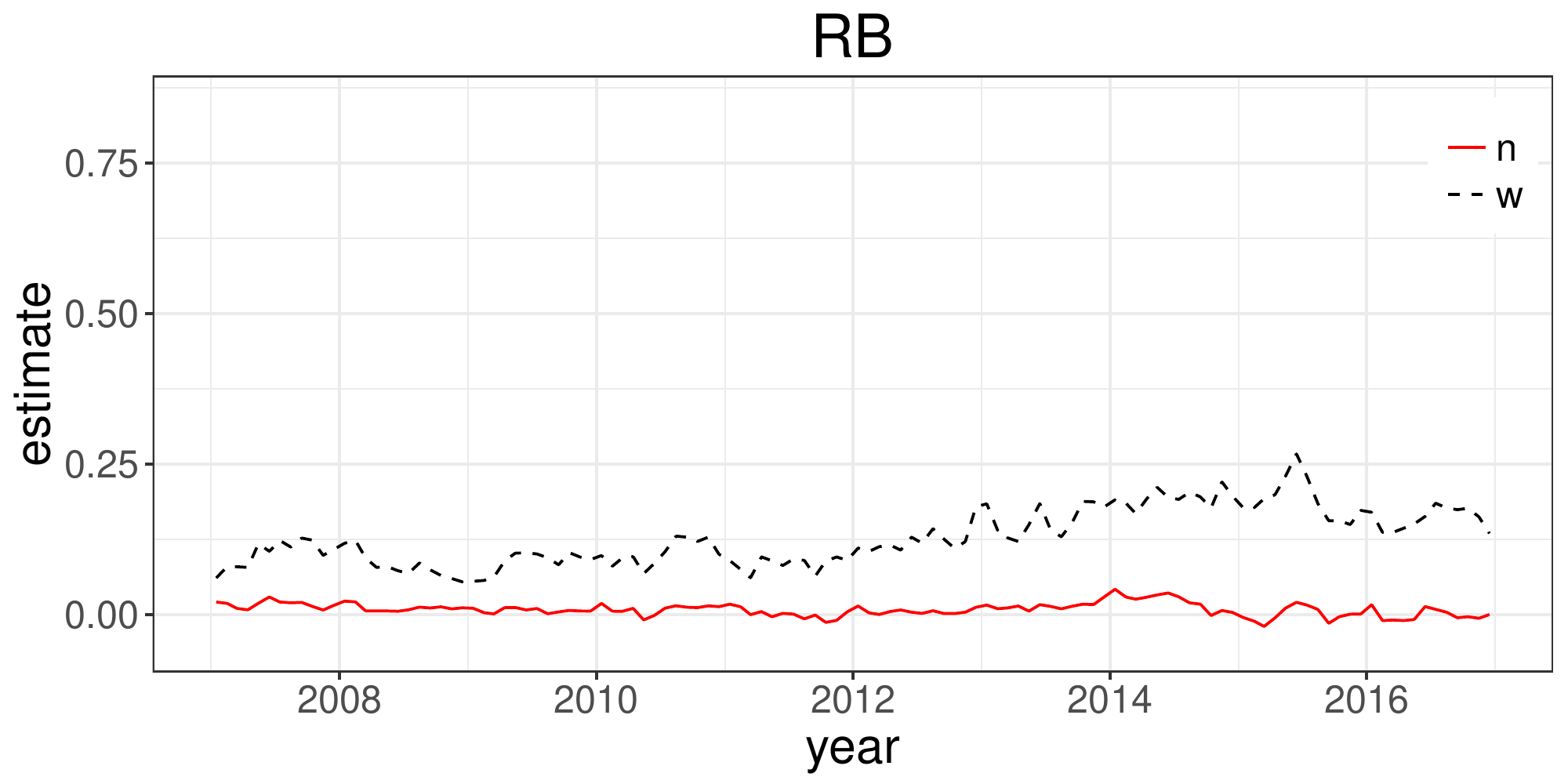}
\end{center}
\caption{Comparison of changes in the flocking parameters $\alpha_n$ (red line) and $\alpha_w$ (black dotted line) for CL (left) and RB (right) from January 2007 to December 2016}
\label{Fig:alpha_nw_10}
\begin{center}
\includegraphics[width=0.48\textwidth]{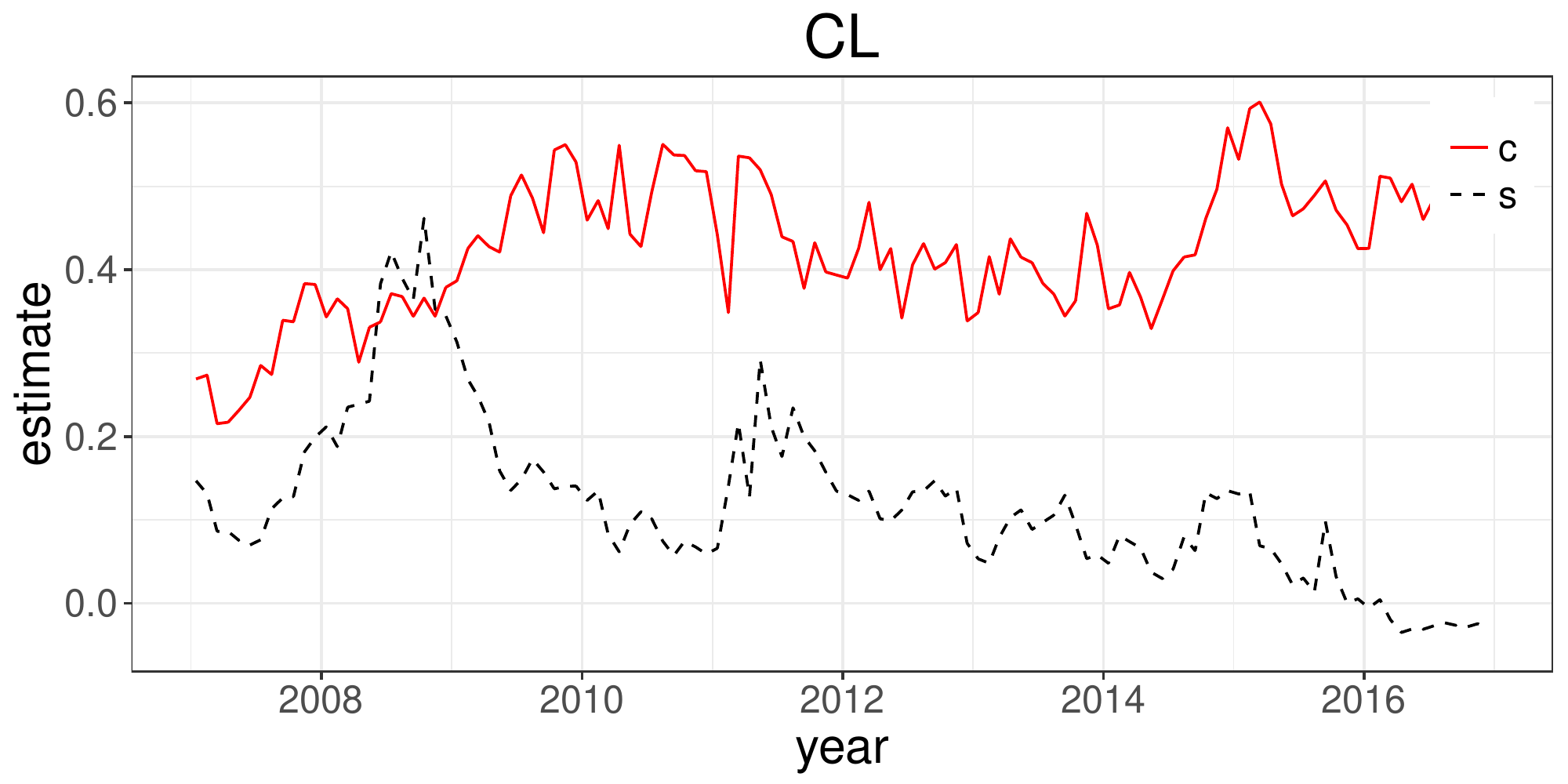}\quad
\includegraphics[width=0.48\textwidth]{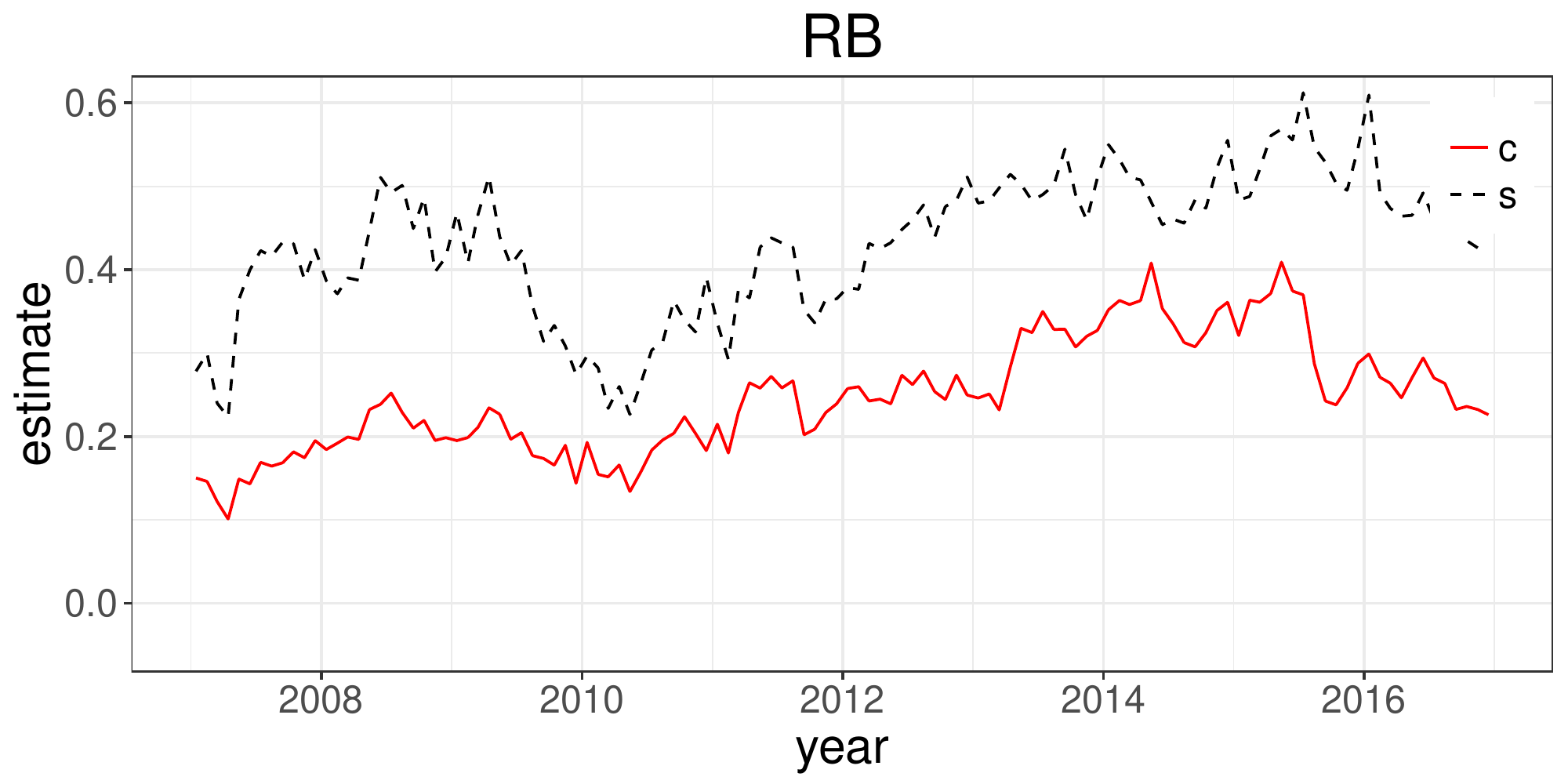}
\end{center}
\caption{Comparison of changes in the self/mutually exciting parameters $\alpha_s$ (red line) and $\alpha_c$ (black dotted line) for CL (left) and RB (right) from January 2007 to December 2016}
\label{Fig:alpha_cs_10}
\begin{center}
\includegraphics[width=0.48\textwidth]{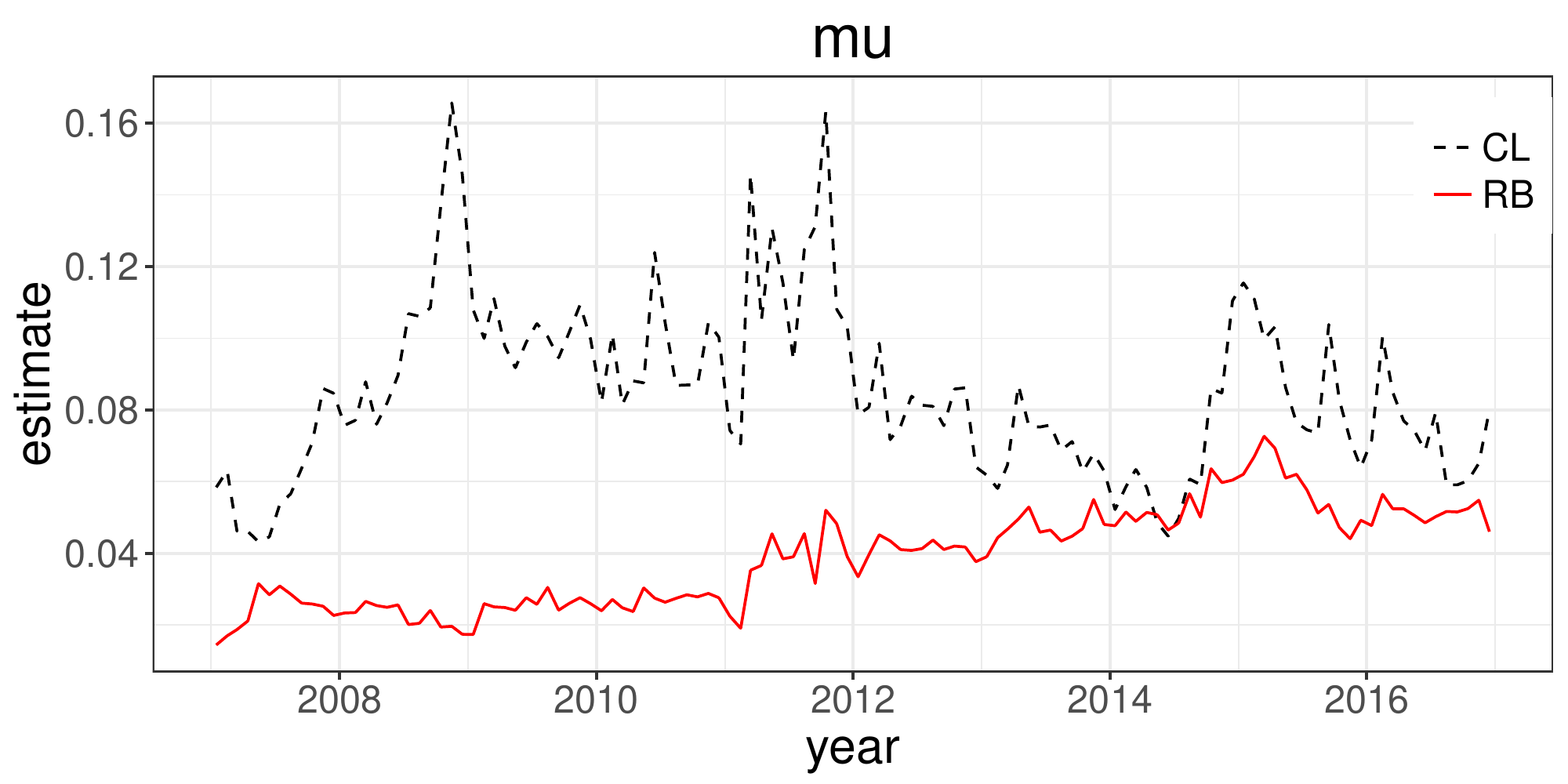}\quad
\includegraphics[width=0.48\textwidth]{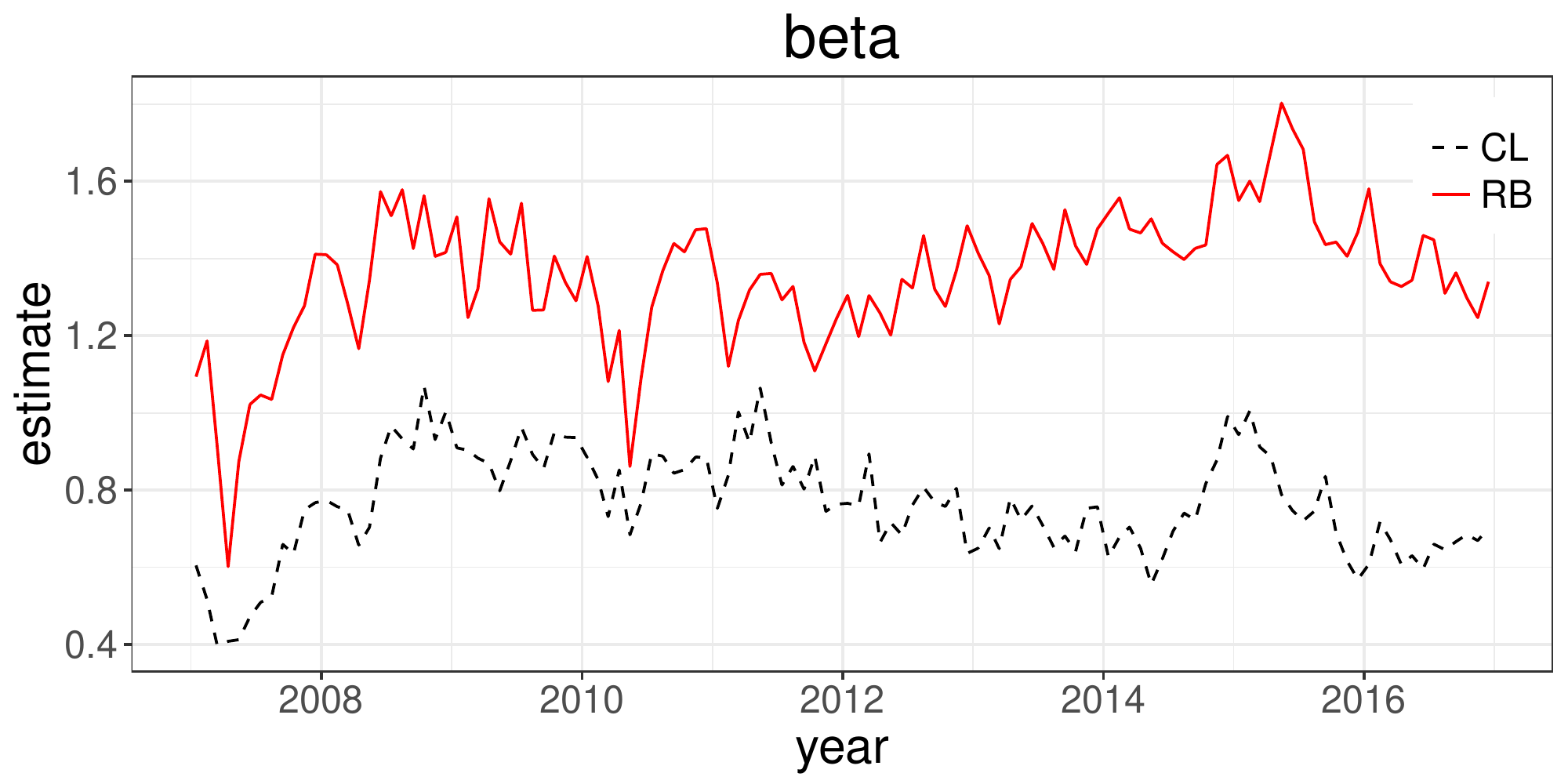}
\end{center}
\caption{Comparison of changes in the exogenous fluctuation parameter $\mu$ (left) and the persistence $\beta$ (right) for CL (black dotted line) and RB (red line) from January 2007 to December 2016}
\label{Fig:mu_beta_10}
\end{figure}

From now, 
we calibrate the Hawkes flocking model by expanding the test period from a sample month (February 2016) to the recent decade from January 2007 to December 2016. 
Using the transaction data of CL and RB futures prices, the estimates are computed on a daily basis and the daily estimates are averaged over a month for better visualization.
The associated results are displayed in Figures~\ref{Fig:alpha_nw_10}, \ref{Fig:alpha_cs_10}, and \ref{Fig:mu_beta_10}.

Figure~\ref{Fig:alpha_nw_10} exhibits the flocking parameters $\alpha_n$ (red solid line) and $\alpha_w$ (black dotted line) for CL (left panel) and RB (right panel).
For each futures price, the level of $\alpha_n$ is much smaller than that of $\alpha_w$ and it is close to zero for a long time period.
This seems reasonable because widening events have a strong role causing the flocking phenomenon, while narrowing events have no or relatively small effects.
In CL, $\alpha_w$ has the maximum value near the fourth quarter of 2008, and it gradually decreases and then increases again around 2015.
In RB, $\alpha_w$ gradually increases.

Figure~\ref{Fig:alpha_cs_10} shows the behavior of $\alpha_s$ (black dotted line) and $\alpha_c$ (red solid line) for CL (left panel) and RB (right panel).
The self-exciting parameter $\alpha_s$ in CL gradually decreases over time and is close to zero in 2016.
All other parameters are far from zero and do not show any particular trend.
In general, $\alpha_c$ is greater than $\alpha_s$ in CL and $\alpha_c$ is less than $\alpha_s$ in RB over the sample period.
It is known that the self-exciting pattern is due to order splitting
and the mutually-exciting pattern is due to microstructure noise.

Figure~\ref{Fig:mu_beta_10} plots the behavior of exogenous fluctuation parameter $\mu$ (left panel) and persistence parameter $\beta$ (right panel).
As mentioned before, the dynamics of $\mu$ seem related to the dynamics of $\alpha_{w}$.
In general, $\mu$ in CL is larger than RB but the gap is closing.
Meanwhile, the persistence parameter $\beta$ is smaller in CL,
and this implies that persistence in CL is stronger than in RB.
In addition, there is no particular trend in $\beta$.

\subsection{Interpretation of the estimation results}
Through the calibration results discussed in Section~\ref{Sec:Calibration}, 
we figure out the stylized behavioral characteristics exhibited in WTI and gasoline futures prices.
First, we observe that the main source for comovement of the two price processes is CL
by the fact that $\alpha_{w}$ in CL is greater than $\alpha_{w}$ in RB as shown in Figure~\ref{Fig:alpha_nw_10}. 
Both prices have comovement propensity if when a widening event occurs driven by either of them. 
However, in terms of the absolute magnitude for power to move, CL is greater than RB.

\begin{figure}[h]
\begin{center}
\includegraphics[width=0.48\textwidth]{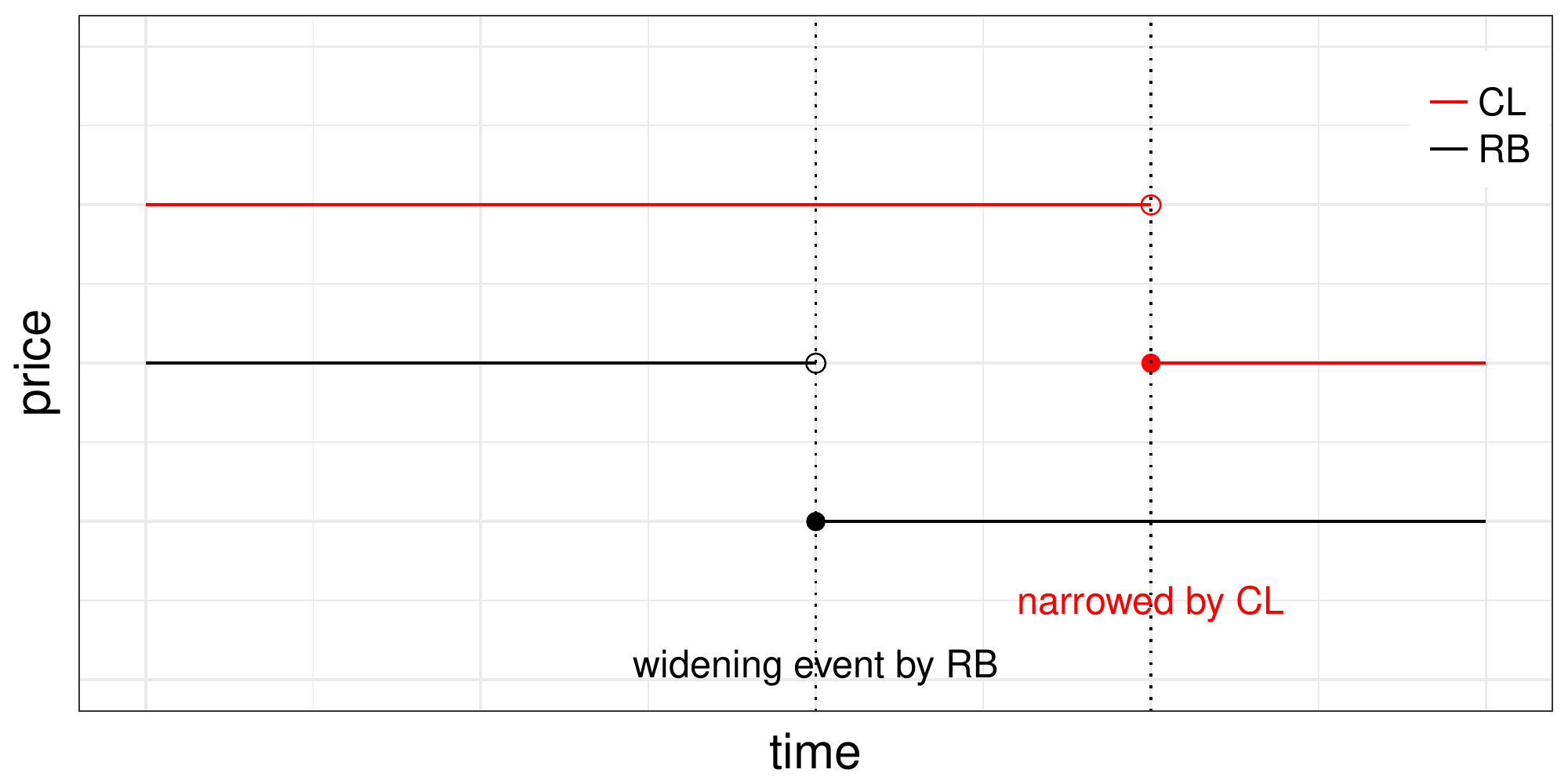}\quad
\includegraphics[width=0.48\textwidth]{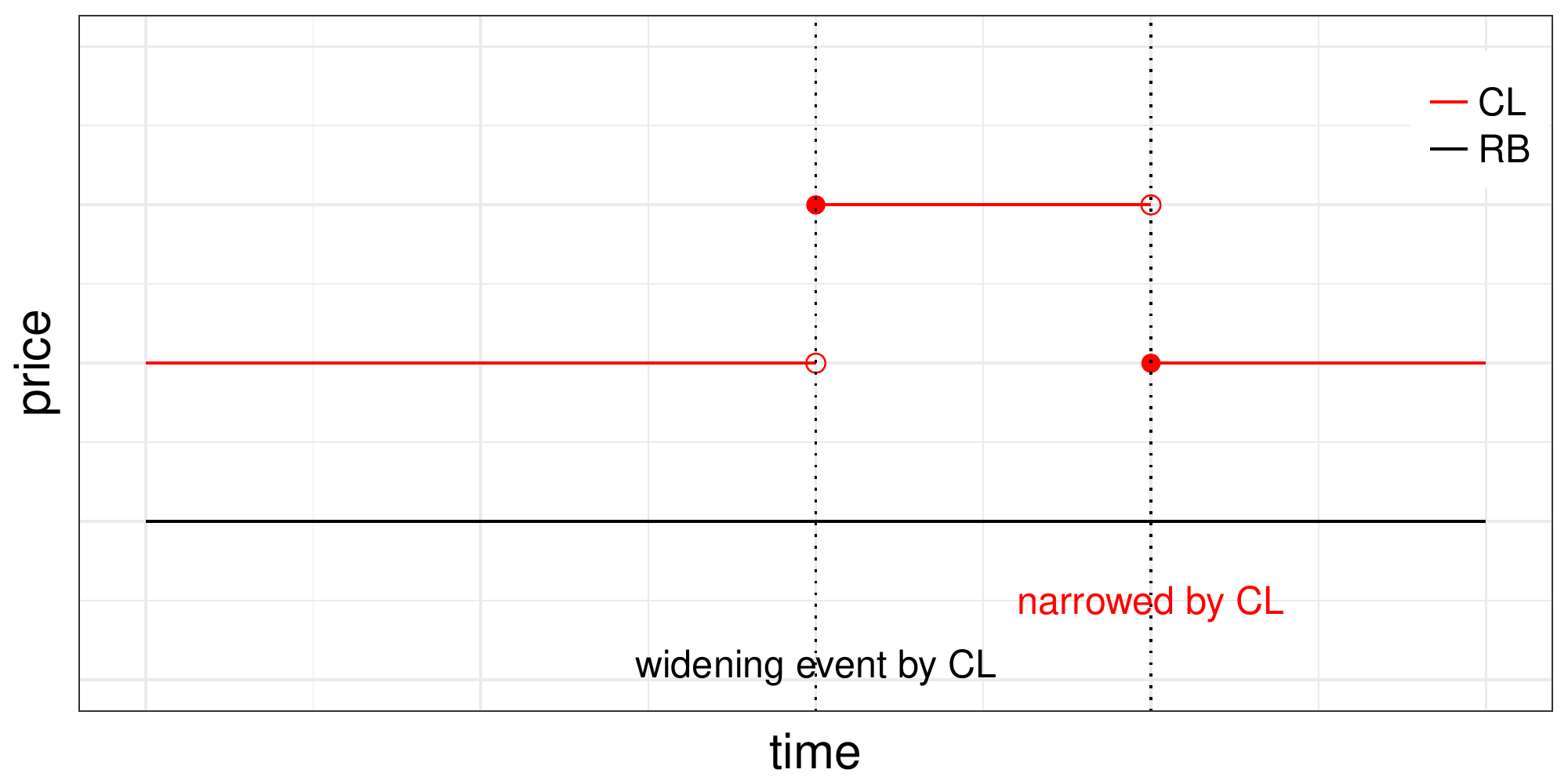}
\end{center}
\caption{Examples of widening events and their possible consequences}\label{Fig:widening}
\end{figure}

Next, in a different viewpoint, we consider the following situation.
Suppose that the CL price is larger than RB, and a widening event happens by a RB's down movement,
as illustrated in the left of Figure~\ref{Fig:widening}.
In this case, the narrowing tendency can be facilitated by 
(i) increase of the up intensity in RB
or (ii) increase of the down intensity in CL.
Increase of the up intensity in RB price is measured by $\alpha_{2c}$,
since it is caused by the previous downward movement in RB which is captured by the individual Hawkes price model within the RB price.
Similarly, increase of the down intensity in CL is represented by $\alpha_{1w}$,
since this jumping is caused by a widening event affecting the intensity of CL.

When comparing $\alpha_{2c}$ in RB and $\alpha_{1w}$ in CL in Figures~\ref{Fig:alpha_nw_10}~and~\ref{Fig:alpha_cs_10},
it is observed that generally $\alpha_{1w}$ is larger than $\alpha_{2c}$.
Although it does not strictly imply that the two price paths are more likely to be narrowed by CL, we deduce that the influence of CL on narrowing is quite significant.

We also suppose that CL is greater than RB and a widening event is activated by a CL's up movement, as in the right of Figure~\ref{Fig:widening}.
In the same manner, the narrowing tendency can be facilitated by 
(i) increase of the up intensity in RB
or (ii) increase of the down intensity in CL.
Increase of the up intensity in RB is captured by $\alpha_{2w}$,
since it is caused by a widening event affecting the intensity of RB.
The increase of the down intensity in CL is represented by $\alpha_{1c}$,
since it is caused by an upward movement of CL.

Comparing $\alpha_{2w}$ in RB and $\alpha_{1c}$ in CL in Figures~\ref{Fig:alpha_nw_10}~and~\ref{Fig:alpha_cs_10},
it is shown that $\alpha_{1c}$ has a much larger value than $\alpha_{w}$ generally.
This means that if the two price levels get widened
then tendency to converge toward each other increases for both prices,
but the magnitude of power to move is much more significant in CL than in RB. 
We can deduce that, in many cases, the flocking feature between CL and RB is more likely to be owing to CL.
One possible interpretation of this result is that participants in WTI crude oil futures market seem to use information from the gasoline market more actively than do participants in the gasoline futures market from the WTI crude oil futures market.


\section{Systemic Risk in Market Microstructure}\label{Sec:Comparison}

So far, we proposed the Hawkes flocking model with an approximative stability condition given as the branching ratio $\rho_{M}$ of the branching matrix $M$ in Section~\ref{Sec:Model},
and we then calibrated the proposed model from the high-frequency data for WTI crude oil and gasoline futures in Section~\ref{Sec:Test}.

In this section, we discuss how to quantify systemic risk existing within and between the two price processes in a microscopic level based on the definition of the branching ratio.
We finally compare the systemic risk levels embedded in the two futures prices by using different measurements through (i) a branching ratio analysis in the proposed model and (ii) CoVaR which is a widely used method for systemic risk.
To implement the CoVaR for the empirical data of WTI crude oil and gasoline futures prices, we adopt 
a CoVaR-copula approach, which is developed by \cite{Reboredo&Ulgolini2015} and \cite{Reboredo2015}.

\subsection{The Hawkes flocking model and its relevance of systemic risk}\label{Sec:Relevance}

Understanding systemic risk is one of major topics in modern financial risk management in terms of definition, quantification, and regulation.
A wide range of literature discusses on systemic risk.
One strand of the literature relates to a market-based risk measure, which are 
CoVaR proposed by \cite{Adrian&Brunnermeier2016}; marginal expected shortfall by \cite{Acharya}; SRISK by \cite{Brownlees&Engle2017};
distress insurance measure by \cite{Huang2009}.
Another strand is under a network-based approach developed by \cite{Elliott}, \cite{Rogers}, \cite{Capponi} based on the idea by \cite{Noe}.
In addition, ones empirically observe systemic risk in a time-varying perspective through various data set. For example,
\cite{Lucas2014}, \cite{Oh2017} use CDS spread; \cite{Reboredo2015} takes stock prices; \cite{Jammazi2015} uses stock-bond returns; \cite{Choros-Tomczyk2014}, \cite{Okhrin2017} employ portfolio credit derivative prices.

In terms of defining systemic risk, this study is inspired by the idea proposed by \cite{Danielsson&Shin2003}, \cite{Danielssonetal2012}; and it is also related to \cite{Filimonov&Sornette2012}, \cite{Hardimanetal2013}.
\cite{Danielssonetal2012} firstly characterize systemic risk by introducing the concept of ``endogenous risk'', which is defined as 
the additional risk that the financial system adds on top of equilibrium risk as commonly understood.
In addition, price movements would be consistent with price efficiency if they were entirely driven by payoff-relevant fundamental news. However, a large part of volatility/correlation is due to a number of feedback effects.
Although the volatility/correlation stem from exogenous factors, a large part of eventual realized magnitude is due to the amplification within the system by the exogenous news.

Taking this argument into the market microstucture,
the Hawkes intensity process in (\ref{Eq:excited}) may have the following meanings:
the base intensity is related to a portion due to the incorporation of fundamental news; and 
the feedback kernel has a role of an endogenous feedback due to the trading patterns of market participants over the incorporation of fundamental news.
In this context, we employ the stability condition estimated with the branching ratio in (\ref{Eq:spectral}) as an indicator of systemic risk in two price processes.
In other words, we may assess that a systemic risk level stays higher as the branching ratio is closer to one and does lower as it is closer to zero.
Thus, we observe how empirical values of (\ref{Eq:spectral}) change over time using the estimated parameters in Section~\ref{Sec:Test}.

With more precise investigation for the proposed feedback kernel, 
it is composed of two parts, the self/mutually-exciting kernel defined in (\ref{Eq:phi}) and the flocking kernel in (\ref{eq:K}) and (\ref{eq:phi_nw}).
We analyze the systemic risk level by distinguishing it into two factors -- endogeneity within a single price process and  interaction between two price processes.
As a systemic risk indicator for microstructure dynamics, we compute a {\it quarter-wise branching ratio} from the feedback kernel, which is related to each factor of endogeneity and interaction in each WTI and gasoline futures price.

The quarter-wise branching ratio is described as follows.
The branching matrix ${M}$ is of four-by-four size that contains 16 components as follows.
\begin{equation}\label{Eq:branchmatrix2}
M=\left[
\begin{array}{cc;{2pt/2pt}cc}
\frac{\alpha_{1s}}{\beta_1} & \frac{\alpha_{1c}}{\beta_1} & \frac{\alpha_{1w}}{2\beta_1} & \frac{\alpha_{1n}}{2\beta_1} \\ 
\frac{\alpha_{1c}}{\beta_1} & \frac{\alpha_{1s}}{\beta_1} & \frac{\alpha_{1n}}{2\beta_1} & \frac{\alpha_{1w}}{2\beta_1} \\ \hdashline[2pt/2pt]
\frac{\alpha_{2w}}{2\beta_2} & \frac{\alpha_{2n}}{2\beta_2} & \frac{\alpha_{2s}}{\beta_2} & \frac{\alpha_{2c}}{\beta_2} \\
\frac{\alpha_{2n}}{2\beta_2} & \frac{\alpha_{2w}}{2\beta_2} & \frac{\alpha_{2c}}{\beta_2} & \frac{\alpha_{2s}}{\beta_2}
\end{array}
\right]
\end{equation}
The matrix can be divided by a four distinct quadrants (indicated by dash lines) according to the role of parameters, as shown in (\ref{Eq:branchmatrix2}).
The components in the first and fourth quadrants represent the branching ratios that affect the self/mutually-exciting factors in CL and RB price processes, respectively. 
Similarly, the components in the second and third quadrants represent the branching ratios that affect the flocking behavior (widening and narrowing events) in CL and RB price processes, respectively.

To measure the extent of amplification caused by the self/mutually-exciting factor explained by $\alpha_{is}, \alpha_{ic}$ and flocking factor by $\alpha_{in}, \alpha_{iw}$ separately,
we examine the branching ratios by component. 
By taking average of four components belonging to each quadrant, we obtain
\begin{equation}\label{Eq:ratios}
\frac{\alpha_{1s} + \alpha_{1c}}{\beta_1},\,\,\,\,\frac{\alpha_{2s} + \alpha_{2c}}{\beta_2} ,\,\,\,\,\frac{\alpha_{1n} + \alpha_{1w}}{2\beta_1},\,\,\,\,\frac{\alpha_{2n} + \alpha_{2w}}{2\beta_2}. 
\end{equation}
Each value has the following interpretation:
First,
$(\alpha_{is} + \alpha_{ic})/{\beta_i}$ indicates the average frequency of the occurrence of offspring events due to price upward or downward movements out of total arrivals for each CL or RB futures price, for $i=1, 2$.
This can be interpreted as {the level of endogeneity} that exists in WTI crude oil futures market for $i=1$ and gasoline future market for $i=2$. 
Next,
$(\alpha_{in} + \alpha_{iw})/(2\beta_i)$ corresponds to the average frequency of the occurrence of offspring events due to widening and narrowing events between the two prices out of total arrivals.
This can be interpreted as {the level of interaction} from gasoline to WTI crude oil futures markets for $i=1$ and the opposite direction for $i=2$.

\subsection{A CoVaR-copula approach and its implementation}\label{subsec:CoVaR}

A systemic risk measure focuses on a tail distribution for potential losses of given portfolios in order to investigate a spillover effect from one to another.
Among the aforementioned systemic measures in practice, CoVaR is a most widely used systemic risk measure, proposed by \cite{Adrian&Brunnermeier2016}.

The CoVaR is the VaR for the financial system conditional on the fact that an individual financial institution is under stress.
Let $R_t^i$ be the return for the financial market as a whole at time $t$ and let $R_t^j$ be the return for market $j$ at time $t$.
The original definition of CoVaR is given by 
\begin{equation}\label{Eq:Def_CoVaR_AB}
\mathbb{P}\left(R_t^i \leq \text{CoVaR}_{\beta, t}^{i|j, \alpha = q} | R_t^j = \text{VaR}_{q, t}^j\right) = \beta.
\end{equation}
This is the VaR when the return of market $j$ stands at the VaR with the $q$-percent confidence level. After that, by replacing the condition to make it more realistic, the definition was extended to the following form.

\begin{equation}\label{Eq:Def_CoVaR}
\mathbb{P}\left(R_t^i \leq \text{CoVaR}_{\beta, t}^{i|j} | R_t^j\leq \text{VaR}_{\alpha, t}^j\right) = \beta,
\end{equation}
where $\text{VaR}_{\alpha, t}^j$ is the VaR for market $j$, measuring the maximum loss that market $j$ may experience for confidence level $1-\alpha$ and a specific time horizon, that is, the $\alpha$-quantile of the return distribution for the market $j$: $\mathbb{P}(R_t^j\leq \text{VaR}_{\alpha,t}^j)=\alpha$.

Using the CoVaR, the systemic risk can be measured by the delta CoVaR ($\Delta\text{CoVaR}$), which is the difference between the VaR of whole market conditional on the distressed state of market $j$, that is, $R_{t}^j\leq \text{VaR}_{\alpha,t}^j$,
and the VaR of the the whole market conditional on the normal state of market $j$, that is, $R_{t}^j = \text{VaR}_{\alpha=50\%,t}^j$. Note that usually the median quantile is considered. 

To implement the defined $\Delta\text{CoVaR}$ as presented,
we consider a copula function approach to implement $\Delta\text{CoVaR}$. 
By consolidating the definition of CoVaR proposed by the relevant literature, 
we present the formula as an analytic form using a copula function. 
The following proposition shows the formula for computing $\Delta\text{CoVaR}$ with
the relevant proof in Appendix~\ref{App:A}.

\begin{proposition}\label{Prof:CoVaR}
For a uniform vector $(U, V)$ with a copula function $C$, 
let $\zeta_v(u)$ denote the conditional distribution by
\begin{equation}\label{Eq:h-function}
\zeta_v(u) = \mathbb{P}(U\leq u |V=v) = \frac{\partial C(u, v)}{\partial v},
\end{equation} 
and $C_{\alpha}^{-1}(\cdot)$ denote the inverse of $C_{\alpha}: x\rightarrow C(\cdot, \alpha)$.
Then, the $\beta$-quantile $\Delta\text{CoVaR}_t^{i|j}$ of asset $i$'s return $R^i$ conditional on asset $j$'s return $R^j$ is given as an analytic form:
\begin{equation}\label{Eq:DeltaCoVaR}
\Delta\text{CoVaR}_t^{i|j} = \text{CoVaR}_{\beta, t}^{i|j} - \text{CoVaR}_{\beta, t}^{i|j, \alpha = 0.5}.
\end{equation}
Each part of (\ref{Eq:DeltaCoVaR}) is computed by
\begin{equation}\label{Eq:CoVaR}
\text{CoVaR}_{\beta, t}^{i|j} = F^{-1}_{R_t^i}\left(C^{-1}_{\alpha}(\alpha\beta)\right) \,\,\,\text{and}\,\,\, \text{CoVaR}_{\beta, t}^{i|j, \alpha = q} = F_{R_t^i}^{-1}\left(h^{-1}_{q}(\beta)\right),
\end{equation}
where $F_{R_t^i}$ is the marginal distribution function of $R_t^i$.

\end{proposition}

To apply the notion of $\Delta \text{CoVaR}$ to our study, 
we compute it based on daily returns for two price dynamics.
This measure captures the level of systemic risk in a day.
We replicate the computation presented in Proposition~\ref{Prof:CoVaR} to the transaction data over a regular time stamp $t$. 

Let $R_t^1$ and $R_t^2$ be the returns for daily observations of $C_1(t)$ and $C_2(t)$, respectively, that is,
\begin{equation}\label{Eq:R1R2}
R_{t}^1 = \frac{C_1(t + \Delta t) - C_1(t)}{C_1(t)},\,\,\,\, R_{t}^2 = \frac{C_2(t + \Delta t) - C_2(t)}{C_2(t)},
\end{equation}
where $\Delta t$ is given by a one-day length.

We consider the following four kinds of copula with different tail dependencies and symmetries: 
the Gaussian copula with tail independence; Student $t$ copula with symmetric tail dependence; Gumbel copula with upper tail dependence and lower tail independence; Clayton copula with upper tail independence and lower tail dependence.
The details are specified in Table~\ref{Table:Copula}.
\begin{table}
\begin{center}
\begin{tabular} {cccccc}
\hline
Copula & Distribution $C(u, v ; \cdot)$& Range of $\theta$ & $\lambda_L$ & $\lambda_U$ & Generator $\psi(\cdot)$\\
\hline
Guassian & $\Phi_{\theta}(\Phi^{-1}(u), \Phi^{-1}(v))$ & $(-1, 1)$ & 0 & 0 & $\times$\\
Student $t$& $T_{\nu, \theta}(t_{\nu}^{-1}(u), t_{\nu}^{-1}(v))$ & $(-1, 1)$ & \multicolumn{2}{c}{$2t_{\nu+1}\left(-\frac{\sqrt{\nu+1}\sqrt{1-\theta}}{\sqrt{1+\theta}}\right)$} &$\times$ \\
Gumbel & $\exp\left(-[(\ln u)^{\theta} + (\ln v)^{\theta}]^{1/\theta}\right)$ & $[1,\infty)$ & 0 & $2 - 2^{1/\theta}$& $(- \ln t)^{\theta}$\\
Clayton & $\left(u^{-\theta} + v^{-\theta} - 1\right)^{-\frac{1}{\theta}}$ & $(0,\infty)$ & $2 - 2^{1/\theta}$ & 0& $(t^{-\theta} - 1)/\theta$\\
\hline
\end{tabular}
\end{center}
\caption{Bivariate copula models with correlation parameter $\theta$, upper tail dependence $\lambda_L$, lower tail $\lambda_U$ dependence parameters}
\label{Table:Copula}
\end{table}
Appendix~\ref{Sec:CoVaR} presents how to implement time-varying CoVaR using the copula functions.

For application of the CoVaR's definition to our data set, we set $R_t^1$ and $R_t^2$ by the daily returns of CL and RB futures prices, respectively, such as defined in (\ref{Eq:R1R2}). 
We compute one-day $\Delta \text{CoVaR}^{1|2}_t$ and one-day $\Delta \text{CoVaR}^{2|1}_t$ based on $R_t^1, R_t^2$\footnote{Since profit returns (not loss returns) are used in the computation of CoVaR and VaR, the CoVaR and VaR values with the minus sign are considered throughout the test. The minus VaR is usually given as a positive value when the quantile level is greater than 50\%. }.
The value of one-day $\Delta \text{CoVaR}^{1|2}_t$ is interpreted as the extent to which extreme downward changes in gasoline futures price (conditioned variable) contribute to the systemic risk in WTI crude oil futures price for a day at time t. 
Conversely, the value of $\Delta \text{CoVaR}^{2|1}_t$ indicates the contribution of extreme downward changes in WTI crude oil futures price (conditioned variable) to systemic risk in gasoline futures price.

By finding the best fitting copula among the aforementioned ones (the detail procedure is explained in Appendix~\ref{App:D}), we compute time-varying $\text{CoVaR}_{t}$ using Proposition~\ref{Prof:CoVaR} with an analytic form of the $\zeta_{\alpha}$ function. 
Since the Student $t$ copula is chosen as having the best fit to our data over the whole test period by the AIC and BIC tests, function (\ref{Eq:hfunction_T}) is employed only in our analysis. 
We pick the 95\% quantile level for computing the CoVaR and VaR used in the conditioned part of CoVaR, that is, $\alpha=5\%$ and $\beta=5\%$.

Figure~\ref{Fig:CoVaR} (Appendix~\ref{App:C}) shows the dynamics of one-day 95\% $\text{CoVaR}^{1|2}_t$ and $\text{CoVaR}^{2|1}_t$ when the conditional variable is given as the distressed situation and when it is given as the normal situation $\alpha=50\%$ as presented in (\ref{Eq:Def_CoVaR_AB}) and (\ref{Eq:Def_CoVaR}), respectively, 
from 2007 to 2016.
The shaded areas represent the three large drops in the WTI crude oil and gasoline futures prices as mentioned in Figure~\ref{Fig:Price_history}.
We find that both CoVaR and $\Delta\text{CoVaR}$ values significantly increase in distressed time periods compared with other normal times.
Moreover, relative contributions of the WTI crude oil futures to the systemic risk in gasoline futures, and vice versa, change almost similarly over time.

\subsection{Comparison of branching ratios with CoVaRs}

\begin{figure}[]
\begin{center}
\includegraphics[width=0.48\textwidth]{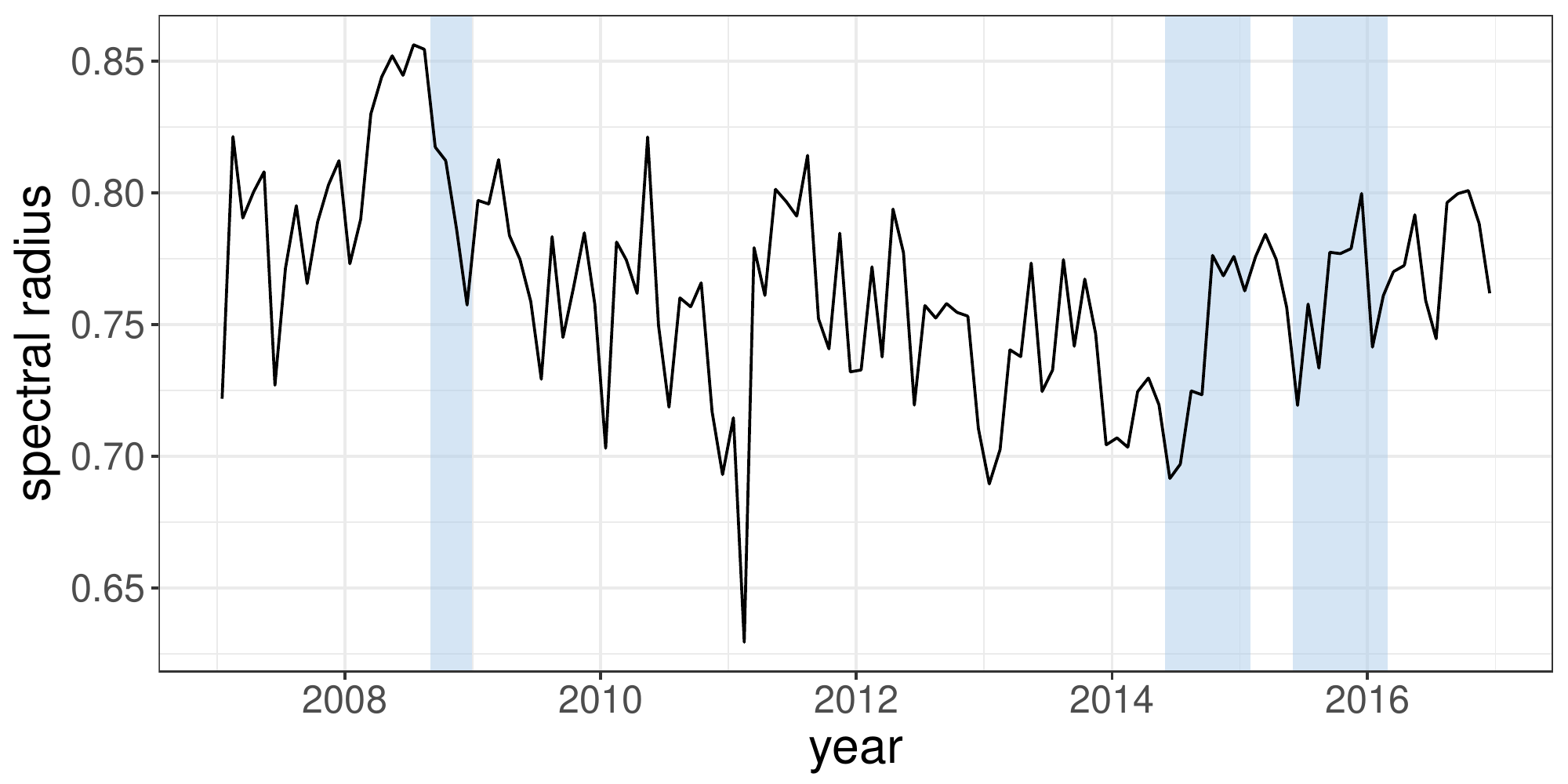}
\end{center}
\caption{Illustration of spectral radius from January 2007 to December 2016}
\label{Fig:ratio_total}
\end{figure}

\begin{figure}[]
\begin{center}
\includegraphics[width=0.48\textwidth]{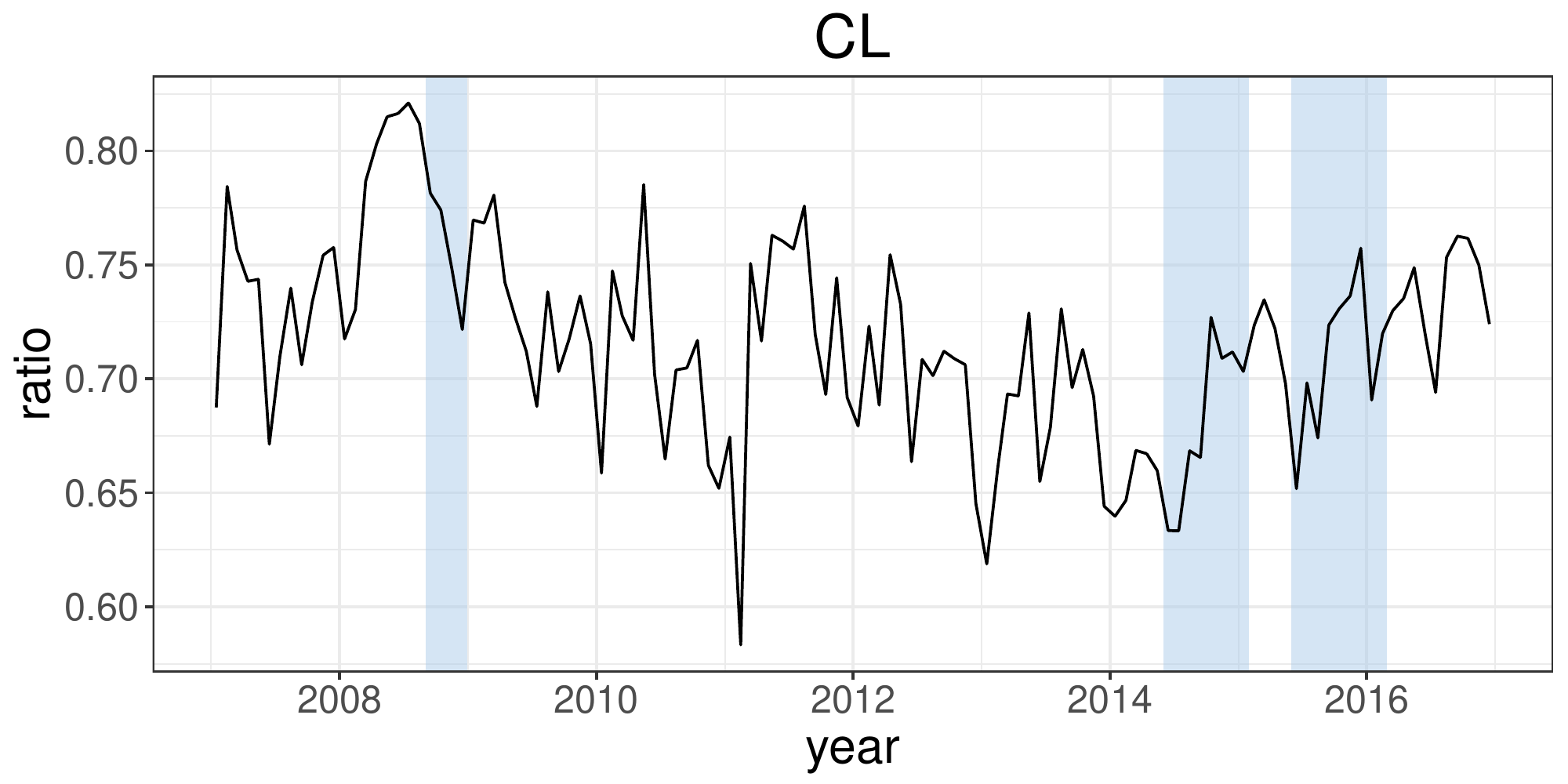}\quad
\includegraphics[width=0.48\textwidth]{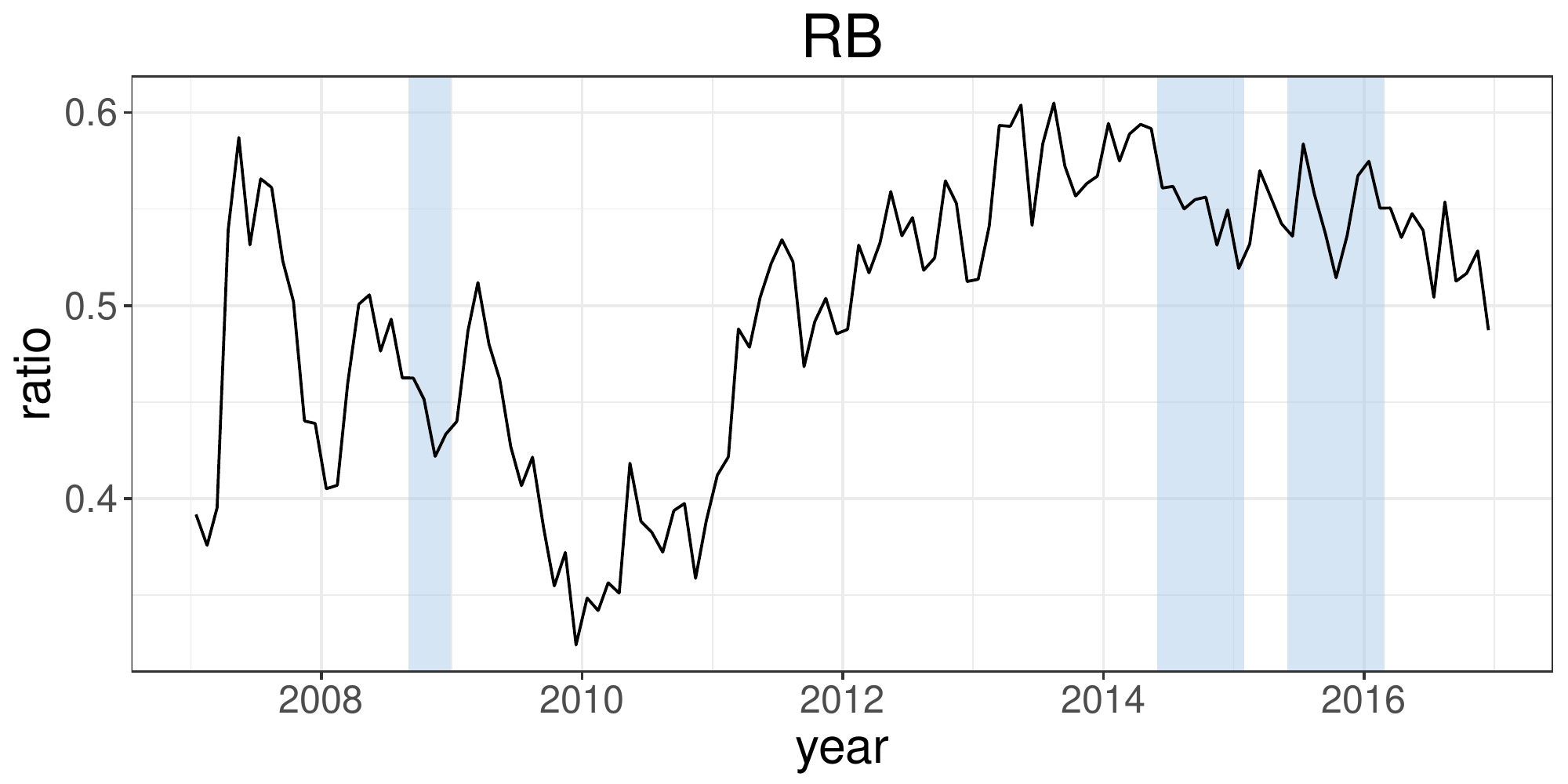}
\end{center}
\begin{center}
\includegraphics[width=0.48\textwidth]{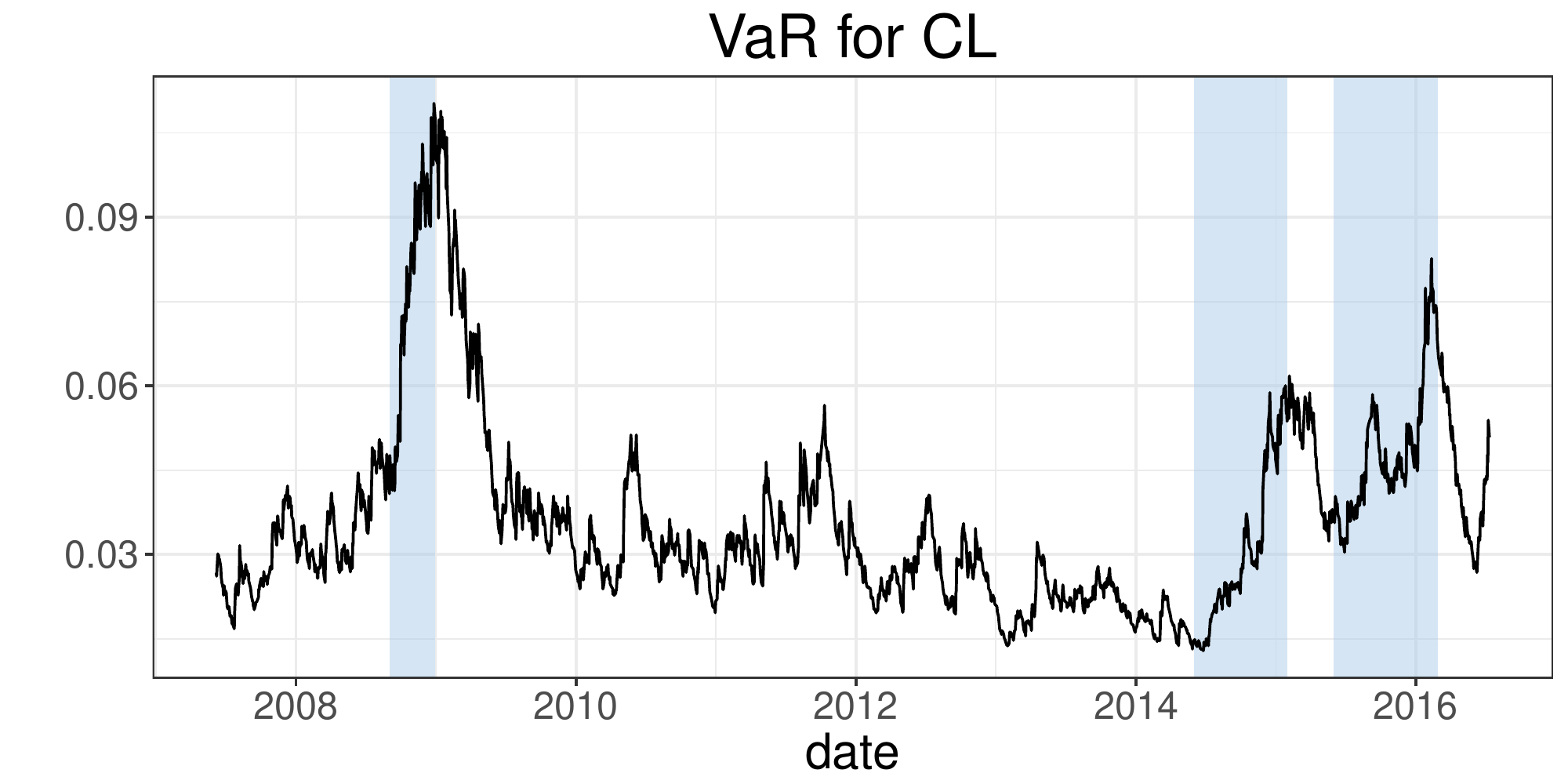}\quad
\includegraphics[width=0.48\textwidth]{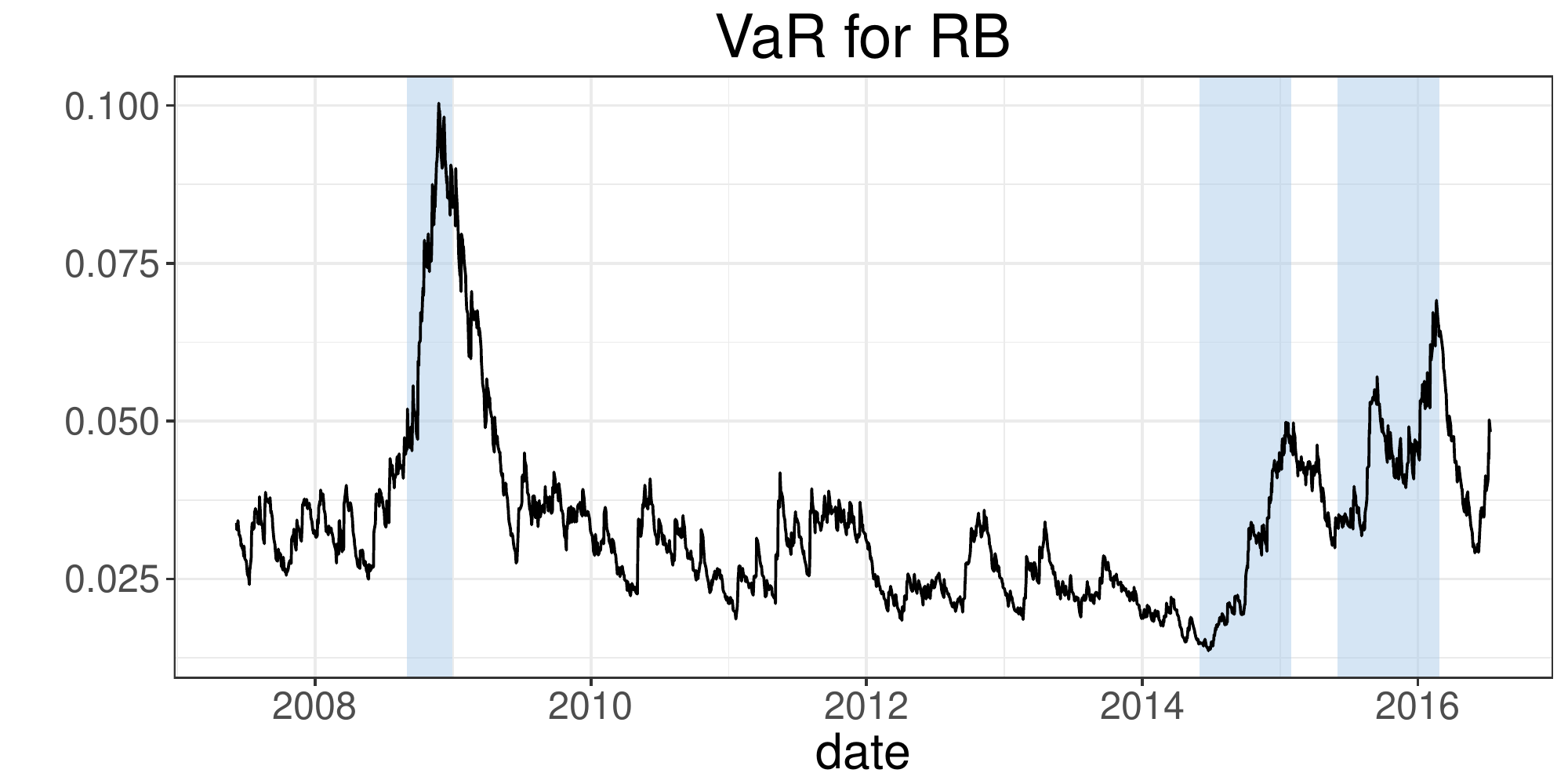}
\end{center}
\caption{Illustration of evolution of $(\alpha_{1s} + \alpha_{1c})/\beta_1$ for CL (top, left), $(\alpha_{2s} + \alpha_{2c})/\beta_2$ for RB (top, right), time-varying one-day 95\% $\mathrm{VaR}$ for CL (bottom, left), and $\mathrm{VaR}$ for RB (bottom, right)
from January 2007 to December 2016}
\label{Fig:ratio_sc}
\end{figure}

\begin{figure}[h]
\begin{center}
\includegraphics[width=0.48\textwidth]{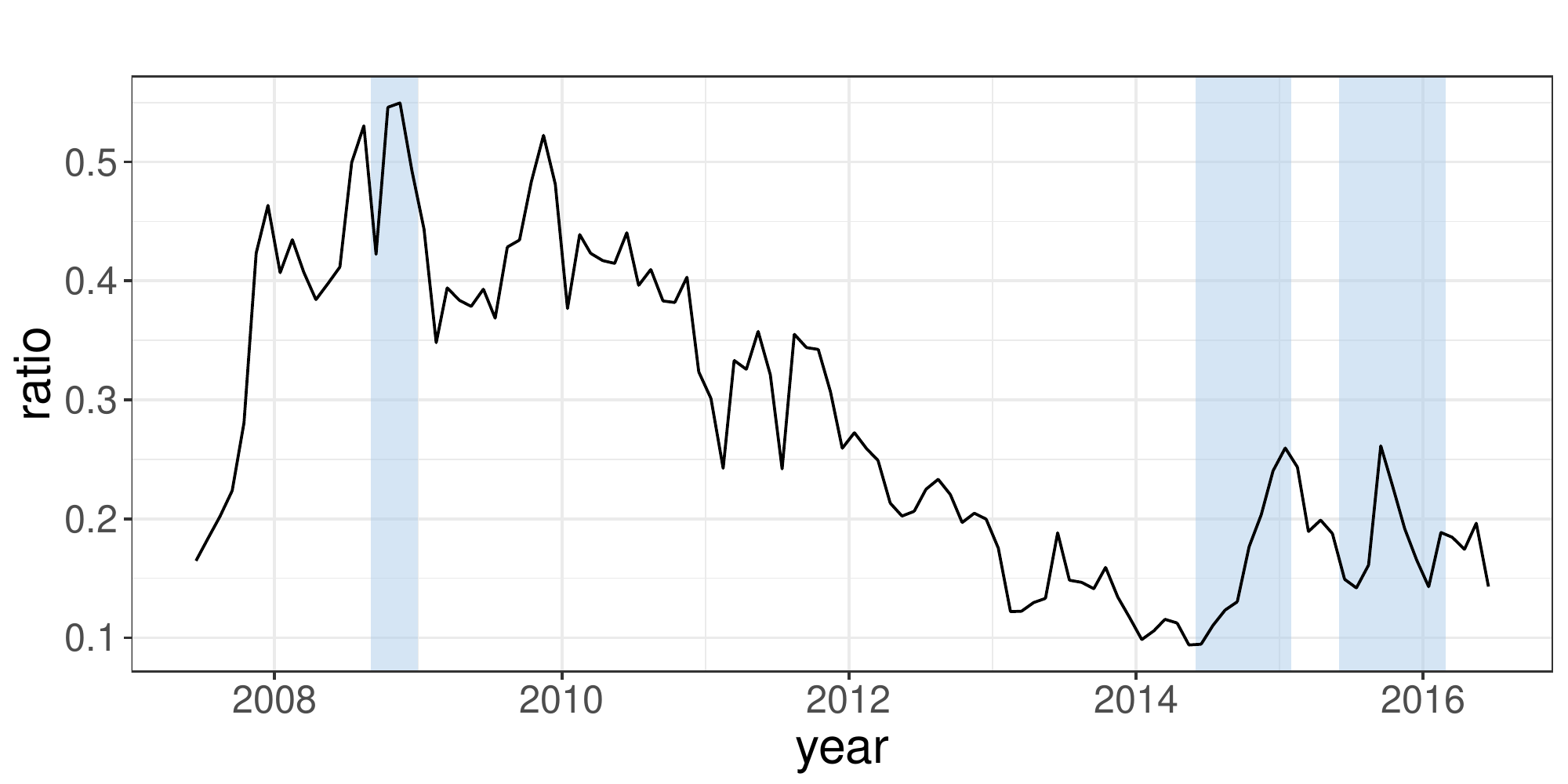}\quad
\includegraphics[width=0.48\textwidth]{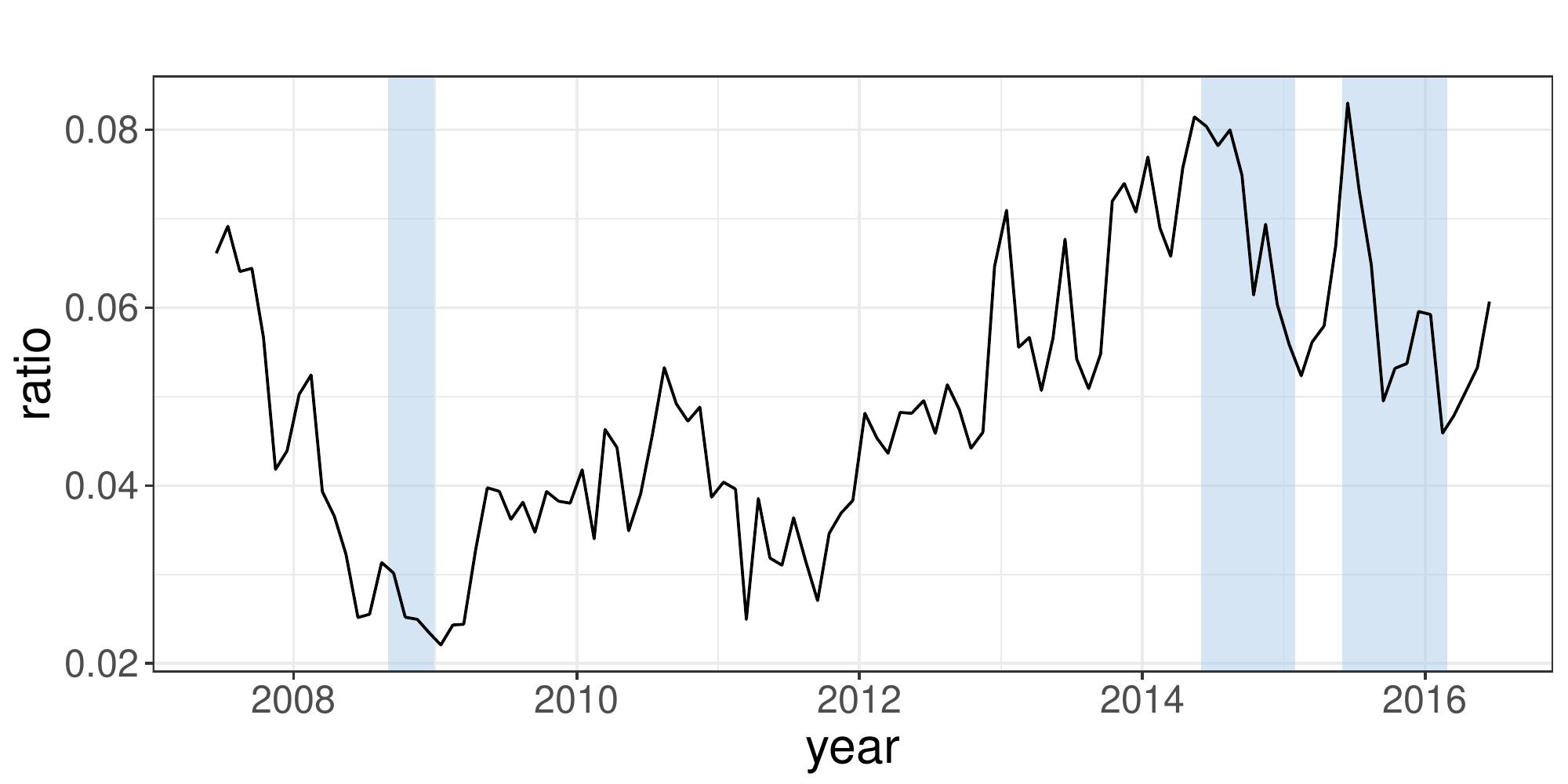}
\end{center}
\begin{center}
\includegraphics[width=0.48\textwidth]{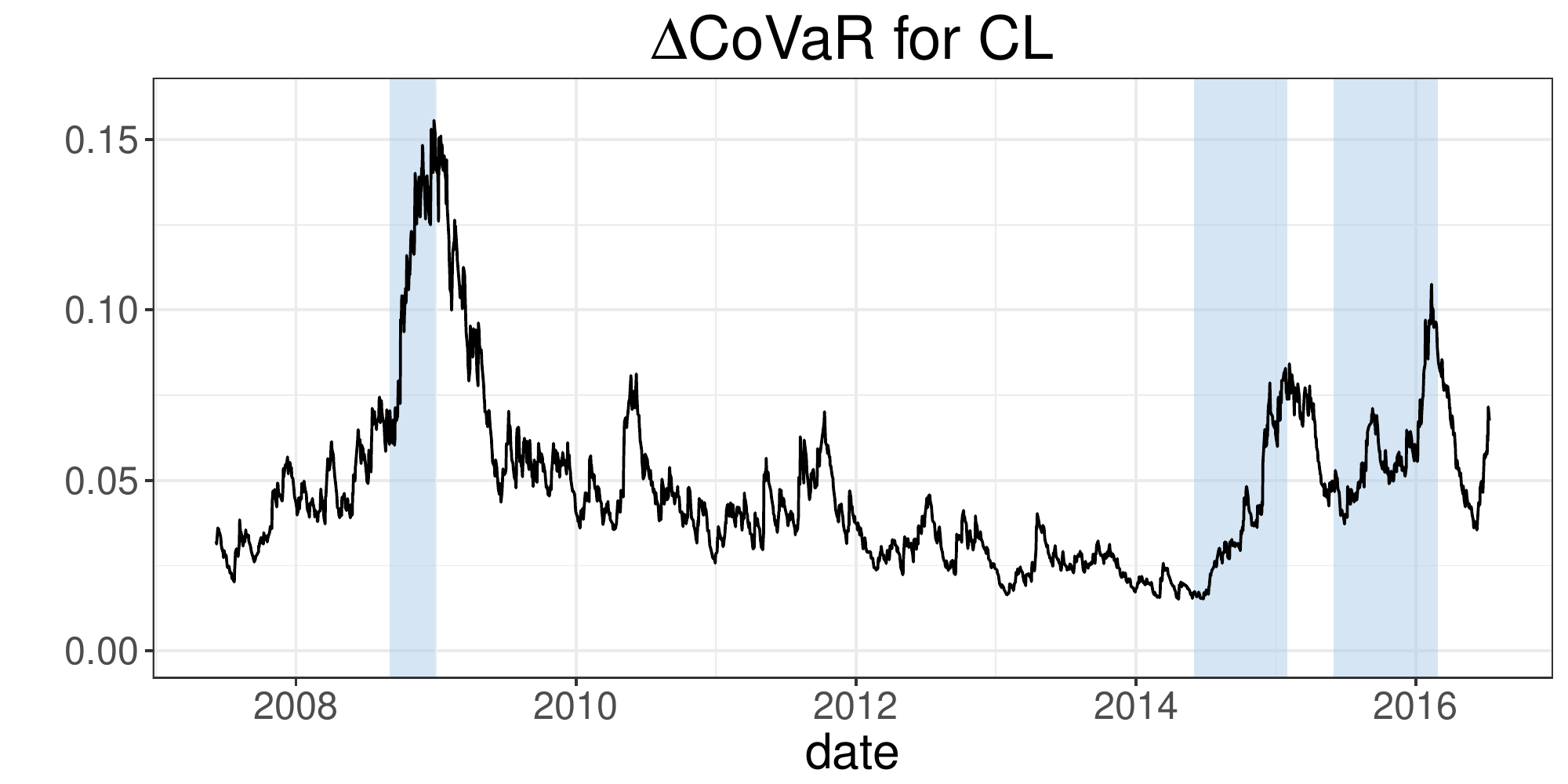}\quad
\includegraphics[width=0.48\textwidth]{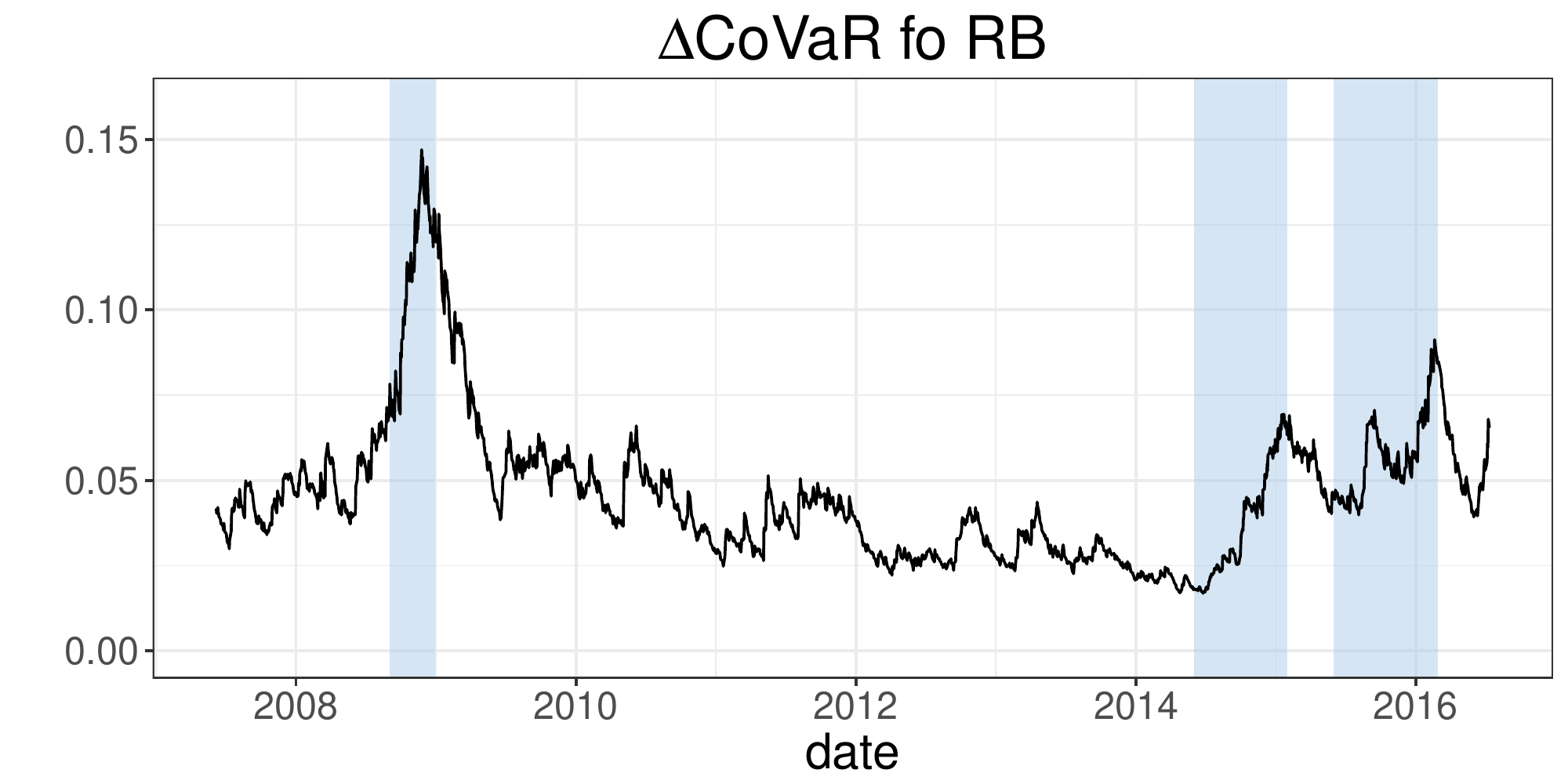}
\end{center}
\caption{Illustration of evolution of $(\alpha_{1n} + \alpha_{1w})/2\beta_1$ (top, left), $(\alpha_{2n} + \alpha_{2w})/2\beta_2$ (top, right), 
$\Delta\mathrm{CoVaR}^{1|2}_t$ (bottom, left), and $\Delta\mathrm{CoVaR}^{2|1}_t$ (bottom, right) where $R_t^1$ and $R_t^2$ are given by the CL and RB daily returns at time $t$, respectively, from January 2007 to December 2016}
\label{Fig:ratio_nw}
\end{figure}

We simulate the branching ratio $\rho_M$ in \eqref{Eq:spectral} and the quarter-wise branching ratios in \eqref{Eq:ratios} using the best fitting kernel parameters $\alpha_{is}, \alpha_{ic}, \alpha_{in}, \alpha_{iw},$ and $\beta_i$ for $i=1, 2$ in the Hawkes flocking model with $p=1/2$, as discussed in Subsection~\ref{SubSec:Calibration}. 
Moreover, the branching ratios are compared with VaR and CoVaR as a benchmark of the systemic risk.
The relevant results are displayed in Figures~\ref{Fig:ratio_total}, \ref{Fig:ratio_sc}, and \ref{Fig:ratio_nw}.

Figure~\ref{Fig:ratio_total} illustrates time-varying $\rho_M$.
We observe that mid-2008 had the highest level of spectral radius at around 85\% just before the collapse of Lehman Brothers in September 2008. 
With the onset of the global credit crisis, the overall level decreased until the beginning of 2011 when it was the lowest at around 63\% during the test period up to December 2016.

Figure~\ref{Fig:ratio_sc} presents the evolution of the quarter-wise branching ratios $(\alpha_{1s} + \alpha_{1c})/\beta_1$ for CL futures, $(\alpha_{2s} + \alpha_{2c})/\beta_2$ for RB futures
and their one-day 95\% VaR values from January 2007 to December 2016. 
For the CL futures, it varies between 58\% and 82\%; however, for RB futures, it varies between 32\% and 60\%.
This implies that the level of endogeneity in CL futures market was consistently higher than that in RB futures prices for the past decade. 
In CL futures in mid-2008, the highest level of endogeneity was recorded just before the onset of the global crisis.
Meanwhile, there were no significant changes for this level in the second and third plunge periods in the CL and RB markets in 2014 and 2016, respectively. 
On the other hand, for one-day 95\% VaR values, precipitous rises occurred on mid-2008, 2014, and 2016 in both CL and RB markets, and the overall flows of VaR in CL and RB markets were similar. 

Figure~\ref{Fig:ratio_nw} presents the evolution of the quarter-wise branching ratios $(\alpha_{1n} + \alpha_{1w})/(2\beta_1)$ for the CL process affected by the RB price fluctuation, 
$(\alpha_{2n} + \alpha_{2w})/(2\beta_2)$ for the RB process affected by CL price fluctuation, 
and one-day 95\% $\Delta \text{CoVaR}^{1|2}_t$ and $\Delta \text{CoVaR}^{2|1}_t$ from January 2007 to December 2016.
At a microscopic level, 
the degree to which RB affects CL has values ranging between 10\% and 55\%; 
however, the degree to which CL affects RB is between 2\% and 8\%.
It infers that the level to which RB price affects CL price has been consistently higher than that in its opposite direction over the past 10 years. 
Moreover, a reverse pattern was observed between the relative influences on the two high-frequency price processes. 
For the degree of the impact of the change in RB price on CL price, the highest level was recorded in late 2008, it gradually decreased after that but rose slightly during the second and third plunge periods.
On the other hand, the level to which the CL price affects RB price was the lowest in late 2008, but it increased to the highest level during the second and third plunge periods.
For delta CoVaR values, the extent of their relative contribution to systemic risk due to each futures market was almost symmetric. 

Throughout the test, we find some remarkable facts about the futures markets of WTI crude oil and gasoline in terms of high-frequency structure.
First, the overall systemic risk level in the two futures markets was the highest before the onset of the global credit crisis, and there was no considerable change in overall systemic risk in these markets when the prices plunge occurred in 2014 and 2016. 
Second, when we compare the levels of endogeneity embedded in each futures market, the level of the WTI market was a significantly higher than that in the gasoline market. 
Moreover, since the WTI crude oil market is more actively affected by change in the gasoline market, it is more likely to react promptly to a delicate change in the gasoline market than in the opposite case.
Last, the levels of the risk interaction between the two markets, that is, from WTI crude oil to gasoline and vice versa, were very asymmetric. However, the degree of the difference has been reducing steadily over the past decade.

\section{Concluding Remark}\label{Sec:Conclusion}

We propose the Hawkes flocking model to quantify systemic risk in high-frequency markets.
The model is designed to capture self/mutually-exciting features as well as cross-exciting on the intensity processes depending on the relative position of asset prices
as the price difference is narrowed or widened.
In the empirical study, we observe a micro-level behavior between the two futures markets of WTI crude oil and gasoline.
We see that when the difference of the two prices narrows, no additional flocking phenomenon occurs, but, when they get widened, a strong flocking phenomenon occurred.
The Hawkes flocking model-based assessment is highly suitable for application of tick-by-tick data, 
and it is also feasible to capture a delicate change in the level of systemic risk that appears in highly correlated data. 

In terms of the assessment of systemic risk, we compare the results of the branching ratios derived from the Hawkes flocking model with the delta CoVaR, which is
introduced as a benchmark for the proposed metric of the systemic risk.
Estimating the best fit kernel using a ML estimator from the given data set, we obtain the following empirical results. 
The systemic risk level in the WTI crude oil futures price has been consistently higher than that in the gasoline futures price for the test period. 
Furthermore, the change in gasoline futures price has a significantly greater impact on WTI crude oil futures price than in the opposite case, 
which implies that the relative contribution of each price is asymmetric at the microscopic level of price structure.

\section*{Data Availability Statement}
The part of data that support the findings of this study are openly available in figshare at \url{https://doi.org/10.6084/m9.figshare.9114383.v4}, refer to~\citet{Lee2019}.

\section*{Acknowledgements}
This work was supported by “Human Resources Program in Energy Technology” of the Korea Institute of Energy Technology Evaluation and Planning (KETEP), granted financial resource from the Ministry of Trade, Industry \& Energy, Republic of Korea. (No. 20184010201680); and the National Research Foundation of Korea(NRF) grant funded by the Korea government(MSIT) (No.2017R1C1B5017338).

\bibliography{flocking}
\bibliographystyle{chicago}

\appendix

\section{A CoVaR-Copula Approach}\label{Sec:CoVaR}

Under the copula specifications in Table~\ref{Table:Copula}, we follow a three-step procedure to implement the time-varying CoVaR.

\smallskip\noindent
{\it Step 1. Estimating marginal distributions for returns}.

To estimate the marginal distributions for each return $R_t^{\ell}$ for $\ell=1,2$, we use an $\text{ARMA}(p, q)-\text{TGARCH}(r, m)$ model, that is,
\begin{equation}
R_t = \phi_0 + \sum_{j=1}^p\phi_j R_{t-j} + \epsilon_t - \sum_{i=1}^{q}\theta_i\epsilon_{t-i},
\end{equation}
where $p$ and $q$ are non-negative integers and $\phi$ and $\theta$ are the ARMA parameters, respectively.
Here, $\epsilon_t = \sigma_t z_t,$ and $\sigma_t^2$ is the conditional variance given by a TGARCH specification:
\begin{equation}
\sigma_t^2 = \omega + \sum_{k=1}^r \beta\sigma_{t-k}^2 + \sum_{h=1}^{m}\alpha_h\epsilon_{t-h}^2 + \sum_{h=1}^{m}\lambda_h\epsilon_{t-h}^2 \mathbbm{1}_{\{t-h>0\}},
\end{equation}
where $\omega$ is a constant, $\sigma_{t-k}^2$ is the GARCH component, $\epsilon_{t-h}$ is the ARCH component, and $\lambda$ captures asymmetric effects. If $\lambda>0$, then the future conditional variance will increase more following a negative shock than following a positive shock of the same magnitude. 
Here, $z_t$ is an independent and identically distributed random variable with zero mean and unit variance that follows 
a skewed-$t$ distribution, given by \cite{Fernandez}:
\begin{equation*}
f(z_{j,t}; \gamma) = \frac{2}{\gamma + \frac{1}{\gamma}} \left\{ f_\nu \left(\frac{z_{j,t}}{\gamma} \right) \mathbbm{1}_{[0,\infty)}(z_{j,t}) + f_\nu (\gamma z_{j,t}) \mathbbm{1}_{(-\infty,0)}(z_{j,t}) \right\}
\end{equation*}
where $\gamma$ is a skew parameter, and $f_\nu$ is the density of the $t$ distribution with $\nu$ degree of freedom.

\smallskip\noindent
{\it Step 2. Finding the best fitting copula}.

Among the copulas, we find one with the best fit in Table~\ref{Table:Copula} with empirical marginal distributions of $R_t^{\ell}$. We use an ML estimation method, that is,
$$\theta_i = \argmax_{\theta} \sum_{t=i}^{i+d}\ln c(\hat{u}_t, \hat{v}_t; \theta)$$
where $c(\cdot, \cdot; \theta) $ is a copula density obtained by $\partial^2 C(u, v; \theta)/\partial u\partial v$, $\hat{u}_t$, and $\hat{v}_t$ are samples transformed from observations $R_t^{\ell}$ by their empirical distributions obtained in step (1). To estimate $\theta_i$, $d$ days of samples $(\hat{u}_t, \hat{v}_t)$ are used.

\smallskip\noindent
{\it Step 3. Computing $\Delta \text{CoVaR}$}.

We compute the time-varying $\Delta \text{CoVaR}_t$ with the best fitting copula obtained in step (2) and marginal distributions in step (1), as stated in Proposition~\ref{Prof:CoVaR}.
To quantify the systemic impact of an asset price return on another asset price return, we compute the $\beta$-quantile
$\text{CoVaR}_{\beta, t}^{1|2}$ and $\text{CoVaR}_{\beta, t}^{2|1}$ that are defined by
$$\mathbb{P}\left(R_t^1 \leq \text{CoVaR}_{\beta, t}^{1|2} | R_t^2\leq \text{VaR}_{\alpha, t}^2\right) = \beta \,\, \text{and}\,\,\mathbb{P}\left(R_t^2 \leq \text{CoVaR}_{\beta, t}^{2|1} | R_t^1\leq \text{VaR}_{\alpha, t}^1\right) = \beta,$$
respectively. 
Then, we compute $\text{CoVaR}_{\beta, t}^{1|2, \alpha=0.5}$ and $\text{CoVaR}_{\beta, t}^{2|1, \alpha=0.5}$ . 

To implement the first term of $\Delta \text{CoVaR}$ in Proposition~\ref{Prof:CoVaR}, the numerical inversion is needed for the Gaussian and Student $t$ copulas.
For the Gumbel and Clayton copulas with generator $\psi$, $C^{-1}_{\alpha}$ can be easily derived in an explicit form as
$\psi^{-1}(\psi(x) - \psi(\alpha))$ for $x\in(0,\alpha)$.
Thus, (\ref{Eq:CoVaR_closed}) is written as
\begin{equation}\label{Eq:CoVaR_closed_Archi}
\text{CoVaR}_{\beta, t}^{1|2} = F^{-1}_{R_t^1}\left(\psi^{-1}(\psi(\alpha\beta) - \psi(\alpha))\right),
\end{equation}
where $\alpha$ and $\beta$ are the given levels.

Next, for the second term of $\Delta \text{CoVaR}$ in Proposition~\ref{Prof:CoVaR}, computing the conditional copula function $\zeta_{\alpha}(u)$ with a given level $\alpha$ is required as defined in (\ref{Eq:h-function}). 
Depending on a choice of copulas, the functions $\zeta_{\alpha}(u)$ are obtained as analytic forms, which are derived by \cite{Aasetal2009} and \cite{Schepsmeier&Stober}, as follows.
\begin{itemize}
	\item Gaussian copula:
	\begin{equation}
	\zeta_{\alpha}(u) = \Phi\left(\frac{\Phi^{-1}(u) - \theta\Phi^{-1}(\alpha)}{\sqrt{1-\theta^2}} \right)
	\end{equation}
	\item Student $t$ copula with the degree of freedom $\nu$:
	\begin{equation}\label{Eq:hfunction_T}
	\zeta_{\alpha}(u) = t_{\nu+1}\left(\frac{t_{\nu}^{-1}(u) - \theta t_{\nu}^{-1}(\alpha)}{\sqrt{\left(\nu + [t_{\nu}^{-1}(\alpha)]^2 \right)(1-\theta^2)/(\nu + 1)}} \right)
	\end{equation}
	\item Gumbel copula: for $x = (-\ln u)^{\theta}$ and $y = (-\ln \alpha)^{\theta}$
	\begin{equation}
	\zeta_{\alpha}(u) = -\frac{\exp\left(-(x + y)^{\frac{1}{\theta}} \right)\cdot\left(x + y\right)^{\frac{1}{\theta} - 1}\cdot y}{\alpha \ln \alpha}
	\end{equation}
	\item Clayton copula
	\begin{equation}
	\zeta_{\alpha}(u) = \alpha^{-\theta - 1}\cdot \left(u^{-\theta} + \alpha^{-\theta} - 1\right)^{-1-\frac{1}{\theta}}
	\end{equation}
\end{itemize}
Dependence parameter $\theta$ is the value estimated in step (2), and the level of $\alpha$ is chosen as the median (i.e., $\alpha=0.5$). 

By conducting the three-step procedure described above, we compute the time-varying $\Delta \text{CoVaR}$'s.
Then, we compare them with the calibrated parameters from the proposed model in Section 2 to examine the consistency of systemic risk measures for high frequency data.

\section{Proof of Proposition~\ref{Prof:CoVaR}}\label{App:A}

As defined in (\ref{Eq:DeltaCoVaR}), computing $\Delta\text{CoVaR}$ consists of determining two types of CoVaR specified in (\ref{Eq:Def_CoVaR}) and ({\ref{Eq:Def_CoVaR_AB}}). Each part is derived in the following step (i) and (ii).

\smallskip\noindent
{\it (i) The CoVaR defined in (\ref{Eq:Def_CoVaR_AB})}.

It can be computed by using the property of a copula function.
Then (\ref{Eq:Def_CoVaR_AB}) can be written as
\begin{equation}\label{Eq:hfunction}
\zeta_q\left(F_{R_t^j}\left(\text{CoVaR}_{\beta, t}^{d|j, \alpha = q}\right)\right) = \beta.
\end{equation} 
Thus, the CoVaR is given by
\begin{equation}\label{Eq:CoVaR2}
\text{CoVaR}_{\beta, t}^{i|j, \alpha = q} = F_{R_t^j}^{-1}\left(\zeta^{-1}_{q}(\beta)\right).
\end{equation}

\smallskip\noindent
{\it (ii) The CoVaR defined in (\ref{Eq:Def_CoVaR})}.

In this case, the quantile value of a conditional distribution, or, alternatively, of an unconditional bivariate distribution is needed if we express in (\ref{Eq:Def_CoVaR}) as
\begin{equation}\label{Eq:Def_CoVaR2}
\frac{\mathbb{P}\left(R_t^i \leq \text{CoVaR}_{\beta, t}^{i|j}, R_t^j\leq \text{VaR}_{\alpha,t}^j \right)}{\mathbb{P}(R_t^j\leq \text{VaR}_{\alpha,t}^j)} = \beta.
\end{equation}
Given that $\mathbb{P}(R_t^j\leq \text{VaR}_{\alpha,t}^j)=\alpha$, the CoVaR in (\ref{Eq:Def_CoVaR2}) can be expressed as:
\begin{equation}
\mathbb{P}\left(R_t^i \leq \text{CoVaR}_{\beta, t}^{i|j}, R_t^j\leq \text{VaR}_{\alpha,t}^j \right) = \alpha\beta.
\end{equation}
Here, the form (\ref{Eq:Def_CoVaR2}) can be expressed in terms of the joint distribution function of $R_t^i$ and $R_t^j$, $F_{R_t^i, R_t^j}$, as
\begin{equation}\label{Eq:CoVaR_Copula}
F_{R_t^i, R_t^j}\left(\text{CoVaR}_{\beta, t}^{i|j}, \text{VaR}_{\alpha, t}^j\right) = \alpha\beta,
\end{equation}
and that, according to Sklar's theorem (\citealp{Sklar}), the joint distribution function of two continuous variables can be expressed in terms of a copula function.
Hence, (\ref{Eq:CoVaR_Copula}) can be written as
\begin{equation}\label{Eq:CoVaR_Copula2}
C(u, v) = \alpha\beta,
\end{equation}
where $C(\cdot, \cdot)$ is a copula function, $u = F_{R_t^i}(\text{CoVaR}_{\beta, t}^{i|j})$ and
$v = F_{R_t^j}(\text{VaR}_{\alpha, t}^j)$, and where $F_{R_t^i}$ and $F_{R_t^j}$ are the marginal distribution function of $R_t^i$ and $R_t^j$, respectively.
Given its copula representation in (\ref{Eq:CoVaR_Copula2}), the CoVaR can be computed from that equation through copulas in a two-step procedure.
First, we obtain the value of $u = F_{R_t^i}(\text{CoVaR}_{\beta, t}^{i|j})$. 
Since $C(u, v) = \alpha \beta$, where $\alpha, \beta$, and $v$ are given (note that $ v=\alpha$), from the copula function specification we can solve to determine the value of $u$.
Next, taking $u$, we can obtain the CoVaR value as the quantile of distribution $R_t^i$, with a cumulative probability equal to $u$, by inverting the marginal distribution function of $R_t^i$: $\text{CoVaR}_{\beta, t}^{i|j} = F^{-1}_{R_t^i}(u)$.
Letting $C_{\alpha}^{-1}(\cdot)$ be the inverse of $C_{\alpha}: x\rightarrow C(\cdot, \alpha)$, then the CoVaR can be expressed as an analytic form
\begin{equation}\label{Eq:CoVaR_closed}
\text{CoVaR}_{\beta, t}^{i|j} = F^{-1}_{R_t^i}\left(C^{-1}_{\alpha}(\alpha\beta)\right).
\end{equation}

\section{Selection of the best fitting copula in Section~\ref{subsec:CoVaR}}\label{App:D}
\begin{figure}[]
	\begin{center}
		\includegraphics[width=0.4\textwidth]{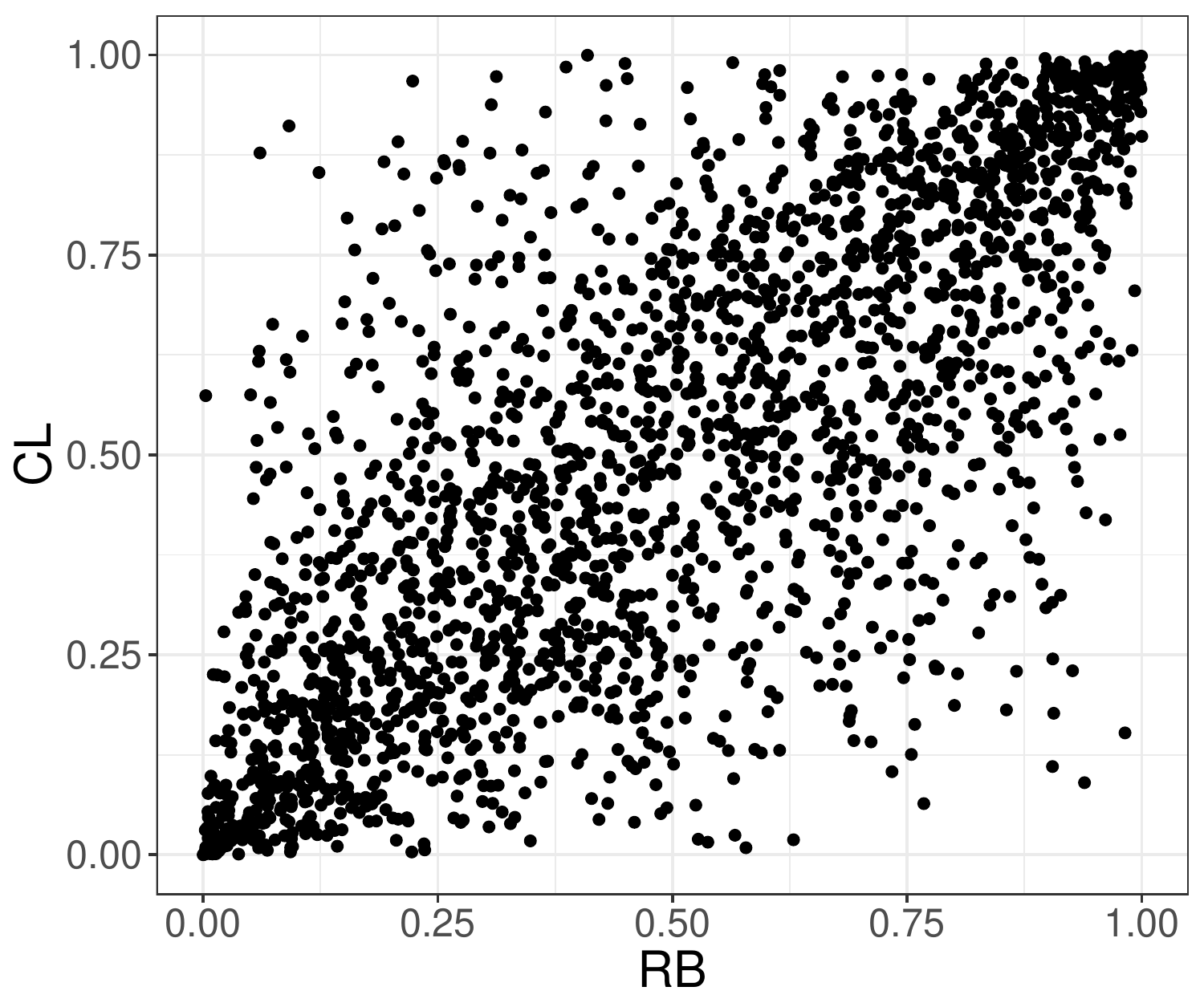}
	\end{center}
	\caption{Scatter plot of $\hat{u}_t, \hat{v}_t$ using the ARMA(1,1)-TGARCH(1,1) model with skewed-$t$ distribution}
	\label{Fig:scatter}
\end{figure}

We estimated marginal distributions for each return series of CL and RB futures prices using the ARMA(1,1)-TGARCH(1,1) model with skewed-$t$ distribution.
Then, we transformed the daily return series into uniform variables such that $\hat{u}_t = \hat{F}_{1}(x_t)$ and $\hat{v}_t = \hat{F}_{2}(y_t)$ where $\hat{F}_i$ is the estimated distribution and $x_t, y_t$ are the standardized returns of $R_t^1, R_t^2$, respectively.
Figure~\ref{Fig:scatter} displays the scatter plot of 2,000 days of pseudo-sample observations, $\hat{u}_t, \hat{v}_t$ under the marginal distribution model. 

\begin{figure}[]
	\begin{center}
		\includegraphics[width=0.48\textwidth]{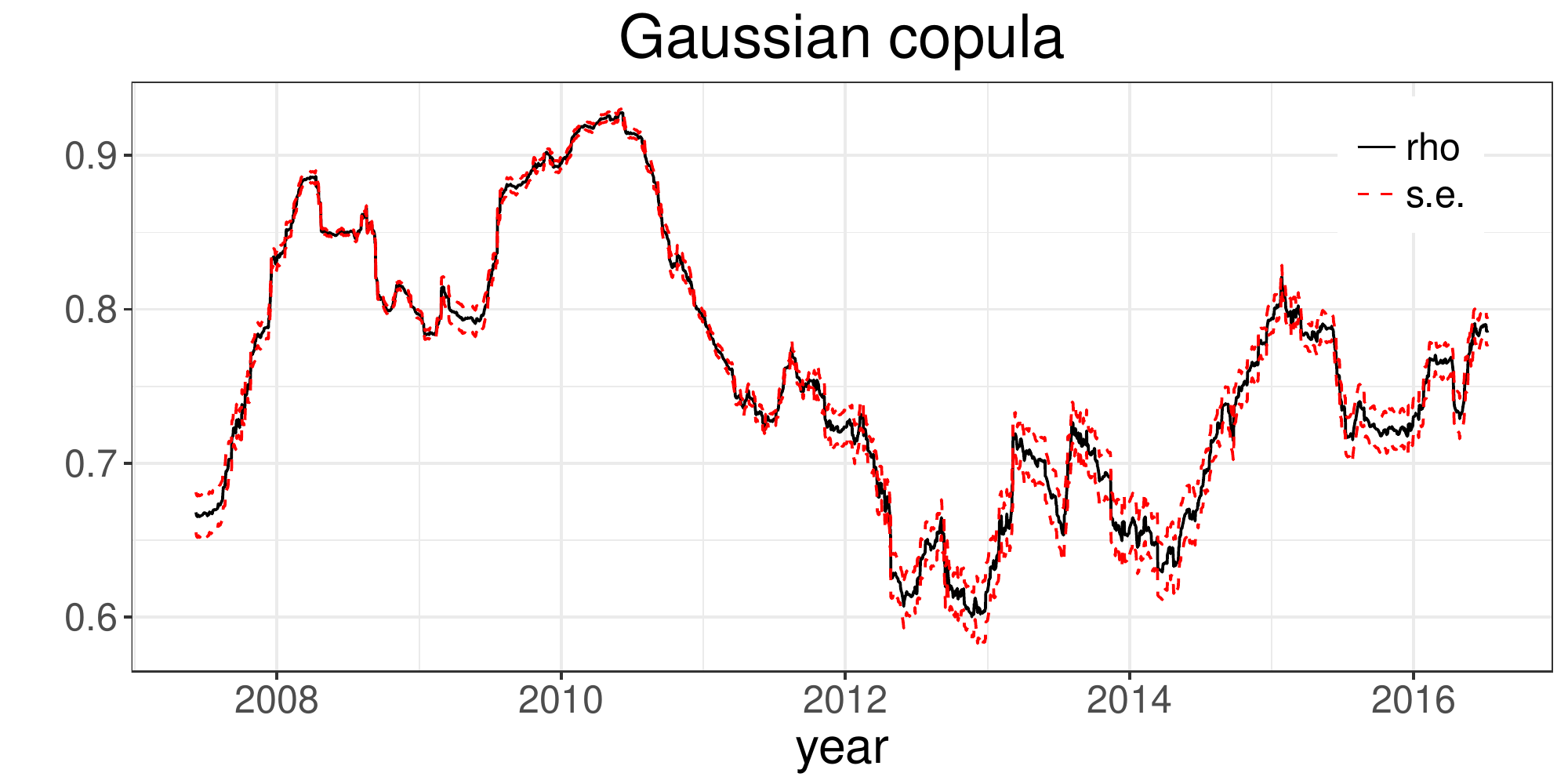}\quad
		\includegraphics[width=0.48\textwidth]{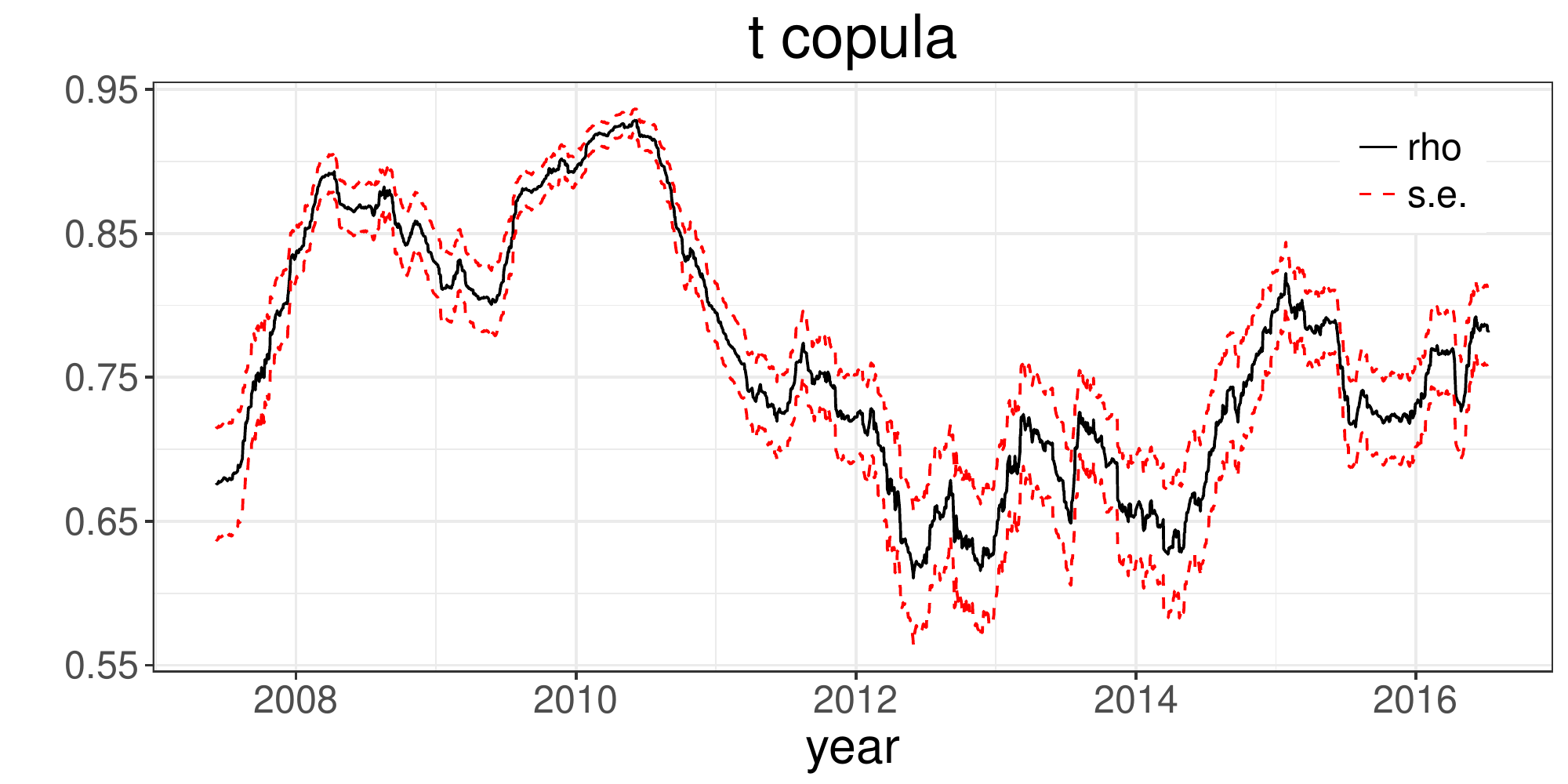}
		\includegraphics[width=0.48\textwidth]{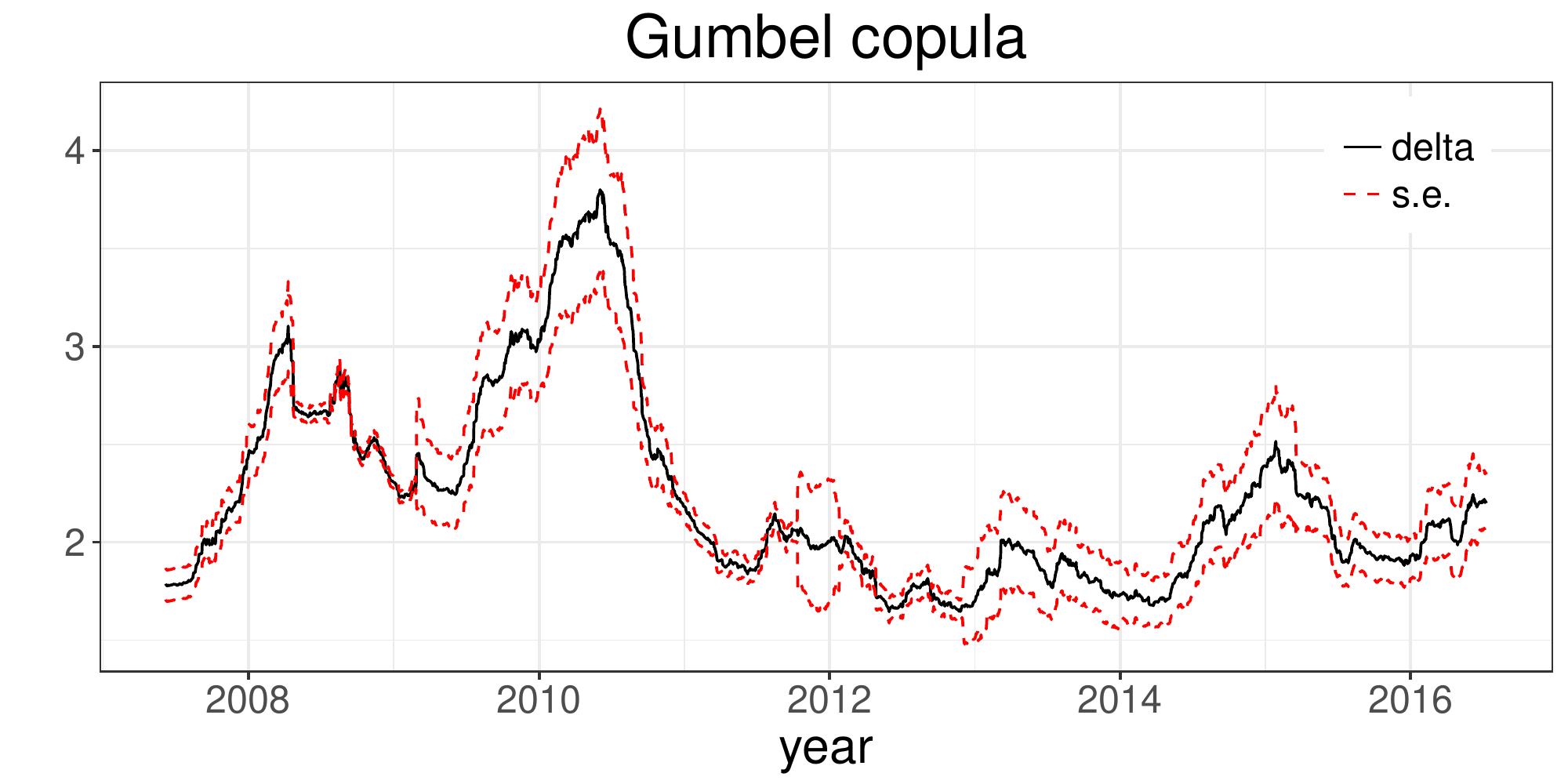}\quad
		\includegraphics[width=0.48\textwidth]{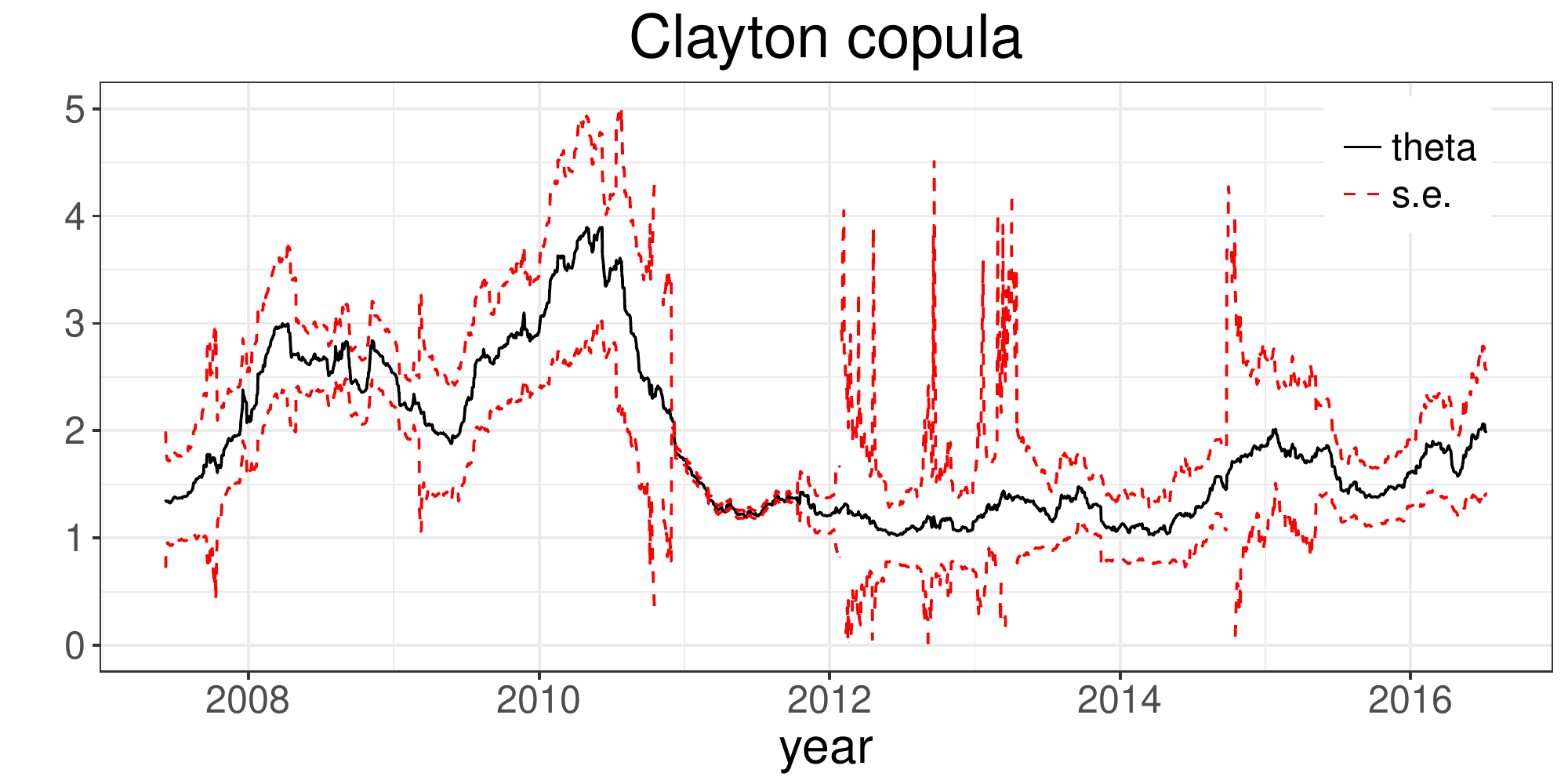}
	\end{center}
	\caption{The evolution of the estimated dependence parameter $\theta$ of CL and RB futures prices using Gaussian copula (top, right), Student $t$ copula (top, left), Gumbel copula (bottom, left), and Clayton copula (bottom, right) with the standard error (red dotted line) from January 2007 to December 2016}
	\label{Fig:compare_alpha_nw}
\end{figure}

We extracted dependence parameters of the copula functions reported in Table~\ref{Table:Copula} using the consecutive $d$ days of series pairs.
Figure~\ref{Fig:compare_alpha_nw} (Appendix~\ref{App:C}) illustrates the dynamics of the estimated parameter $\theta$ of CL and RB futures prices using given copulas with the standard error during the test period from 2007 to 2016.

\begin{figure}[]
	\begin{center}
		\includegraphics[width=0.48\textwidth]{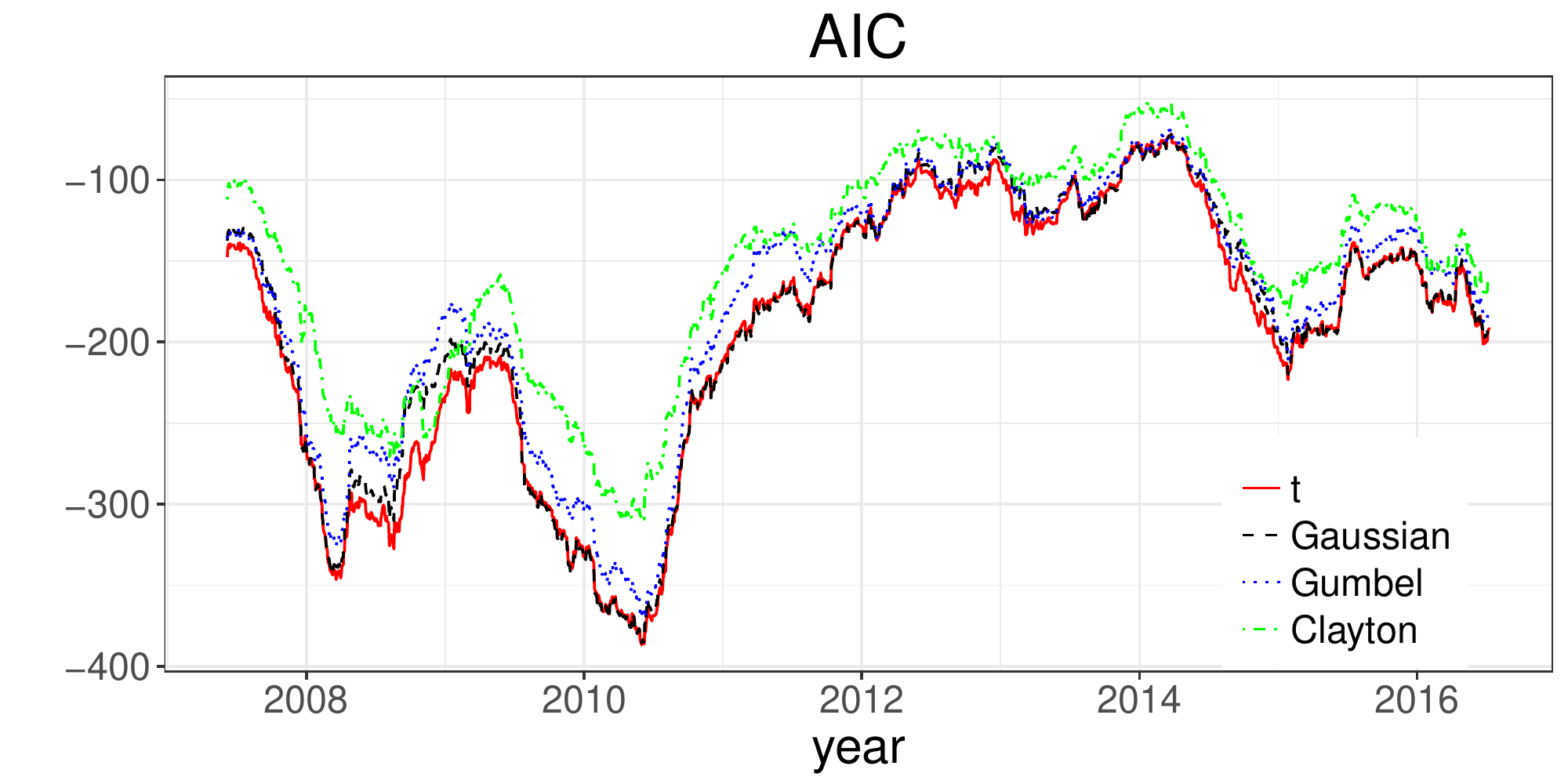}\quad
		\includegraphics[width=0.48\textwidth]{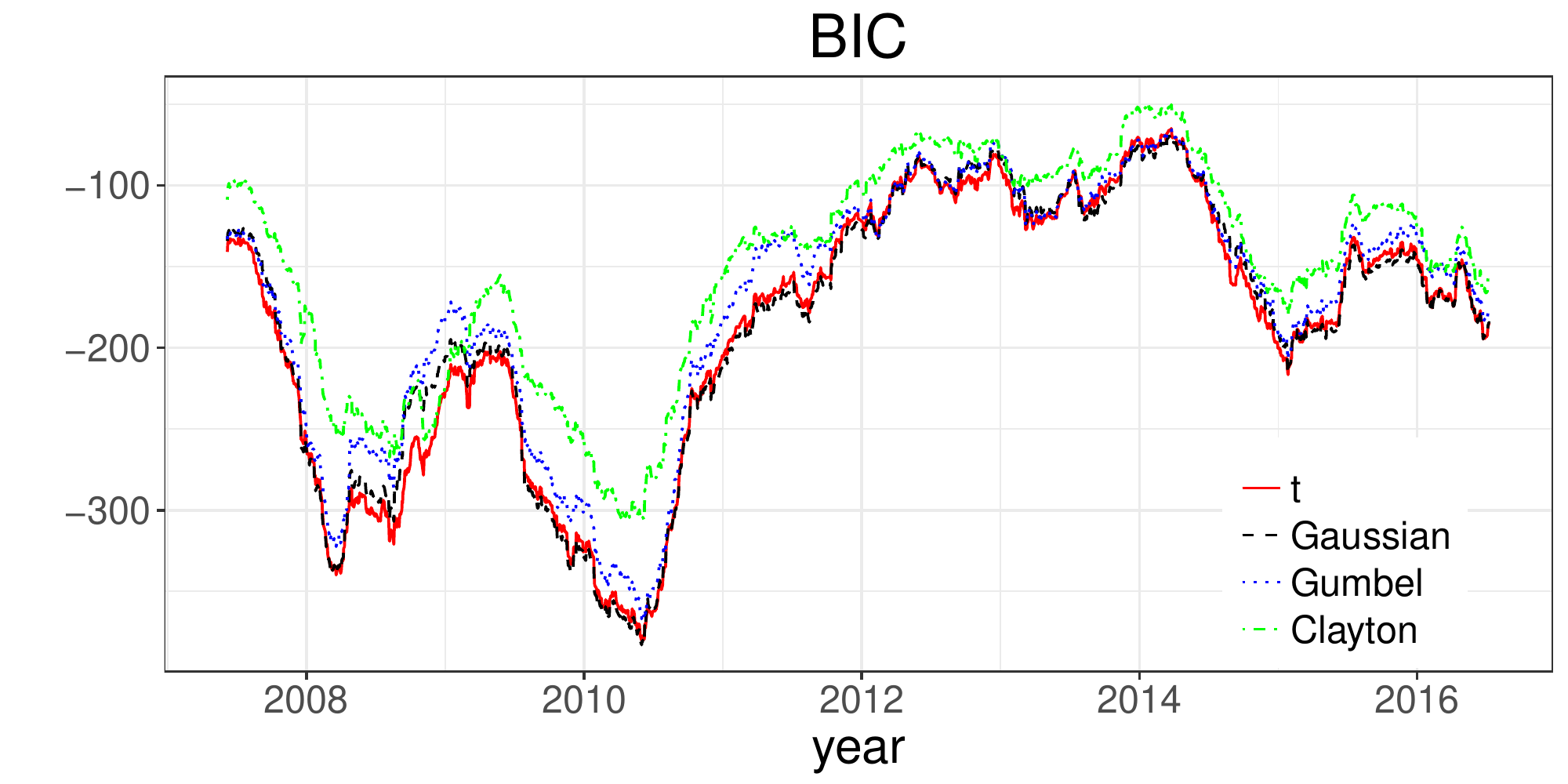}
	\end{center}
	\caption{Results of AIC (left) and BIC (right) measures for Gaussian, Student $t$, Gumbel, and Clayton copulas from January 2007 to December 2016}
	\label{Fig:AICBIC}
\end{figure}

To select the best fitting copula among them, we compare different copula specifications using the commonly used error measures of Akaike information criterion (AIC) and Bayesian information criterion (BIC) in the model selection based on an ML estimation. 
Figure~\ref{Fig:AICBIC} shows the results of AIC and BIC values from estimations of the Gaussian, Student $t$, Gumbel, and Clayton copulas over time.
We find that the Student $t$ copula provides the lowest values by both measures on the entire timeline.

\begin{figure}[]
	\begin{center}
		\includegraphics[width=0.48\textwidth]{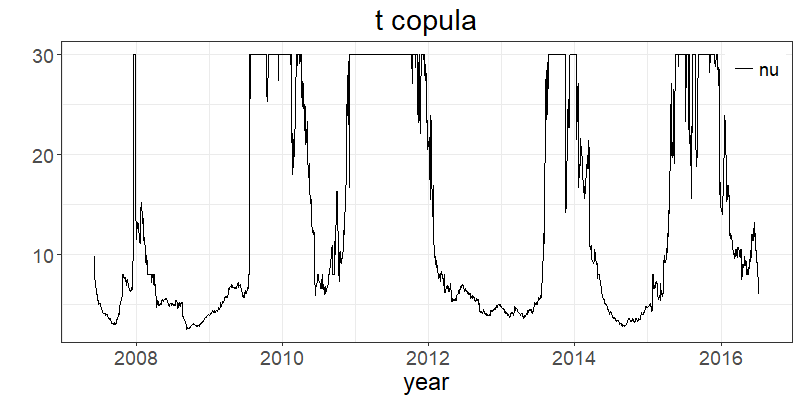}
	\end{center}
	\caption{The evolution of the estimated degree of freedom parameter $\nu$ in the Student $t$ copula for the return pairs of CL and RB futures prices from January 2007 to December 2016}
	\label{Fig:nu}
\end{figure}

In addition, the degrees of freedom $\nu$ for the Student $t$ copula are estimated over time and displayed in Figure~\ref{Fig:nu} (Appendix~\ref{App:C}), which provides evidence of fat tails of the joint distribution of the return pair of CL and RB futures prices.
{The empirical result indicates the existence of positive and symmetric dependence with fat tails between the two futures prices.
	The extent of the overall positive dependency and extreme tail dependency has varied over time.}

\section{More tests for calibration in Section~\ref{Sec:Test}}\label{App:B}

Some of selected estimates are presented in Table~\ref{Table:Hawkes_RBCL}
with the numerically computed standard errors in the parentheses.
In this period of time, the self-exciting term $\alpha_{1s}$ of CL is close to zero and all other parameters are significant.
Since we performed non-constraint parameter estimation, sometimes negative $\alpha_s$ are observed,
but it is better to be considered as zero by the Hawkes-based model definition.
In general, $\mu$ in CL is larger than that in RB implying larger trade frequency in CL
and $\beta$ in RB is larger than that in CL implying longer persistence in RB.

\begin{table}[]
	\centering
	\caption{Estimates for CL (left) and RB (right) under the Hawkes-based model without the flocking-related parameters}
	\label{Table:Hawkes_RBCL}
	\begin{tabular}{ccccc|cccc}
		\hline
		&  \multicolumn{4}{c|}{WTI crude oil} &  \multicolumn{4}{c}{RBOB gasoline} \\
		Date/Maturity   & $\mu_1$  & $\alpha_{1s}$ & $\alpha_{1c}$ & $\beta_1$ &  $\mu_2$  & $\alpha_{2s}$ & $\alpha_{2c}$ & $\beta_2$ \\
		\hline
		2016-03-16 & 0.0780 & 0.0400 & 0.3710 & 0.5120 & 0.0577 & 0.3763 & 0.2001 & 0.9109  \\
		March 2016 & (0.0025) & (0.0050) & (0.0122) & (0.0176) & (0.0017) & (0.0183) & (0.0110) & (0.0439) \\
		2016-04-15 &  0.0731 & 0.0128 & 0.3455 & 0.4827 & 0.0593 & 0.5337 & 0.2126 & 1.3083 \\
		April 2016 & (0.0025) & (0.0044) & (0.0119) & (0.0184) & (0.0014) & (0.0204) & (0.0118) & (0.0446) \\
		2016-05-17 &  0.0872 & -0.0206 & 0.3527 & 0.5106 & 0.0568 & 0.4200 & 0.2712 & 1.1314  \\
		May 2016 & (0.0025) & (0.0037)  & (0.0112) & (0.0195) & (0.0015) & (0.0204) & (0.0144) & (0.0510) \\
		2016-06-16 &  0.1001 & -0.0033 & 0.4315 & 0.6022 & 0.0574 & 0.3405 & 0.1954 & 0.8668 \\
		June 2016 & (0.0027) & (0.0041) & (0.0124) & (0.0189) & (0.0016) & (0.0151) & (0.0101) & (0.0367) \\
		\hline
	\end{tabular}
\end{table}

The estimates and numerically computed standard errors in parenthesis are presented in Table~\ref{Table:flock_RBCL}.
Note that $\alpha_{is}$ and $\alpha_{ic}$ are similar to the estimates in Table~\ref{Table:Hawkes_RBCL} for $i=1,2$,
which means that the additionally introduced $\alpha_{in}$ and $\alpha_{iw}$ do not affect the self/mutually-exciting terms.
The result also shows that $\alpha_{in}$ are close to zero in the selected time period, which means that the price difference for narrowing events does not affect the intensities.
We expand the time range, nonzero positive $\alpha_{1n}$ of CL is also observed, 
but overall $\alpha_{in}$ is quite small and close to zero.
In addition, $\alpha_{iw}$ are significant, which means that the price difference widening events 
increase the intensities so that the two price processes tend to converge to each other.

\begin{table}[]
	\centering
	\caption{Estimates for CL (top) and RB (bottom) with the Hawkes flocking model}
	\label{Table:flock_RBCL}
	\begin{tabular}{ccccccc}
		\hline
		&  \multicolumn{6}{c}{WTI crude oil} \\
		date/maturity & $\mu_1$    & $\alpha_{1n}$ & $\alpha_{1w}$ & $\alpha_{1s}$ & $\alpha_{1c}$  & $\beta_1$    \\
		\hline
		2016-03-16 & 0.0755 & -0.0120  & 0.2265  & 0.0105  & 0.4333   & 0.6079   \\
		March 2016 & (0.0039) & (0.0086)  & (0.0286)  & (0.0053)  & (0.0321)   & (0.0523)   \\
		2016-04-15 & 0.0345 & 0.0046  & 0.1993  & 0.0076  & 0.2992   & 0.4084   \\
		April 2016 & (0.0021) & (0.0078)  & (0.0118)  & (0.0038)  & (0.0108)   & (0.0160)    \\
		2016-05-17 & 0.0670  & -0.0085 & 0.2053  & -0.0318 & 0.3761   & 0.5699   \\
		May 2016 & (0.0024) & (0.0069)  & (0.0119)  & (0.0032)  & (0.0121)   & (0.0217)   \\
		2016-06-16 & 0.0706 & -0.0189 & 0.2748  & -0.0121 & 0.4348   & 0.6142   \\
		June 2016 & (0.0026) & (0.0093)  & (0.0160)   & (0.0032)  & (0.0154)   & (0.0245)   \\
		\hline
		&  \multicolumn{6}{c}{RBOB gasoline} \\
		date/maturity & $\mu_2$    & $\alpha_{2n}$ & $\alpha_{2w}$ & $\alpha_{2s}$ & $\alpha_{2c}$  & $\beta_2$    \\
		\hline
		2016-03-16 & 0.0499 & -0.0109 & 0.1335   & 0.4384   & 0.2306   & 1.2513   \\
		March, 2016 & (0.0037) & (0.0083)  & (0.0269)   & (0.1005)   & (0.0471)   & (0.3109)   \\
		2016-04-15 & 0.0403 & 0.0098  & 0.1719   & 0.4450    & 0.1836   & 1.5166   \\
		April, 2016 & (0.0013) & (0.0073)  & (0.0112)   & (0.0212)   & (0.0125)   & (0.0598)   \\
		2016-05-17 & 0.0413 & 0.0347  & 0.2229   & 0.4555   & 0.3085   & 1.4429   \\
		May, 2016 & (0.0013) & (0.0096)  & (0.0121)   & (0.0200)     & (0.0151)   & (0.0492)   \\
		2016-06-16 & 0.0348 & 0.0186  & 0.1302   & 0.2950    & 0.1882   & 0.9146   \\
		June, 2016 & (0.0014) & (0.0065)  & (0.0077)   & (0.0133)   & (0.0096)   & (0.0352)   \\
		\hline
	\end{tabular}
\end{table}

By data visualization, we confirm the previous discussion.
For the graph, the selected maturity for the futures is February 2016 and estimates are computed on a daily basis over the sample period from January 4 to February 22, 2016.
Figure~\ref{Fig:alpha_nw} compares $\alpha_n$ and $\alpha_w$.
For both CL and RB, the estimated $\alpha_n$ are almost zero, but $\alpha_w$ are far from zero
and $\alpha_w$ in CL is larger than that in RB.

\begin{figure}
	\begin{center}
		\includegraphics[width=0.48\textwidth]{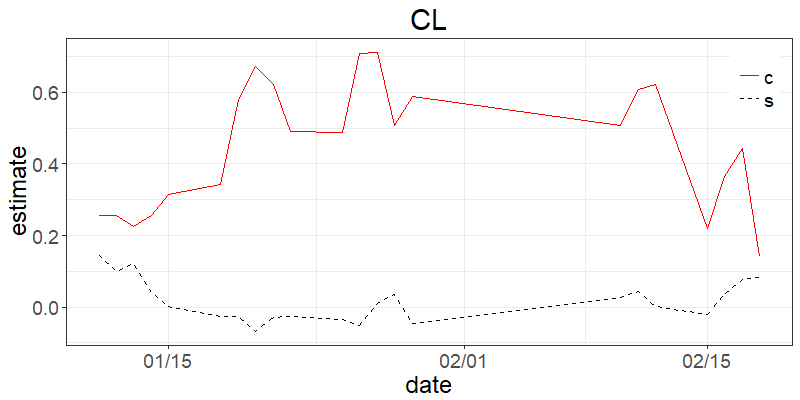}\quad
		\includegraphics[width=0.48\textwidth]{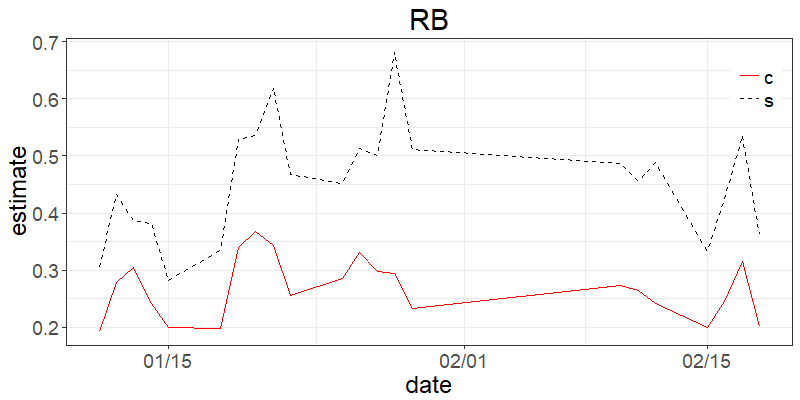}
	\end{center}
	\caption{Comparison of $\alpha_s$ and $\alpha_c$ for CL (left) and RB (right) futures prices with maturity in February 2016}
	\label{Fig:alpha_sc}
\end{figure}

Figure~\ref{Fig:alpha_sc} compares $\alpha_s$ and $\alpha_c$.
For CL, $\alpha_s$ are close to zero and for other cases, $\alpha_c$ in CL and $\alpha_s$, $\alpha_c$ in RB, the estimates are significantly positive.
In addition, $\alpha_s$ is less than $\alpha_c$ in CL, but $\alpha_s$ is greater than $\alpha_c$ in RB over the sample period.

In Table~\ref{Table:mean_estimates}, for each maturity, 
we calculated the averages of estimates on a daily basis using 20 days' data.
In this result, we also observe that $\alpha_{in}$ are close to zero for all maturities.

\begin{table}[]
	\centering
	\caption{Means of estimates for CL and RB}
	\label{Table:mean_estimates}
	\begin{tabular}{ccccc|cccc}
		\hline
		&   \multicolumn{4}{c|}{WTI crude oil} &  \multicolumn{4}{c}{RBOB gasoline} \\
		maturity & $\alpha_{1n}$ & $\alpha_{1w}$ & $\alpha_{1s}$ & $\alpha_{1c}$ & $\alpha_{2n}$ & $\alpha_{2w}$ & $\alpha_{2s}$ & $\alpha_{2c}$  \\
		\hline
		2016-01 & -0.0229 & 0.1828 & 0.0003  & 0.3919 & 0.0197  & 0.1683 & 0.5835 & 0.2900 \\
		2016-02 & -0.0332 & 0.2782 & 0.0106  & 0.4662 & -0.0059 & 0.1290 & 0.4887 & 0.2605 \\
		2016-03 & -0.0237 & 0.2481 & -0.0129 & 0.4621 & -0.0059 & 0.1321 & 0.4625 & 0.2491 \\
		2016-04 & -0.0234 & 0.2161 & -0.0266 & 0.4359 & -0.0020 & 0.1517 & 0.4600 & 0.2333 \\
		2016-05 & -0.0229 & 0.2497 & -0.0233 & 0.4611 & -0.0026 & 0.1480 & 0.4570 & 0.2577 \\
		2016-06 & -0.0187 & 0.1725 & -0.0250 & 0.4136 & 0.0174  & 0.1618 & 0.4737 & 0.2731 \\
		2016-07 & -0.0157 & 0.2321 & -0.0199 & 0.4300 & 0.0185  & 0.1872 & 0.4588 & 0.2579 \\
		2016-08 & -0.0205 & 0.2013 & -0.0149 & 0.4707 & 0.0026  & 0.1787 & 0.4498 & 0.2497 \\
		2016-09 & -0.0145 & 0.2223 & -0.0178 & 0.4870 & -0.0015 & 0.1834 & 0.4574 & 0.2237 \\
		2016-10 & -0.0115 & 0.2205 & -0.0222 & 0.4970 & 0.0005  & 0.1738 & 0.4053 & 0.2212 \\
		2016-11 & -0.0137 & 0.2123 & -0.0164 & 0.4798 & -0.0049 & 0.1576 & 0.4033 & 0.2123 \\
		2016-12 & -0.0073 & 0.2756 & -0.0242 & 0.4879 & 0.0028  & 0.1363 & 0.4169 & 0.2217 \\
		\hline
	\end{tabular}
\end{table}

\section{Figures related to Section~\ref{Sec:Comparison}}\label{App:C}

\begin{figure}[]
	\begin{center}
		\includegraphics[width=0.48\textwidth]{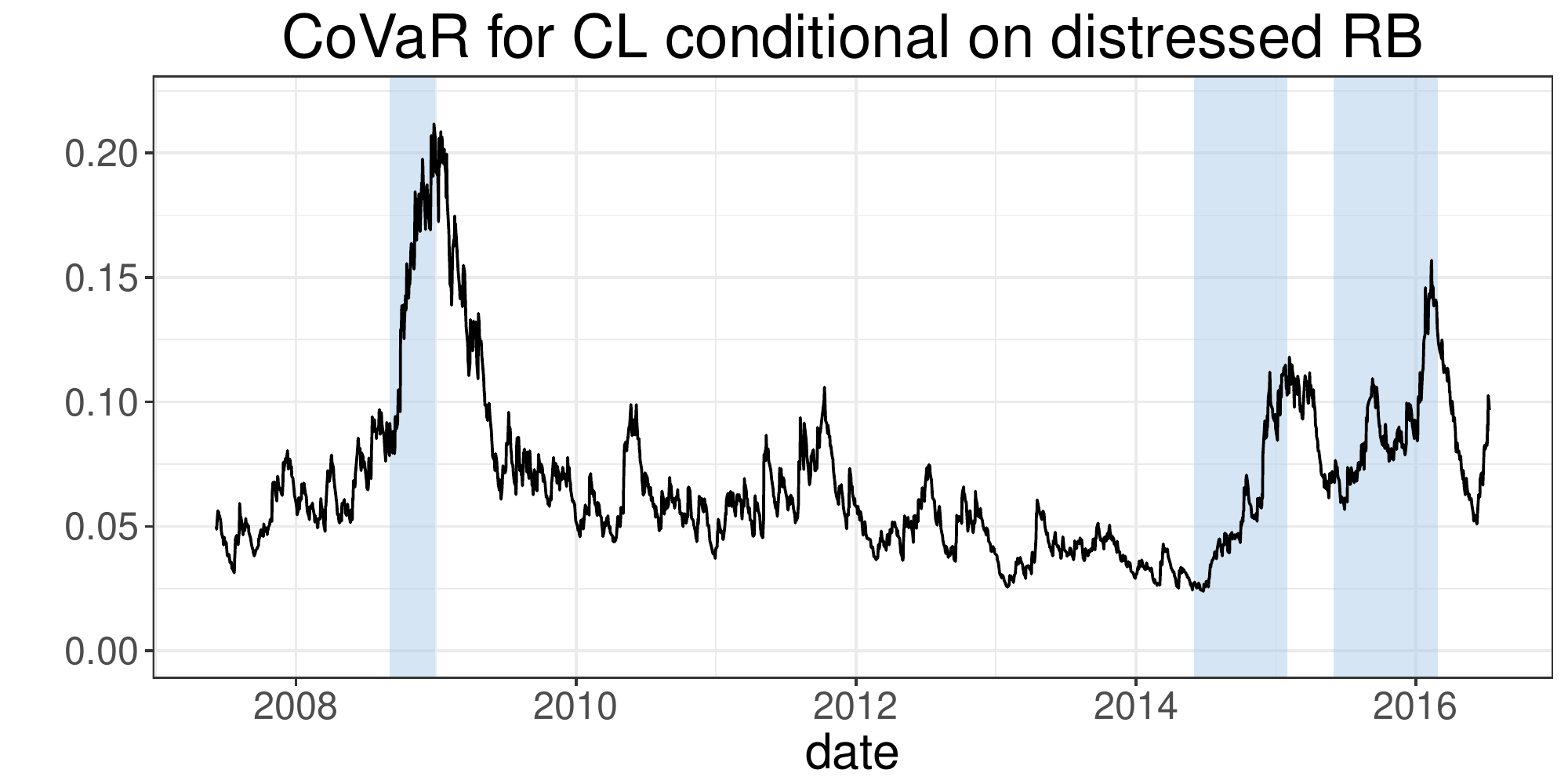}\quad
		\includegraphics[width=0.48\textwidth]{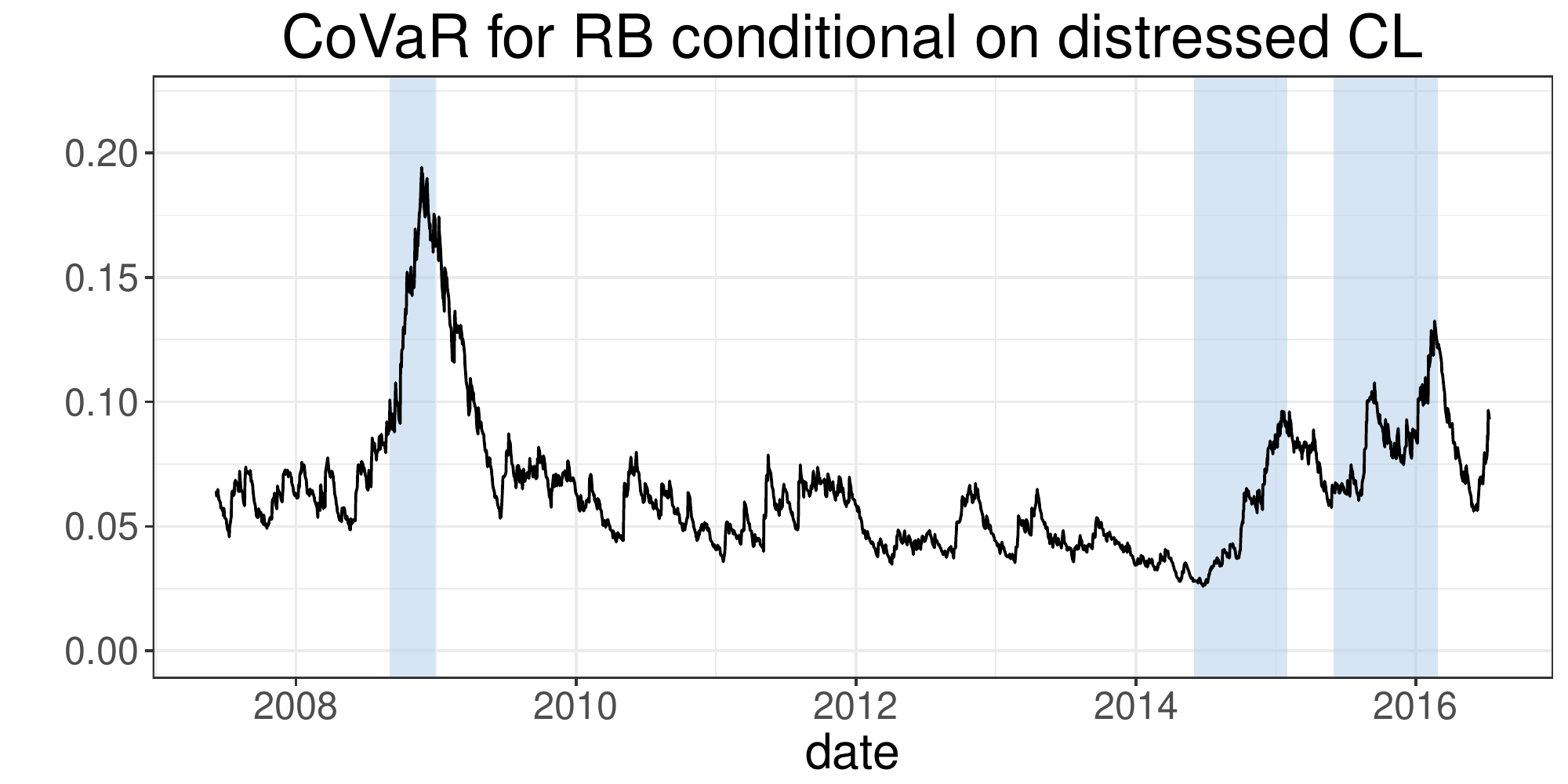}
	\end{center}
	\begin{center}
		\includegraphics[width=0.48\textwidth]{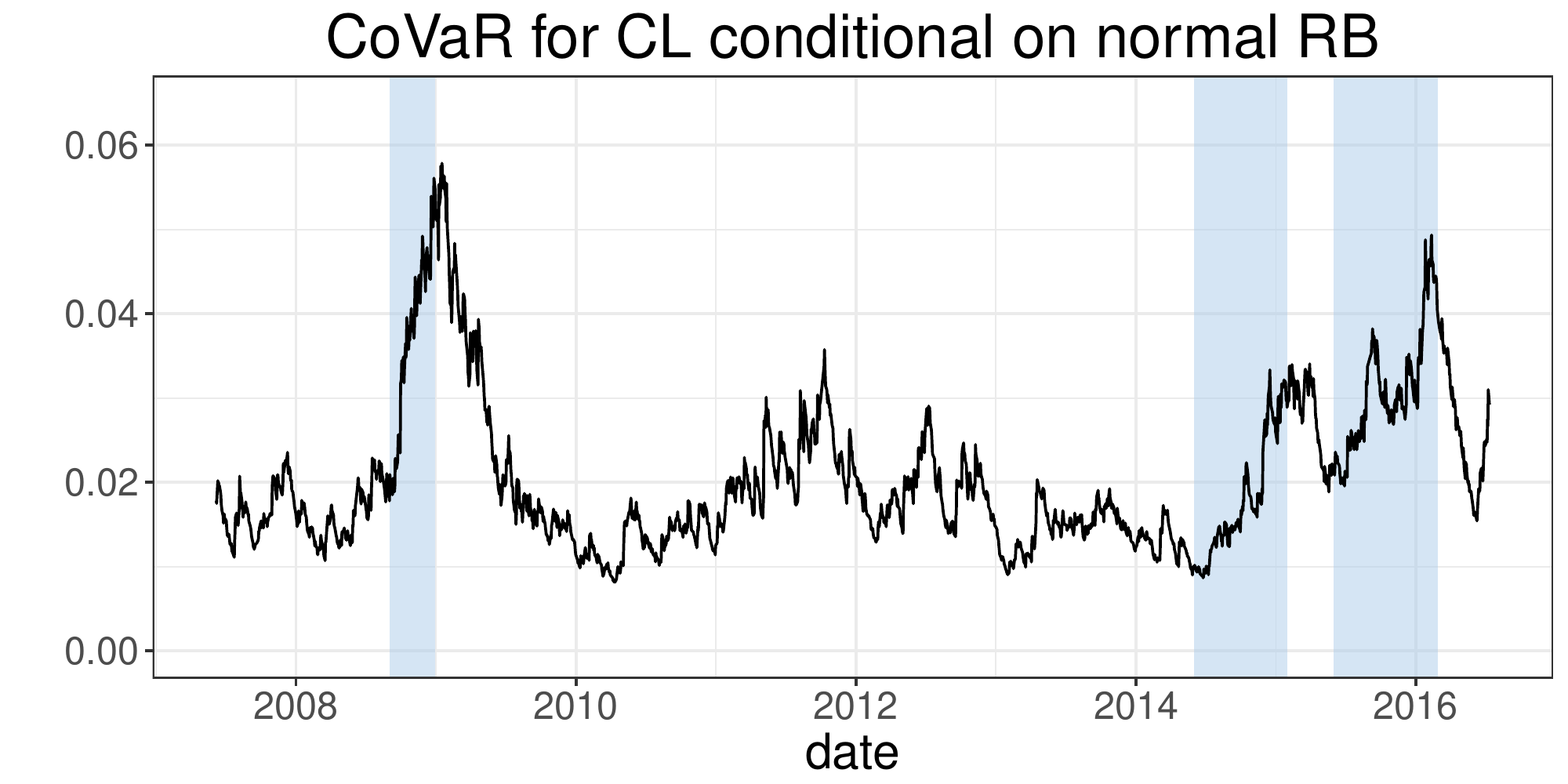}\quad
		\includegraphics[width=0.48\textwidth]{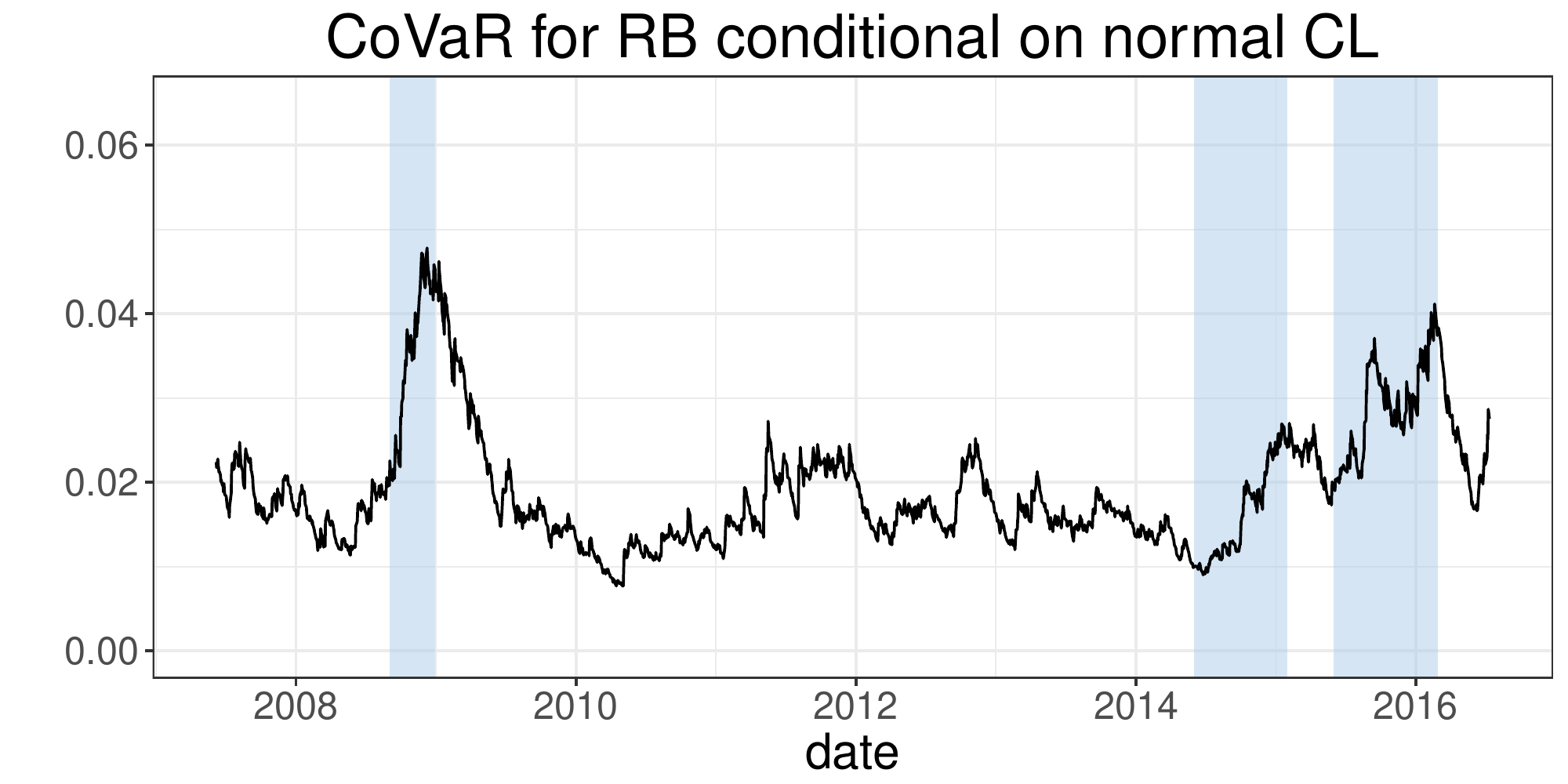}
	\end{center}
	\caption{Illustration of time-varying one-day $\mathrm{CoVaR}^{1|2}_{95\%, t}$ (top, left), $\Delta\mathrm{CoVaR}^{2|1}_{95\%, t}$ (top, right), 
		$\mathrm{CoVaR}^{1|2, \alpha=50\%}_{95\%, t}$ (bottom, left), and $\Delta\mathrm{CoVaR}^{2|1, \alpha=50\%}_{95\%, t}$ (bottom, right)
		where $R_t^1$ and $R_t^2$ are given by the CL and RB daily returns at time $t$, respectively, from January 2007 to December 2016}
	\label{Fig:CoVaR}
\end{figure}

\end{document}